\documentclass[lettersize,journal]{IEEEtran}
\usepackage{amsmath,amsfonts}
\usepackage{algorithmic}
\usepackage{algorithm}
\usepackage{array}
\usepackage[caption=false,font=normalsize,labelfont=sf,textfont=sf]{subfig}
\usepackage{textcomp}
\usepackage{stfloats}
\usepackage{url}
\usepackage{verbatim}
\usepackage{graphicx}
\usepackage{booktabs}
\usepackage{tabularx}
\usepackage{cite}
\usepackage{comment}
\usepackage{multirow}
\usepackage{makecell}
\global\long\def\FuncIndicator#1{\iota_{#1}}

\global\long\def\BallFidel#1{B_{2, \varepsilon}^{#1}}

\global\long\def\BallSparse{B_{1, \eta}}

\global\long\def\ParamStepsize#1{\gamma_{#1}}

\global\long\def\Z{\mathbf{Z}}

\global\long\def\D{\mathbf{D}}
\global\long\def\T{\mathbf{T}}

\global\long\def\z{\mathbf{z}}
\global\long\def\x{\mathbf{x}}
\global\long\def\y{\mathbf{y}}
\global\long\def\U{\mathbf{U}}
\global\long\def\V{\mathbf{V}}

\global\long\def\mode{\boldsymbol{\phi}}
\global\long\def\dynamics{\lambda}

\global\long\def\dataovs{\widetilde{\mathbf{x}}}
\global\long\def\Dataclean{\mathbf{X}}
\global\long\def\dataclean{\mathbf{x}}

\global\long\def\outliers{\mathbf{s}}

\global\long\def\gaussnoise{\mathbf{n}}
\global\long\def\Amplitude{\mathbf{\boldsymbol{\xi}}}
\global\long\def\Mode{\mathbf{\Phi}}
\global\long\def\Dynamics{\mathbf{C}}
\global\long\def\Balance{\mathbf{\boldsymbol{\nu}}}
\global\long\def\balance{\nu}
\global\long\def\Importance{p}
\global\long\def\IndexAlg{t}
\global\long\def\Realnumber#1{\in\mathbb{R}^{#1}}

\global\long\def\A{\mathbf{A}}
\global\long\def\T{\mathbf{T}}
\global\long\def\X{\mathbf{X}}

\global\long\def\argmin#1{\underset{#1}{\operatorname{argmin}}}

\DeclareMathOperator{\prox}{prox}
\DeclareMathOperator{\diag}{diag}

\DeclareMathOperator{\Real}{Re}
\DeclareMathOperator{\Imag}{Im}

\begin{document}
\bstctlcite{IEEEexample:BSTcontrol}

\title{Comprehensive Robust Dynamic Mode Decomposition from Mode Extraction to Dimensional Reduction}

\author{Yuki~Nakamura,
        Shingo~Takemoto,~\IEEEmembership{Student~Member,~IEEE,},
        and Shunsuke~Ono,~\IEEEmembership{Senior~Member,~IEEE,}
\thanks{Y. Nakamura, S. Takemoto, and S. Ono are with the Department of Computer Science, Institute of Science Tokyo, Yokohama, Kanagawa, 226-8501, Japan (e-mail: nakamura.y.47af@m.isct.ac.jp, takemoto.s.e908@m.isct.ac.jp, ono@comp.isct.ac.jp).}
\thanks{This work was supported in part by JST FOREST under Grant JPMJFR232M and JST AdCORP under Grant JPMJKB2307, and in part by JSPS KAKENHI under Grant 22H03610, 22H00512, 23H01415, 23K17461, 24K03119, 24K22291, 25H01296, and 25K03136, and in part by Grant-in-Aid for JSPS Fellows Grant Number JP24KJ1068.}}

\markboth{Journal of \LaTeX\ Class Files,~Vol.~14, No.~8, August~2021}
{Shell \MakeLowercase{\textit{et al.}}: A Sample Article Using IEEEtran.cls for IEEE Journals}

\IEEEpubid{0000--0000/00\$00.00~\copyright~2021 IEEE}

\maketitle

\begin{abstract}
  We propose Comprehensive Robust Dynamic Mode Decomposition (CR-DMD), a novel framework that robustifies the entire DMD process---from mode extraction to dimensional reduction---against mixed noise. Although standard DMD widely used for uncovering spatio-temporal patterns and constructing low-dimensional models of dynamical systems, it suffers from significant performance degradation under noise due to its reliance on least-squares estimation for computing the linear time evolution operator. Existing robust variants typically modify the least-squares formulation, but they remain unstable and fail to ensure faithful low-dimensional representations. First, we introduce a convex optimization-based preprocessing method designed to effectively remove mixed noise, achieving accurate and stable mode extraction. Second, we propose a new convex formulation for dimensional reduction that explicitly links the robustly extracted modes to the original noisy observations, constructing a faithful representation of the original data via a sparse weighted sum of the modes. Both stages are efficiently solved by a preconditioned primal-dual splitting method. Experiments on fluid dynamics datasets demonstrate that CR-DMD consistently outperforms state-of-the-art robust DMD methods in terms of mode accuracy and fidelity of low-dimensional representations under noisy conditions.
\end{abstract}


\section{Introduction}
\label{sec:introduction}
\IEEEPARstart{A}{nalyzing} high-dimensional signals is fundamental for uncovering latent structures and motion patterns in complex temporal data. Such analyses are applied to diverse tasks, including background-foreground separation\cite{markowitz2022multimodal, wang2023double} and Robust time-series forecasting\cite{shu2025guaranteed} to anomaly detection in time-varying graphs\cite{wang2013locality, buciulea2025polynomial}.
In this context, Dynamic Mode Decomposition (DMD)\cite{schmid2010dynamic,rowley2009spectral,tu2014dynamic,schmid2022dynamic,kutz2016dynamic} has emerged as a powerful technique for extracting characteristic patterns and their underlying dynamics from complex signals.
Initially introduced in fluid mechanics\cite{rowley2009spectral},
DMD is now widely applied in various fields, such as robotics\cite{abraham2019active,berger2015estimation}, power grids\cite{susuki2011nonlinear}, neuroscience\cite{brunton2016extracting}, finance\cite{mann2016dynamic}, plasma physics\cite{taylor2018dynamic,kaptanoglu2020characterizing}, and climate science\cite{kutz2016multiresolution}.

The core idea of DMD is to approximate the evolution of a dynamical system with a linear time-evolution operator \(\A\).
This operator maps each snapshot \(\x_t\) to the next snapshot \(\x_{t+1}\) as follows:
\begin{align}
  \x_{t+1} \approx \A \x_t.
\end{align}
The optimal operator \(\A\) is estimated by solving the following least-squares problem with 
the data matrices \(\X = (\x_1, \ldots, \x_{M-1})\) and \(\X^{'} = (\x_2, \ldots, \x_{M})\):
\begin{align}
  \label{eq:linear_operator}
  \min_{\A} \left\| \X^{'} - \A \X \right\|_F^2.
\end{align}
By analyzing the spectral properties of the operator \(\A\), DMD can achieve two main objectives:
mode extraction, which involves extracting characteristic spatio-temporal patterns (modes) and their temporal dynamics, 
and dimensional reduction, which involves obtaining a low-dimensional representation of the original high-dimensional data.
Specifically, the eigenvectors and eigenvalues of \(\A\) correspond to the modes of the system and their temporal dynamics.
A low-dimensional representation is then constructed by weighted summation of these modes,
where the amplitudes are calculated as these weights by fitting these modes to the data.
Dimensional reduction is achieved by selecting a subset of these modes based on their calculated amplitudes.
\IEEEpubidadjcol

Despite its utility, standard DMD relies on least-squares estimation, making it highly sensitive to noise in the data\cite{duke2012error,bagheri2014effects,zhang2020evaluating,lu2020prediction}.
In real-world applications, measurement errors and impulsive outliers from sensor failures or missing values \cite{maceas2021methodology,sciacchitano2019uncertainty} are unavoidable, 
and such contamination distorts the least-squares estimation of the operator \(\A\), leading to inaccurate analysis.
To address this, recent robust methods\cite{dawson2016characterizing,hemati2017biasing,askham2018variable,azencot2019consistent,scherl2020robust,sashidhar2022bagging,baddoo2023physics}
have attempted to improve the robustness of the operator estimation process in Prob.~\eqref{eq:linear_operator}.
For instance, Consistent DMD (CDMD)\cite{azencot2019consistent} seeks a more dynamically consistent operator by introducing forward-backward projections
to accurately capture the bidirectional evolution of the system.
Physics-informed DMD (PiDMD)\cite{baddoo2023physics} incorporates known physical laws, such as conservation principles,
as constraints into Prob.~\eqref{eq:linear_operator} to ensure the resulting modes and dynamics are physically consistent.
By incorporating such prior knowledge, these methods improve robustness, particularly under Gaussian noise.

However, these operator-centric methods still suffer from three fundamental limitations.
First, they remain vulnerable to high-magnitude noise, such as impulsive outliers.
This vulnerability arises from the continued reliance on the least-squares framework,
which causes large-magnitude noise to dominate the objective function, skewing the estimation of the operator \(\A\).
Second, the resulting optimization problems are often non-convex\cite{azencot2019consistent,askham2018variable}.
This means that convergence to the global optimal solution is not guaranteed, leading to local minima and sensitivity to initialization.
As a result, the mode extraction process can be unstable, yielding inconsistent results.
Third, they fail to address the robustness of the subsequent dimensional reduction stage.
Even if the modes are accurately extracted, calculating their amplitudes by projecting them back onto the original noisy data
inevitably propagates errors into the final low-dimensional representation.

Based on the above discussion, a question arises:
{\em How can we design a DMD framework that achieves robustness throughout the entire pipeline---from mode extraction to dimensional reduction---under mixed-noise conditions?}

In this article, we propose a novel robust DMD framework, named Comprehensive Robust DMD (CR-DMD), 
to extract accurate modes and dynamics and achieve high-fidelity dimensional reduction,
even when the data is corrupted by a mixture of white Gaussian noise and outliers.
The main contributions of this work are threefold:
\begin{enumerate}
  \item {\em Robust and Stable Mode Extraction:}  We propose a novel preprocessing method that achieves stable and accurate mode extraction from data corrupted by a mixture of white Gaussian noise and outliers. By formulating the denoising task as a convex optimization problem, our approach guarantees convergence to a global optimal solution independent of initialization. This effectively overcomes the instability issues present in the existing non-convex operator-centric robust DMD methods, ensuring reliable mode extraction.
  \item {\em High-Fidelity Dimensional Reduction:} We formulate a novel convex optimization problem that constructs a high-fidelity low-dimensional representation of the true underlying data. This formulation explicitly couples the reliable modes obtained from our preprocessing to the original noisy observations, providing access to fine structures that the preprocessing stage may have filtered out. This approach achieves three important tasks simultaneously: robust amplitude estimation, sparse mode selection, and interpolation of structures lost during preprocessing.
  \item {\em Simplified Parameter Tuning and Efficient Algorithms:} Our formulation simplifies hyperparameter tuning by incorporating data fidelity and outlier characterization as constraints rather than penalty terms. This allows for independent setting of parameters based on the noise levels of Gaussian noise and outliers respectively. Furthermore, to solve the convex optimization problems, we develop efficient algorithms based on a Preconditioned Primal-Dual Splitting Method (P-PDS)\cite{pock2011diagonal,naganuma2023variable} that can automatically determine the stepsize parameters from the structure of the problem, ensuring the practical usability of our method.
\end{enumerate}

This paper is organized as follows. 
Section~\ref{sec:preliminaries} reviews the foundations for the proposed method, namely the DMD algorithm and the P-PDS optimization method.
Section~\ref{sec:proposed_method} details our proposed framework, Section~\ref{sec:experiments} demonstrates its
effectiveness through numerical experiments, and Section~\ref{sec:conclusion} concludes the paper.

\section{Preliminaries}
\label{sec:preliminaries}

\subsection{Notations and Definitions}
\label{subsec:notation}
In this paper, vectors and matrices are denoted by lowercase and uppercase bold letters, respectively (e.g., $\mathbf{x}$ and $\mathbf{X}$). 
We treat a time-series data with spatial resolution \(N_1 \times N_2\) and temporal length \(M\) as a matrix \(\X=(\x_1,\ldots,\x_M)\Realnumber{N\times M}(N=N_1N_2)\) or as a vector \(\x = (\x_1^\top, \ldots, \x_M^\top)^\top\Realnumber{NM}\),
where \(\x_t\Realnumber{N}\) is the vector representing the spatial data at time \(t\).
Here, the \(n\)-th pixel of \(\x_t\) is denoted by \(x_{n,t}\in\mathbb{R}\).
Let \(\Gamma_0(\mathbb{R}^{NM})\) be the set of all proper lower semicontinuous convex functions defined on \(\mathbb{R}^{NM}\).
The \(\ell_1\)-norm and \(\ell_2\)-norm of a vector \(\x\Realnumber{N}\) are defined as \(\|\x\|_1 := \sum_{i=1}^N |x_i|\) and \(\|\x\|_2 := \sqrt{\sum_{i=1}^N |x_i|^2}\), respectively, where \(x_i\) represents the \(i\)-th entry of \(\x\).
We also define the \(\ell_{1,2}\)-norm. For a vector \(\x\Realnumber{NM}\) that is partitioned into a set of non-overlapping groups \(\{\x^{(g)}\Realnumber{N_g}\}_{g=1}^G\), the \(\ell_{1,2}\)-norm is defined as the sum of the \(\ell_2\)-norms of each group:
\begin{align}
  \|\x\|_{1,2} := \sum_{g=1}^G \|\x^{(g)}\|_2.
\end{align}
The Frobenius norm of a matrix $\mathbf{X}$ is defined as $\|\mathbf{X}\|_F := \sqrt{\sum_{i=1}^N\sum_{j=1}^M |x_{i,j}|^2}$.
For a complex scalar \(z\in\mathbb{C}\), vector \(\z\in\mathbb{C}^N\), and matrix \(\Z\in\mathbb{C}^{N\times M}\), the operators \(\Real(\cdot)\) and \(\Imag(\cdot)\) 
denote the real and imaginary parts of each element, respectively.
The absolute value of a complex scalar is defined as \(|z| := \sqrt{\Real(z)^2 + \Imag(z)^2}\).
The norms of complex vectors and matrices are defined in the same manner as those of real vectors and matrices.
For a vector \(\x\Realnumber{NM}\), let \(\D_v\Realnumber{NM\times NM}\), \(\D_h\Realnumber{NM\times NM}\), and \(\D_t\Realnumber{NM\times NM}\)
be the forward difference operators along the vertical, horizontal, and temporal directions, respectively, with the Neumann boundary conditions.
The \(\ell_p\)-norm ball \((p=1,2)\) with center \(\mathbf{c}\) and radius \(\varepsilon\) is denoted by
\begin{align}
  B_{p,\varepsilon}^{\mathbf{c}} := \{\x \in \mathbb{R}^{NM} | \|\x - \mathbf{c}\|_p \leq \varepsilon\}.
\end{align}
The indicator function of a nonempty closed convex set \(C\subset{\mathbb{R}^NM}\), denoted by \(\iota_C\), is defined as
\begin{align}
  \iota_C(\x) := \begin{cases}
    0, & \text{if } \x \in C, \\
    +\infty, & \text{otherwise}.
  \end{cases}
\end{align}

\subsection{Dynamic Mode decomposition (DMD)}
\label{subsec:dmd}
DMD\cite{schmid2010dynamic} is a data-driven technique that analyzes a time-series of data to achieve two 
main objectives:
\begin{enumerate}
  \item \textbf{Mode Extraction}: To extract spatio-temporal patterns (modes) and their temporal dynamics. 
  \item \textbf{Dimensional Reduction}: To obtain a low-dimensional representation of the original high-dimensional data.
\end{enumerate}

To achieve these objectives, DMD is based on the key idea of approximating the evolution of the system with a single linear operator \(\A\).
This is formulated by assuming a linear mapping from a matrix of snapshots \(\X=(\x_1, \ldots, \x_{M-1})\) to the next snapshot matrix \(\X'=(\x_2, \ldots, \x_{M})\) as follows:
\begin{align}
  \X' \approx \A \X.
\end{align}
The first objective, mode extraction, is achieved by analyzing the spectral properties of this operator:
the eigenvalues and eigenvectors of \(\A\) correspond to the temporal dynamics and spatial modes, respectively.

In practice, since the operator \(\A\in\mathbb{R}^{N\times N}\) is often too large to compute directly, 
its spectral properties are typically computed in a reduced subspace using the Singular Value Decomposition (SVD) of the data matrix \(\X\).
Let us now outline the practical procedure for mode extraction.

First, we compute the truncated SVD of the data matrix \(\X\) as follows:
\begin{align}
\X = \U \boldsymbol{\Sigma} \V^*,
\label{eq:mode_extraction_1}
\end{align}
where \( \U \in \mathbb{R}^{N \times r} \), \( \boldsymbol{\Sigma} \in \mathbb{R}^{r \times r} \), and \( \V^* \in \mathbb{R}^{r \times M} \)
are the truncated left singular vectors, singular values, and right singular vectors, respectively.
Here, the truncation rank \( r \) is chosen manually based on the number of significant modes or the energy content of the singular values.

Using the SVD components, we project the operator \( \A \) onto the reduced subspace defined by \( \U \) as follows:
\begin{align}
\widetilde{\A} := \U^* \A \U = \U^* \X' \V \boldsymbol{\Sigma}^{-1}\Realnumber{r \times r}.
\label{eq:mode_extraction_2}
\end{align}
The matrix \( \widetilde{\A} \) captures the dynamics in the reduced subspace.
Its eigenvalues \( \dynamics_i \) directly provide the dynamics, and the original DMD modes \( \mode_i \)
are obtained by projecting its eigenvectors \( \mathbf{w}_i \) back into the full space:
\begin{align}
\mode_i = \X^{'} \V \boldsymbol{\Sigma}^{-1} \mathbf{w}_i.
\label{eq:mode_extraction_3}
\end{align}

Now that the modes and dynamics have been extracted, we can address the second objective of dimensional reduction.
This involves representing the original data \( \X \) using a small number of the extracted modes. 
The key lies in determining the amplitude of each mode. This is typically done by fitting the full decomposition model to the data as follows:
\begin{align}
  \label{eq:dmd_full}
  \X & \approx 
  \begingroup
  \setlength{\arraycolsep}{1pt}
  \begin{pmatrix}
    \mode_1 \cdots \mode_{N}
  \end{pmatrix}
    \begin{pmatrix}
    \Amplitude_1 & & \\
    & \ddots & \\
    & & \Amplitude_N
  \end{pmatrix}
  \begin{pmatrix}
    1 & \dynamics_1 & \cdots & \dynamics_1^{M-1} \\
    \vdots & \vdots & \ddots & \vdots \\
    1 & \dynamics_N & \cdots & \dynamics_N^{M-1}
  \end{pmatrix}
  \endgroup \nonumber \\
  & = \Mode \diag(\Amplitude) \Dynamics,
\end{align}
where \(\mode_i\), \(\Amplitude_i\), and \(\dynamics_i\) are the \(i\)-th spatial mode, its amplitude, and its temporal dynamics, respectively.
Dimensional reduction is then typically accomplished by selecting only the most significant modes, 
which can be done by truncating modes based on their amplitudes or 
by directly promoting sparsity in the amplitude vector \(\Amplitude\).

This fitting is generally formulated as the following least-square problem:
\begin{figure*}[tp]
  \centering
    \begin{minipage}{0.93\hsize}
      \centering
      \includegraphics[width=0.85\textwidth]{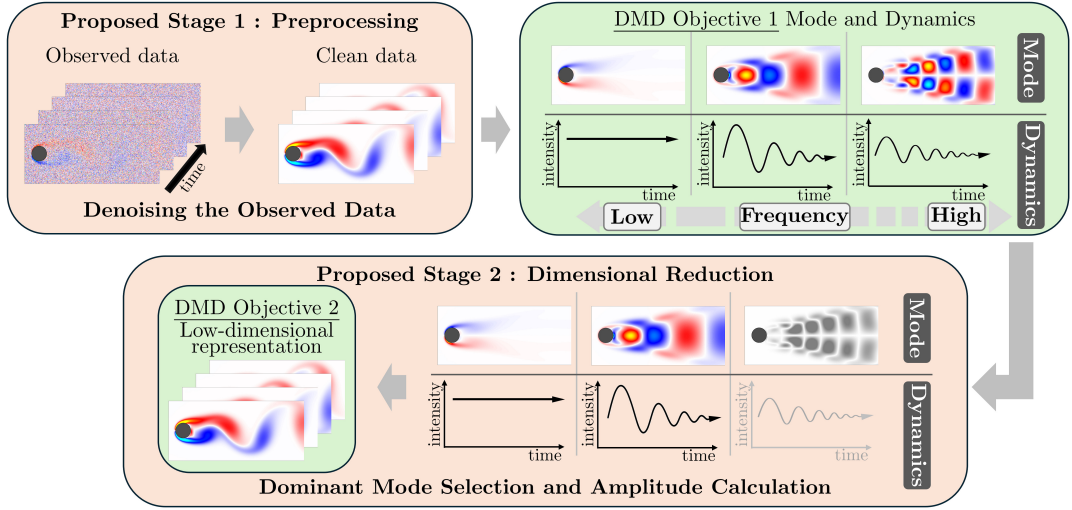}
    \end{minipage}
  \caption{Schematic illustration of the proposed CR-DMD framework.}
  \label{fig:flowchart}
\end{figure*}
\begin{align}
  \label{eq:standard_dmd}
  \min_{\Amplitude} \left\| \Dataclean - \Mode \diag(\Amplitude) \Dynamics \right\|_F^2.
\end{align}
The solution to this problem quantifies the contribution of each mode to the dataset, providing a basis for selecting the dominant modes for dimensional reduction.

\subsection{Proximal Tools}
\label{subsec:prox}
The optimization problem formulated in our proposed method involves nonsmooth convex functions.
To solve this problem, we utilize the proximity operator for a function \(f \in \Gamma_0(\mathbb{R}^N) \) 
and a parameter \(\gamma > 0\) as follows:
\begin{align}
  \prox_{\gamma f}(\x) := \argmin{\y\Realnumber{N}} \: f(\y) + \frac{1}{2\gamma} \|\x - \y\|_2^2.
\end{align}

The Fenchel-Rockafellar conjugate \(f^*\) of a function \(f \in \Gamma_0(\mathbb{R}^N)\) is defined as:
\begin{align}
  f^*(\y) := \sup_{\x\Realnumber{N}} \: \left( \langle \x, \y \rangle - f(\x) \right).
\end{align}
The proximity operator of the conjugate function \(f^*\) is given by the Moreau's identity\cite{combettes2013moreau}, as follows:
\begin{align}
  \prox_{\gamma f^*}(\x) = \x - \gamma \prox_{\frac{1}{\gamma} f}\left(\frac{1}{\gamma}\x\right).
\end{align}

Below, we summarize the proximity operators of the functions used in our proposed method.
The proximity operator of the \(\ell_{1,2}\)-norm is given group-wise for each vector segment \(\x^{(g)}\) as follows:
\begin{align}
  \left[\prox_{\gamma \|\cdot\|_{1,2}}(\x)\right]^{(g)} = \max\left(1 - \frac{\gamma}{\|\x^{(g)}\|_2}, 0\right) \x^{(g)},
\end{align}
where \(g:=\{1,\ldots,G\}\) denotes the group index.
The proximity operator of indicator function of the \(\ell_2\)-norm ball and the \(\ell_1\)-norm ball are given by
\begin{align}
  \prox_{\gamma \FuncIndicator{\BallFidel{\mathbf{c}}}}(\x) = 
  \begin{cases}
    \x, & \text{if}\:\x\in\BallFidel{\mathbf{c}}, \\
    \x + \frac{\varepsilon(\x-\mathbf{c})}{\|\x - \mathbf{c}\|_2}, & \text{otherwise},
  \end{cases}
\end{align}
and a fast \(\ell_1\)-ball projection algorithm\cite{condat2016fast}, respectively.

\subsection{Preconditioned Primal-Dual Splitting Method (P-PDS)}
\label{subsec:PPDS}
The standard PDS\cite{chambolle2011first,condat2013primal} and P-PDS\cite{pock2011diagonal}, on which our algorithm is based,
are efficient algorithms for solving the following generic form of nonsmooth convex optimization problems:
\begin{align}
  \min_{\substack{\y_1,\ldots,\y_N \\ \z_1,\ldots,\z_M}} \: & \sum_{i=1}^N g_i(\y_i) + \sum_{j=1}^M h_j(\z_j) \nonumber \\
  \mathrm{s.t.} \: & 
  \begin{cases}
    \z_1 = \sum_{i=1}^N \mathbf{L}_{1,i} \y_i, \\
    \vdots \\
    \z_M = \sum_{i=1}^N \mathbf{L}_{M,i} \y_i,
  \end{cases}
  \label{eq:PPDS_problem}
\end{align}
where \(g_i \in \Gamma_0(\mathbb{R}^{n_i})(i=1,\ldots,N)\), \(h_j \in \Gamma_0(\mathbb{R}^{m_j})(j=1,\ldots,M)\), 
\(\y_i \in \mathbb{R}^{n_i}(i=1,\ldots,N)\) are primal variables, \(\z_j \in \mathbb{R}^{m_j}(j=1,\ldots,M)\) are dual variables,
and \(\mathbf{L}_{j,i}\in\mathbb{R}^{m_j \times n_i} (i=1,\ldots,N, j=1,\ldots,M)\) are linear operators.
P-PDS solves Prob.~\eqref{eq:PPDS_problem} by iterating the following steps:
\begin{align}
  \left\lfloor\begin{array}{l}
    \y_1^{(t+1)} \leftarrow \prox_{\ParamStepsize{\y_1} g_1} \left( \y_1^{(t)} - \ParamStepsize{\y_1} \sum_{j=1}^M \mathbf{L}_{j,i}^\top \z_j^{(t)} \right), \\
    \vdots \\
    \y_N^{(t+1)} \leftarrow \prox_{\ParamStepsize{\y_N} g_N} \left( \y_N^{(t)} - \ParamStepsize{\y_N} \sum_{j=1}^M \mathbf{L}_{j,N}^\top \z_j^{(t)} \right), \\
    \z_i^{'} = 2\y_i^{(t+1)} - \y_i^{(t)} (\forall i=1,\ldots,N), \\
    \z_1^{(t+1)} \leftarrow \prox_{\ParamStepsize{\z_1} h_1^*}\left(\z_1^{(t)} - \ParamStepsize{\z_1} (\sum_{i=1}^N \mathbf{L}_{1,i} \z_i^{'})\right), \\
    \vdots \\
    \z_M^{(t+1)} \leftarrow \prox_{\ParamStepsize{\z_M} h_M^*}\left(\z_M^{(t)} - \ParamStepsize{\z_M} (\sum_{i=1}^N \mathbf{L}_{M,i} \z_i^{'})\right),
  \end{array}\right. \nonumber
\end{align}
where \(\ParamStepsize{\y_i}(i=1,\ldots,N)\) and \(\ParamStepsize{\z_j}(j=1,\ldots,M)\) are the stepsize parameters.

The standard PDS needs to manually adjust the appropriate stepsize parameters to ensure convergence.
On the other hand, P-PDS can automatically determine the stepsize parameters from the structure of the problem\cite{pock2011diagonal,naganuma2023variable}.
According to \cite{naganuma2023variable}, the stepsize parameters are determined as follows:
\begin{align}
  \ParamStepsize{\y_i} = \frac{1}{\sum_{j=1}^M \|\mathbf{L}_{j,i}\|_{op}^2}, \quad \ParamStepsize{\z_j} = \frac{1}{N},
  \label{eq:stepsize}
\end{align}
where \(\|\cdot\|_{op}\) represents the operator norm defined by:
\begin{align}
  \|\mathbf{L}\|_{op} := \max_{\x \neq \mathbf{0}} \frac{\|\mathbf{L}\x\|_2}{\|\x\|_2}.
\end{align}

\section{Proposed Method}
\label{sec:proposed_method}
In this section, we propose a novel robust DMD framework, named Comprehensive Robust DMD (CR-DMD),
to achieve accurate mode extraction and high-fidelity dimensional reduction, even when an observed data is corrupted by
a mixture of white Gaussian noise and outliers.
The overall process is illustrated in Fig.~\ref{fig:flowchart},
where our proposed method consists of the following two stages:
\begin{itemize}
  \item {\em Preprocessing:} Before mode extraction, our preprocessing method effectively removes both white Gaussian noise and outliers from the observed data, achieving a stable and robust mode extraction.
  \item {\em Dimensional Reduction:} After extracting the modes from the clean data obtained in the preprocessing stage, our dimensional reduction method determines the amplitudes for the extracted modes to obtain a high-fidelity low-dimensional representation of the underlying clean data.
\end{itemize}

\subsection{Preprocessing}
\label{subsec:preprocessing}
\subsubsection{Problem Formulation}
The first stage of our CR-DMD framework is a preprocessing step designed to achieve stable and robust mode extraction.
As discussed in Sec.~\ref{sec:introduction}, existing approaches that directly robustify the operator estimation process
remain vulnerable to high-magnitude noise. To overcome this limitation, our strategy is to remove the noise 
before the core DMD algorithm is applied, which we formulate as a convex optimization problem.

Consider that an observed data \(\dataovs\Realnumber{NM}\) contaminated by mixed noise is modeled by
\begin{align}
    \dataovs = \dataclean + \outliers + \gaussnoise,
    \label{eq:observation_prepro}
\end{align}
where \(\dataclean, \outliers\), and \(\gaussnoise\) represent a clean data, sparse noise such as outliers and missing values, and random noise, respectively.

Based on the above observation model, we formulate a preprocessing mixed noise removal problem as the following convex optimization problem:
\begin{equation}
    \label{eq:erpgn}
    \min_{\dataclean,\outliers\Realnumber{NM}} \|\D_w \dataclean\|_{1,2}, \quad \mathrm{s.t.} \: 
    \begin{cases}
        \dataclean + \outliers \in \BallFidel{\dataovs}, \\
        \outliers \in \BallSparse, \\
    \end{cases}
\end{equation}
where 
\begin{align}
  \D_w\dataclean &:= \begin{pmatrix}
    w\D_v\dataclean \\
    w\D_h\dataclean \\
    (1-w)\D_t\dataclean
  \end{pmatrix}.
\end{align}

The first term is a Total Variation regularization term that defined by grouping the vertical, horizontal, and temporal 
directional differences for each element of \(\dataclean\) using \(\ell_2\)-norm and 
then aggregates them across all elements by \(\ell_1\)-norm.
The balancing parameter \(w\in[0,1]\) controls the relative weight between the spatial and temporal differences.
In contrast to the conventional TV\cite{rudin1992nonlinear} designed for natural images, 
this formulation also incorporates temporal differences, capturing spatial piecewise smoothness and 
temporal continuity of the data.

The first constraint serves as data-fidelity with the \(\dataovs\)-centered \(\ell_2\)-norm ball of the radius \(\varepsilon > 0\).
The second constraint characterizes sparse noise with the zero-centered \(\ell_1\)-norm ball of the radius \(\eta > 0\).
Since we impose these constraints instead of adding terms to the objective function, the hyperparameters associated with each term are independent of one another.
Consequently, appropriate parameters can be set based solely on the noise levels of white Gaussian noise and outliers, respectively.

Such advantage has been addressed, e.g., in \cite{afonso2010augmented, chierchia2015epigraphical, ono2015signal, ono2017primal, ono2019efficient}.

\begin{figure}[!t]
	\vspace{-3mm}
	\begin{algorithm}[H]
	    \caption{
				\begin{tabular}[t]{@{}l@{}}
					Solver for \eqref{eq:erpgn} in Preprocessing
				\end{tabular}
			}
		\label{algo_ERPGN}
		\begin{algorithmic}[1]
			\renewcommand{\algorithmicrequire}{\textbf{Input:}}
			\renewcommand{\algorithmicensure}{\textbf{Output:}}
			\REQUIRE $\dataclean^{(0)}, \outliers^{(0)}, \z_1^{(0)}, \z_2^{(0)}$
			\ENSURE $\dataclean^{(\IndexAlg)}$
			\WHILE {A stopping criterion is not satisfied}
                    \STATE $\dataclean^{(\IndexAlg+1)} \leftarrow \dataclean^{(\IndexAlg)}-\ParamStepsize{\dataclean}(\D_w^\top\z_1^{(\IndexAlg)}+\z_2^{(\IndexAlg)});$
                    \STATE $\outliers^{(\IndexAlg+1)} \leftarrow \prox_{\ParamStepsize{\outliers}\FuncIndicator\BallSparse}(\outliers^{(\IndexAlg)}-\ParamStepsize{\outliers}\z_2^{(\IndexAlg)});$ \label{alg:preprocessing_3}
										\STATE $\dataclean^{'} \leftarrow 2\dataclean^{(\IndexAlg+1)} - \dataclean^{(\IndexAlg)};$
										\STATE $\outliers^{'} \leftarrow 2\outliers^{(\IndexAlg+1)} - \outliers^{(\IndexAlg)};$
										\STATE $\z_1^{'} \leftarrow \z_1^{(\IndexAlg)} + \ParamStepsize{\z_1}\D_w\dataclean^{'};$
										\STATE $\z_1^{(\IndexAlg+1)} \leftarrow \z_1^{'} - \ParamStepsize{\z_1}\prox_{\frac{1}{\ParamStepsize{\z_1}}\|\cdot\|_{1,2}}\left(\frac{1}{\ParamStepsize{\z_1}}\z_1^{'}\right);$ \label{alg:preprocessing_5}
										\STATE $\z_2^{'} \leftarrow \z_2^{(\IndexAlg)} + \ParamStepsize{\z_2}\left(\D_w\dataclean^{'}+\outliers^{'}\right);$
                    \STATE $\z_2^{(\IndexAlg+1)} \leftarrow \z_2^{'} - \ParamStepsize{\z_2}\prox_{\ParamStepsize{\z_2}\FuncIndicator{\BallFidel{\dataovs}}}\left(\frac{1}{\ParamStepsize{\z_2}}\z_2^{'}\right);$ \label{alg:preprocessing_7}
    			\STATE $\IndexAlg \leftarrow \IndexAlg + 1;$
			\ENDWHILE
		\end{algorithmic}
	\end{algorithm}
	\vspace{-8mm}
\end{figure}
\subsubsection{Optimization Algorithm}
To solve Prob.~\eqref{eq:erpgn} by an efficient algorithm based on P-PDS\cite{pock2011diagonal},
we need to reformulate it into the P-PDS applicable form in \eqref{eq:PPDS_problem}.

Using the indicator functions of \(\BallSparse\) and \(\BallFidel{\dataovs}\), Prob.~\eqref{eq:erpgn} can be equivalently transformed as follows:
\begin{align}
    \label{eq:erpgn2}
    \min_{\dataclean,\outliers,\z_1,\z_2} \: &\|\z_1\|_{1,2} + \FuncIndicator{\BallSparse}(\outliers) + \FuncIndicator{\BallFidel{\dataovs}} (\z_2), \nonumber \\
    & \mathrm{s.t.} \:
    \begin{cases}
        \z_1 = \D_w \dataclean, \\
        \z_2 = \dataclean + \outliers. \\
    \end{cases}
\end{align}
Let \(\dataclean,\outliers\) be the primal variables and \(\z_1,\z_2\) be the dual variables.
The operator \(\D_w\) is linear operator.
The indicator functions \(\FuncIndicator{\BallSparse}\) and \(\FuncIndicator{\BallFidel{\dataovs}}\) and the \(\ell_{1,2}\)-norm are proper lower semicontinuous convex functions.
Then, by defining \(g_1(\dataclean) := 0,\: g_2(\outliers) := 0,\: h_1(\z_1) := \|\z_1\|_{1,2}, \: h_2(\z_2) := \FuncIndicator{\BallFidel{\dataovs}} (\z_2),\)

Prob.~\eqref{eq:erpgn2} can be expressed in the form of Prob.~\eqref{eq:PPDS_problem}.
Therefore, P-PDS can be applied to solve Prob.~\eqref{eq:erpgn2}.

We show the detailed algorithm in Alg.~\ref{algo_ERPGN}. 
The proximity operators in Steps \ref{alg:preprocessing_3}, \ref{alg:preprocessing_5}, and \ref{alg:preprocessing_7} are detailed in Sec.~\ref{subsec:prox}.

Based on Eq.~\eqref{eq:stepsize}, the stepsize parameters \(\ParamStepsize{\dataclean}, \ParamStepsize{\outliers}, \ParamStepsize{\z_1}\), and \(\ParamStepsize{\z_2}\) are given by
\begin{align}
    \ParamStepsize{\dataclean} &= \frac{1}{1 + 8w^2 + 4(1-w)^2}, \nonumber \\
    \ParamStepsize{\outliers} &= 1, \ParamStepsize{\z_1} = \ParamStepsize{\z_2} = \frac{1}{2}.
\end{align}

\subsection{Dimensional Reduction}
\label{subsec:SpDMD-RP}
\subsubsection{Problem Formulation}
Through the preprocessing stage, we obtain a clean data \(\dataclean\) that effectively removes mixed noise from the observed data.
Then, we apply the standard DMD algorithm in \eqref{eq:mode_extraction_1}-\eqref{eq:mode_extraction_3} to \(\dataclean\) in order to extract the modes \(\Mode\) and dynamics \(\Dynamics\).

The next crucial stage is dimensional reduction, which determines the amplitudes of these extracted modes to obtain a low-dimensional representation of the underlying clean data.
A key challenge in this stage is which data should be used for reference when estimating the amplitudes.
Using the clean data \(\dataclean\) obtained from the preprocessing stage as a reference may lead to the loss of fine structures, since the regularization in the preprocessing stage inevitably causes over-smoothing of sharp gradients and detailed features.
To address this, we develop a dimensional reduction method that refers back to the original noisy observations.

Consider the observation model as follows:
\begin{align}
  \dataovs = \T\Amplitude + \outliers + \gaussnoise,
  \label{eq:observation_rec}
\end{align}
where \(\T:=\Dynamics^\top \odot \Mode\in\mathbb{C}^{NM \times r}\) is the linear operator to construct a low-dimensional representation from the amplitudes \(\Amplitude\in\mathbb{C}^r\).
Here, \(\odot\) denotes the Khatri-Rao product, which is defined as the column-wise Kronecker product.
The term \(\T\Amplitude\) represents an equivalent form of the matrix decomposition \(\Mode \diag(\Amplitude) \Dynamics\) in \eqref{eq:dmd_full}.

Based on the above observation model, we formulate our dimensional reduction method as the following convex optimization problem:
\begin{align}
  \label{eq:dr_problem}
  \min_{\Amplitude \in \mathbb{C}^r, \outliers \in \mathbb{C}^{NM}} \: & \|\D_w\T\Amplitude\|_{1,2} + \mu \|\Balance \circ \Amplitude\|_1, \nonumber \\ & \mathrm{s.t.} \: 
  \begin{cases}
      \T\Amplitude + \outliers \in \BallFidel{\dataovs}, \\
      \outliers \in \BallSparse. \\
  \end{cases}
\end{align}

The first term is the TV regularization term that promotes the spatio-temporal smoothness of the dimensionally reduced data, as detailed in the preprocessing section (Sec.~\ref{subsec:preprocessing}).
The second term promotes a sparse selection of dominant modes, controlled by the parameter \(\mu > 0\).
Here, \(\circ\) denotes the Hadamard product, and \(\Balance\in\mathbb{R}^r\) is a weighting vector that controls the optimization of the amplitudes
to prioritize modes that are expected to be more significant.
According to the criterion proposed by Kou et al.\cite{kou2017improved}, the importance of the \(i\)-th mode is defined as
\begin{align}
  \Importance_i := \sum_{j=1}^{M} |\Amplitude_{pre,i}| \|\Mode_i\|_2 |\dynamics_i|^{j-1},
\end{align}
where \(\Mode_i\) and \(\dynamics_i\) are the \(i\)-th mode and its dynamics and \(\Amplitude_{pre,i}\) is its amplitude calculated by solving Prob.~\eqref{eq:standard_dmd}, respectively.
This importance quantifies the overall contribution of the \(i\)-th mode to the entire dataset by summing its magnitude across all time steps.
Then, the \(i\)-th element of the weighting vector \(\Balance\) is defined as the inverse of this importance:

\begin{align}
  \balance_i := 
  \begin{cases}
    \frac{\Importance_i^{-1}}{\sum_{j=1}^{r} \Importance_j^{-1}}, & \text{if } \Importance_i \neq 0, \\
    1, & \text{otherwise}.
  \end{cases}
\end{align}

The constraints in Prob.~\eqref{eq:dr_problem} link the low-dimensional representation to the original noisy observations.
This simultaneously achieves robust amplitude estimation, sparse mode selection,
and recovery of structures lost during preprocessing.

\begin{figure}[!t]
	\vspace{-3mm}
	\begin{algorithm}[H]
		\caption{
			\begin{tabular}[t]{@{}l@{}}
				Solver for \eqref{eq:erpgn} in Dimensional Reduction
			\end{tabular}
		}
		\label{algo_dimensional_reduction}
		% \vspace{-1mm}
		\begin{algorithmic}[1]
			\renewcommand{\algorithmicrequire}{\textbf{Input:}}
			\renewcommand{\algorithmicensure}{\textbf{Output:}}
			\REQUIRE $\Amplitude_{\mathbb{R}}^{(0)}, \outliers^{(0)}, \z_1^{(0)}, \z_2^{(0)}$
			\ENSURE $\Amplitude_{\mathbb{R}}^{(\IndexAlg)}$
			\WHILE {A stopping criterion is not satisfied}
										\STATE $\hat{\Amplitude}_{\mathbb{R}} \leftarrow \Amplitude_{\mathbb{R}}^{(\IndexAlg)} - \ParamStepsize{\Amplitude_{\mathbb{R}}}\left((\D_w\T_{\mathbb{R}})^\top\z_1^{(\IndexAlg)}+\T_{\mathbb{R}}^{\top}\z_2^{(\IndexAlg)}\right);$
                    \STATE $\Amplitude_{\mathbb{R}}^{(\IndexAlg+1)} \leftarrow \prox_{\ParamStepsize{\Amplitude_{\mathbb{R}}}\mu\|\underline{\Balance} \circ 	\cdot\|_{1,2}}(\hat{\Amplitude}_{\mathbb{R}});$ \label{alg:dimensional_reduction_2}
                    \STATE $\outliers^{(\IndexAlg+1)} \leftarrow \prox_{\ParamStepsize{\outliers}\FuncIndicator\BallSparse}\left(\outliers^{(\IndexAlg)} - \ParamStepsize{\outliers}\z_2^{(\IndexAlg)}\right);$ \label{alg:dimensional_reduction_3}
                    \STATE $\Amplitude_{\mathbb{R}}^{'} \leftarrow 2\Amplitude_{\mathbb{R}}^{(\IndexAlg+1)} - \Amplitude_{\mathbb{R}}^{(\IndexAlg)};$
                    \STATE $\outliers^{'} \leftarrow 2\outliers^{(\IndexAlg+1)} - \outliers^{(\IndexAlg)};$
                    \STATE $\z_1^{'} \leftarrow \z_1^{(\IndexAlg)}+\ParamStepsize{\z_1}\D_w\T_{\mathbb{R}}\Amplitude_{\mathbb{R}}^{'};$
                    \STATE $\z_1^{(\IndexAlg+1)} \leftarrow \z_1^{'} - \ParamStepsize{\z_1}\prox_{\frac{1}{\ParamStepsize{\z_1}}\|\cdot\|_{1,2}}\left(\frac{1}{\ParamStepsize{\z_1}}\z_1^{'}\right);$ \label{alg:dimensional_reduction_8}
                    \STATE $\z_2^{'} \leftarrow \z_2^{(\IndexAlg)}+\ParamStepsize{\z_2}\left(\T_{\mathbb{R}}\Amplitude_{\mathbb{R}}^{'}+\outliers^{'}\right);$
                    \STATE $\z_2^{(\IndexAlg+1)} \leftarrow \z_2^{'} - \prox_{\ParamStepsize{\z_2}\FuncIndicator{\BallFidel{\dataovs}}}\left(\frac{1}{\ParamStepsize{\z_2}}\z_2^{'}\right);$ \label{alg:dimensional_reduction_10}
    			\STATE $\IndexAlg \leftarrow \IndexAlg + 1$;
			\ENDWHILE
		\end{algorithmic}
	\end{algorithm}
	\vspace{-7mm}
\end{figure}
\subsubsection{Optimization Algorithm}
To solve Prob.~\eqref{eq:dr_problem} by an efficient algorithm based on P-PDS\cite{pock2011diagonal},
we need to reformulate it into the P-PDS applicable form \eqref{eq:PPDS_problem}.
Using the indicator functions of \(\BallSparse\) and \(\BallFidel{\dataovs}\),
we rewrite Prob.~\eqref{eq:dr_problem} into an equivalent form:
\begin{align}
  \label{eq:SpDMD-RP2}
  \min_{\Amplitude_{\mathbb{R}},\outliers,\z_1,\z_2} \: & \|\z_1\|_{1,2} + \mu\|\underline{\Balance} \circ \Amplitude_{\mathbb{R}}\|_{1,2} + \FuncIndicator{\BallSparse}(\outliers) + \FuncIndicator{\BallFidel{\dataovs}}(\z_2), \nonumber \\
  & \mathrm{s.t.} \:
  \begin{cases}
      \z_1 = \D_w \T_{\mathbb{R}} \Amplitude_{\mathbb{R}}, \\
      \z_2 = \T_{\mathbb{R}}\Amplitude_{\mathbb{R}} + \outliers, \\
  \end{cases}
\end{align}
where \(\T\in\mathbb{C}^{NM \times r}\) and \(\Amplitude\in\mathbb{C}^r\) in Prob.~\eqref{eq:dr_problem} are transformed into their real-valued forms as follows:
\begin{align}
  \T_{\mathbb{R}} &:= \begin{pmatrix}
    \Real(\T) & -\Imag(\T)
  \end{pmatrix}, \\
  \Amplitude_{\mathbb{R}} &:= \begin{pmatrix}
    \Real(\Amplitude) \\
    \Imag(\Amplitude)
  \end{pmatrix},
\end{align}
and the corresponding weighting vector \(\Balance\in\mathbb{R}^r\) is transformed as follows:
\begin{align}
  \underline{\Balance} &:= \begin{pmatrix}
    \Balance \\
    \Balance
  \end{pmatrix}.
\end{align}
Since the underlying data is assumed to be real-valued, the DMD modes and eigenvalues exhibit conjugate symmetry.
This property allows us the expanded forms of \(\T_{\mathbb{R}}\) and \(\Amplitude_{\mathbb{R}}\) without any loss of information.
Following this transformation, the \(\ell_1\)-norm on \(\Amplitude\) is replaced by the \(\ell_{1,2}\)-norm on \(\Amplitude_{\mathbb{R}}\).
This norm is calculated by treating the real and imaginary parts of each element as a group.

Let \(\Amplitude_{\mathbb{R}},\outliers\) be the primal variables and \(\z_1,\z_2\) be the dual variables.
The operator \(\D_w \T_{\mathbb{R}}\) is linear operator.
The indicator functions \(\FuncIndicator{\BallSparse}\) and \(\FuncIndicator{\BallFidel{\dataovs}}\), the \(\ell_{1,2}\)-norm, and the weighted \(\ell_1\)-norm are proper lower semicontinuous convex functions.
Then, by defining \(g_1(\Amplitude_{\mathbb{R}}) := 0,\: g_2(\outliers) := 0,\: h_1(\z_1) := \|\z_1\|_{1,2}, \: h_2(\z_2) := \FuncIndicator{\BallFidel{\dataovs}} (\z_2),\)

Prob.~\eqref{eq:SpDMD-RP2} can be expressed in the form of Prob.~\eqref{eq:PPDS_problem}.
Therefore, P-PDS can be applied to solve Prob.~\eqref{eq:SpDMD-RP2}.

We show the detailed algorithm in Alg.~\ref{algo_dimensional_reduction}.
The proximity operators of \(\|\underline{\Balance} \circ \cdot\|_{1,2}\) in Step \ref{alg:dimensional_reduction_2} is calculated by
\begin{align}
  \left[\prox_{\gamma \mu \|\underline{\Balance} \circ \cdot\|_{1,2}}(\x)\right]^{(g)} = \max\left(1 - \frac{\gamma \mu \underline{\balance}_g}{\|\x^{(g)}\|_2}, 0\right) \x^{(g)},
\end{align}
where the operation is applied to each group \(g=1,\ldots,r\).
The proximity operators in Steps \ref{alg:dimensional_reduction_3}, \ref{alg:dimensional_reduction_8}, and \ref{alg:dimensional_reduction_10} are detailed in Sec.~\ref{subsec:prox}.

Based on Eq.~\eqref{eq:stepsize}, the stepsize parameters \(\ParamStepsize{\Amplitude}, \ParamStepsize{\outliers}, \ParamStepsize{\z_1}\), and \(\ParamStepsize{\z_2}\) are given by
\begin{align}
  \ParamStepsize{\Amplitude} &= \frac{1}{\sigma_1(\Dynamics)^2\sigma_1(\Mode)^2(1 + 8w^2 + 4(1-w)^2)}, \nonumber \\
  \ParamStepsize{\outliers} &= 1, \ParamStepsize{\z_1} = \ParamStepsize{\z_2} = \frac{1}{2},
\end{align}
where \(\sigma_1(\cdot)\) is the maximum singular value of a matrix.

\subsection{Comparison with Existing Methods}
\label{subsec:comparison}
\subsubsection{Reasonable Preprocessing}
The distinct feature of CR-DMD is the introduction of an efficient preprocessing stage.
Conventional operator-centric methods maintain the least-squares framework, which risks instability in the presence of strong outliers.
In contrast, our method stabilizes the estimation of Prob.~\eqref{eq:linear_operator} via standard DMD by providing the clean data, recovered through the explicit separation of outliers and Gaussian noise as shown in Sec.~\ref{subsec:preprocessing}.
Furthermore, unlike the existing RPCA-based preprocessing method\cite{scherl2020robust}, which relies on nuclear-norm regularization, we employ TV-based regularization as formulated in Prob.~\eqref{eq:erpgn}, which avoids expensive SVD computations and reduces excessive smoothing of dynamic structures.

\subsubsection{Observation-referencing Dimensional Reduction}
Another distinct feature of our approach is the formulation that directly references the original noisy observations \(\dataovs\) in the dimensional reduction stage.
Conventional amplitude estimation, as formulated in Prob.~\eqref{eq:standard_dmd}, typically faces a critical trade-off depending on the data used for fitting.
First, fitting extracted modes directly to the original noisy data causes the estimated amplitudes to be significantly distorted by noise.
Second, fitting only to preprocessed clean data risks losing fine-scale structural details, since the denoising process—even with TV-based regularization—inevitably introduces some level of smoothing.
To address this, we integrate the robustly extracted modes directly into the original observation model in \eqref{eq:observation_rec}, and transform the task into the dimensional reduction problem in \eqref{eq:dr_problem}.
This strategy combines the stability of the robust modes with the rich structural information from raw data, effectively recovering the detailed structures lost during preprocessing and yielding a more accurate low-dimensional model.

\begin{table*}[t]
  \centering
  \scriptsize
  \renewcommand{\arraystretch}{0.9}
  \caption{MSEs of Estimated Dynamics from Noisy Fluid Flow Data}
  \label{table:mse_eigen}
  \setlength{\tabcolsep}{2pt}
  \begin{tabularx}{\linewidth}{c c c *{6}{>{\centering\arraybackslash}X}}
    \toprule
    Data & Noise & Mode & TLS-DMD\cite{hemati2017biasing} & OptDMD\cite{askham2018variable} & CDMD\cite{azencot2019consistent} & PiDMD\cite{baddoo2023physics} & RPCA\cite{scherl2020robust} & CR-DMD (Ours) \\
    \cmidrule(lr){1-9}
    \multirow{18}{*}{Cylinder Wake}
      & \multirow{6}{*}{\shortstack{Low Noise\\$\left(\makecell{\sigma=0.05 \\ \eta=0.05}\right)$}}
      & Mode 1 & $1.12 \times 10^{-6}$ & $1.91 \times 10^{-4}$ & $3.94 \times 10^{-6}$ & $3.13 \times 10^{-6}$ & $\mathbf{1.74 \times 10^{-8}}$ & \underline{$1.02 \times 10^{-6}$} \\
      & & Mode 2 & $1.14 \times 10^{-5}$ & $3.02 \times 10^{-5}$ & $4.10 \times 10^{-6}$ & \underline{$2.32 \times 10^{-6}$} & $\mathbf{4.70 \times 10^{-7}}$ & $4.53 \times 10^{-6}$ \\
      & & Mode 3 & $5.28 \times 10^{-4}$ & $1.18 \times 10^{-3}$ & $3.28 \times 10^{-4}$ & $4.16 \times 10^{-4}$ & \underline{$1.32 \times 10^{-4}$} & $\mathbf{1.15 \times 10^{-5}}$ \\
      & & Mode 4 & $2.18 \times 10^{-3}$ & $6.13 \times 10^{-3}$ & $1.83 \times 10^{-3}$ & $1.89 \times 10^{-3}$ & \underline{$7.27 \times 10^{-4}$} & $\mathbf{7.91 \times 10^{-5}}$ \\
      & & Mode 5 & $4.43 \times 10^{-2}$ & \underline{$1.95 \times 10^{-2}$} & $4.35 \times 10^{-2}$ & $3.96 \times 10^{-2}$ & $6.84 \times 10^{-1}$ & $\mathbf{3.15 \times 10^{-4}}$ \\
      & & Mode 6 & $4.40 \times 10^{-1}$ & \underline{$7.99 \times 10^{-2}$} & $4.34 \times 10^{-1}$ & $4.45 \times 10^{-1}$ & $4.04 \times 10^{-1}$ & $\mathbf{7.03 \times 10^{-4}}$ \\
    \cmidrule(lr){2-9}
      & \multirow{6}{*}{\shortstack{Medium Noise\\$\left(\makecell{\sigma=0.1 \\ \eta=0.1}\right)$}}
      & Mode 1 & $5.73 \times 10^{-6}$ & $7.72 \times 10^{-5}$ & $8.92 \times 10^{-6}$ & $8.51 \times 10^{-6}$ & \underline{$2.05 \times 10^{-7}$} & $\mathbf{2.93 \times 10^{-8}}$ \\
      & & Mode 2 & $3.96 \times 10^{-5}$ & $1.93 \times 10^{-4}$ & $2.51 \times 10^{-5}$ & \underline{$2.26 \times 10^{-5}$} & $2.22 \times 10^{-5}$ & $\mathbf{1.35 \times 10^{-5}}$ \\
      & & Mode 3 & $8.35 \times 10^{-3}$ & \underline{$3.54 \times 10^{-3}$} & $5.66 \times 10^{-3}$ & $7.57 \times 10^{-3}$ & $2.61 \times 10^{-2}$ & $\mathbf{1.61 \times 10^{-4}}$ \\
      & & Mode 4 & $2.73 \times 10^{-2}$ & \underline{$6.40 \times 10^{-3}$} & $1.86 \times 10^{-2}$ & $2.50 \times 10^{-2}$ & $3.81 \times 10^{-1}$ & $\mathbf{3.97 \times 10^{-3}}$ \\
      & & Mode 5 & $1.02 \times 10^{-1}$ & $\mathbf{1.50 \times 10^{-2}}$ & \underline{$8.49 \times 10^{-2}$} & $1.02 \times 10^{-1}$ & $4.61 \times 10^{-1}$ & $5.78 \times 10^{-1}$ \\
      & & Mode 6 & $5.42 \times 10^{-1}$ & $\mathbf{4.67 \times 10^{-2}}$ & \underline{$5.37 \times 10^{-1}$} & $5.57 \times 10^{-1}$ & $6.94 \times 10^{-1}$ & $6.19 \times 10^{-1}$ \\
    \cmidrule(lr){2-9}
      & \multirow{6}{*}{\shortstack{High Noise\\$\left(\makecell{\sigma=0.15 \\ \eta=0.15}\right)$}}
      & Mode 1 & $1.38 \times 10^{-5}$ & $3.95 \times 10^{-4}$ & $1.63 \times 10^{-5}$ & $1.52 \times 10^{-5}$ & \underline{$1.55 \times 10^{-6}$} & $\mathbf{1.11 \times 10^{-7}}$ \\
      & & Mode 2 & $9.72 \times 10^{-5}$ & $4.47 \times 10^{-4}$ & $1.04 \times 10^{-4}$ & \underline{$8.42 \times 10^{-5}$} & $4.76 \times 10^{-4}$ & $\mathbf{8.59 \times 10^{-6}}$ \\
      & & Mode 3 & $1.67 \times 10^{-2}$ & \underline{$5.68 \times 10^{-3}$} & $1.55 \times 10^{-2}$ & $1.42 \times 10^{-2}$ & $4.68 \times 10^{-1}$ & $\mathbf{5.28 \times 10^{-4}}$ \\
      & & Mode 4 & $4.62 \times 10^{-2}$ & $\mathbf{7.76 \times 10^{-3}}$ & $4.84 \times 10^{-2}$ & \underline{$4.17 \times 10^{-2}$} & $4.47 \times 10^{-1}$ & $1.12 \times 10^{-1}$ \\
      & & Mode 5 & $1.23 \times 10^{-1}$ & $\mathbf{1.74 \times 10^{-2}}$ & \underline{$1.11 \times 10^{-1}$} & $1.18 \times 10^{-1}$ & $5.95 \times 10^{-1}$ & $5.01 \times 10^{-1}$ \\
      & & Mode 6 & \underline{$5.76 \times 10^{-1}$} & $\mathbf{3.20 \times 10^{-2}}$ & $5.93 \times 10^{-1}$ & $5.83 \times 10^{-1}$ & $6.71 \times 10^{-1}$ & $6.28 \times 10^{-1}$ \\
    \midrule
    \multirow{18}{*}{Channel Flow}
      & \multirow{6}{*}{\shortstack{Low Noise\\$\left(\makecell{\sigma=0.05 \\ \eta=0.05}\right)$}}
      & Mode 1 & $9.61 \times 10^{-9}$ & $8.90 \times 10^{-5}$ & $1.18 \times 10^{-7}$ & $1.95 \times 10^{-8}$ & \underline{$1.88 \times 10^{-9}$} & $\mathbf{1.57 \times 10^{-10}}$ \\
      & & Mode 2 & $\mathbf{4.37 \times 10^{-5}}$ & $5.80 \times 10^{-4}$ & $8.09 \times 10^{-5}$ & $7.29 \times 10^{-5}$ & $1.44 \times 10^{-4}$ & \underline{$4.87 \times 10^{-5}$} \\
      & & Mode 3 & $7.95 \times 10^{-5}$ & $1.32 \times 10^{-3}$ & $6.19 \times 10^{-5}$ & $5.34 \times 10^{-5}$ & \underline{$4.78 \times 10^{-5}$} & $\mathbf{4.77 \times 10^{-5}}$ \\
      & & Mode 4 & $8.08 \times 10^{-5}$ & $2.45 \times 10^{-3}$ & $9.11 \times 10^{-5}$ & \underline{$7.38 \times 10^{-5}$} & $\mathbf{6.57 \times 10^{-5}}$ & $3.43 \times 10^{-4}$ \\
      & & Mode 5 & \underline{$8.79 \times 10^{-5}$} & $3.78 \times 10^{-3}$ & $1.29 \times 10^{-4}$ & $9.42 \times 10^{-5}$ & $\mathbf{6.06 \times 10^{-5}}$ & $1.85 \times 10^{-4}$ \\
      & & Mode 6 & $1.45 \times 10^{-4}$ & $4.90 \times 10^{-3}$ & \underline{$1.34 \times 10^{-4}$} & $1.44 \times 10^{-4}$ & $\mathbf{9.44 \times 10^{-5}}$ & $2.90 \times 10^{-4}$ \\
    \cmidrule(lr){2-9}
      & \multirow{6}{*}{\shortstack{Medium Noise\\$\left(\makecell{\sigma=0.1 \\ \eta=0.1}\right)$}}
      & Mode 1 & $8.17 \times 10^{-9}$ & $1.15 \times 10^{-4}$ & $8.03 \times 10^{-8}$ & $1.76 \times 10^{-8}$ & \underline{$1.98 \times 10^{-9}$} & $\mathbf{4.93 \times 10^{-10}}$ \\
      & & Mode 2 & $\mathbf{4.19 \times 10^{-5}}$ & $5.38 \times 10^{-4}$ & $7.95 \times 10^{-5}$ & $7.06 \times 10^{-5}$ & $7.16 \times 10^{-5}$ & \underline{$5.64 \times 10^{-5}$} \\
      & & Mode 3 & $6.16 \times 10^{-5}$ & $1.63 \times 10^{-3}$ & $7.67 \times 10^{-5}$ & $\mathbf{5.45 \times 10^{-5}}$ & $6.30 \times 10^{-5}$ & \underline{$5.62 \times 10^{-5}$} \\
      & & Mode 4 & $1.70 \times 10^{-3}$ & $3.21 \times 10^{-3}$ & $1.05 \times 10^{-3}$ & $1.01 \times 10^{-3}$ & $\mathbf{6.21 \times 10^{-4}}$ & \underline{$7.29 \times 10^{-4}$} \\
      & & Mode 5 & $1.17 \times 10^{-3}$ & $4.71 \times 10^{-3}$ & $\mathbf{9.58 \times 10^{-4}}$ & \underline{$1.01 \times 10^{-3}$} & $2.73 \times 10^{-3}$ & $1.12 \times 10^{-3}$ \\
      & & Mode 6 & $7.84 \times 10^{-4}$ & $8.56 \times 10^{-3}$ & $\mathbf{5.45 \times 10^{-4}}$ & \underline{$6.17 \times 10^{-4}$} & $3.21 \times 10^{-2}$ & $1.47 \times 10^{-3}$ \\
    \cmidrule(lr){2-9}
      & \multirow{6}{*}{\shortstack{High Noise\\$\left(\makecell{\sigma=0.15 \\ \eta=0.15}\right)$}}
      & Mode 1 & $2.17 \times 10^{-8}$ & $2.45 \times 10^{-4}$ & $1.11 \times 10^{-7}$ & $1.20 \times 10^{-8}$ & \underline{$2.04 \times 10^{-9}$} & $\mathbf{1.19 \times 10^{-10}}$ \\
      & & Mode 2 & $\mathbf{4.23 \times 10^{-5}}$ & $7.31 \times 10^{-4}$ & $8.51 \times 10^{-5}$ & $7.29 \times 10^{-5}$ & $7.93 \times 10^{-5}$ & \underline{$6.38 \times 10^{-5}$} \\
      & & Mode 3 & $5.84 \times 10^{-4}$ & $1.38 \times 10^{-3}$ & $4.51 \times 10^{-4}$ & \underline{$3.15 \times 10^{-4}$} & $4.20 \times 10^{-4}$ & $\mathbf{6.57 \times 10^{-5}}$ \\
      & & Mode 4 & $1.10 \times 10^{-2}$ & \underline{$2.90 \times 10^{-3}$} & $9.27 \times 10^{-3}$ & $9.79 \times 10^{-3}$ & $6.34 \times 10^{-3}$ & $\mathbf{3.71 \times 10^{-4}}$ \\
      & & Mode 5 & $2.61 \times 10^{-3}$ & $5.84 \times 10^{-3}$ & $2.39 \times 10^{-3}$ & \underline{$2.37 \times 10^{-3}$} & $2.69 \times 10^{-2}$ & $\mathbf{2.07 \times 10^{-4}}$ \\
      & & Mode 6 & $1.19 \times 10^{-3}$ & $5.61 \times 10^{-3}$ & \underline{$1.03 \times 10^{-3}$} & $\mathbf{1.01 \times 10^{-3}}$ & $1.07 \times 10^{-1}$ & $1.44 \times 10^{-2}$ \\
    \bottomrule
  \end{tabularx}
\end{table*}
\begin{table*}[t]
  \centering
  \scriptsize
  \renewcommand{\arraystretch}{0.9}
  \caption{STDs of Estimated Dynamics from Noisy Fluid Flow Data}
  \label{table:std_eigen}
  \setlength{\tabcolsep}{2pt}
  \begin{tabularx}{\linewidth}{c c c *{6}{>{\centering\arraybackslash}X}}
    \toprule
    Data & Noise & Mode & TLS-DMD\cite{hemati2017biasing} & OptDMD\cite{askham2018variable} & CDMD\cite{azencot2019consistent} & PiDMD\cite{baddoo2023physics} & RPCA\cite{scherl2020robust} & CR-DMD (Ours) \\
    \cmidrule(lr){1-9}
    \multirow{18}{*}{Cylinder Wake}
      & \multirow{6}{*}{\shortstack{Low Noise\\$\left(\makecell{\sigma=0.05 \\ \eta=0.05}\right)$}}
      & Mode 1 & \underline{$6.17 \times 10^{-4}$} & $1.34 \times 10^{-2}$ & $1.87 \times 10^{-3}$ & $1.50 \times 10^{-3}$ & $\mathbf{9.76 \times 10^{-5}}$ & $1.01 \times 10^{-3}$ \\
      & & Mode 2 & $1.43 \times 10^{-3}$ & $5.33 \times 10^{-3}$ & $2.02 \times 10^{-3}$ & $1.49 \times 10^{-3}$ & $\mathbf{2.95 \times 10^{-4}}$ & \underline{$9.42 \times 10^{-4}$} \\
      & & Mode 3 & $2.17 \times 10^{-2}$ & $2.85 \times 10^{-2}$ & $1.81 \times 10^{-2}$ & $2.04 \times 10^{-2}$ & \underline{$1.68 \times 10^{-3}$} & $\mathbf{5.89 \times 10^{-4}}$ \\
      & & Mode 4 & $4.43 \times 10^{-2}$ & $5.11 \times 10^{-2}$ & $4.26 \times 10^{-2}$ & $4.35 \times 10^{-2}$ & \underline{$3.73 \times 10^{-3}$} & $\mathbf{1.40 \times 10^{-3}}$ \\
      & & Mode 5 & $2.01 \times 10^{-1}$ & \underline{$7.29 \times 10^{-2}$} & $2.07 \times 10^{-1}$ & $1.89 \times 10^{-1}$ & $1.35 \times 10^{-1}$ & $\mathbf{2.79 \times 10^{-3}}$ \\
      & & Mode 6 & $6.13 \times 10^{-1}$ & \underline{$1.36 \times 10^{-1}$} & $6.14 \times 10^{-1}$ & $6.32 \times 10^{-1}$ & $1.55 \times 10^{-1}$ & $\mathbf{4.42 \times 10^{-3}}$ \\
    \cmidrule(lr){2-9}
      & \multirow{6}{*}{\shortstack{Medium Noise\\$\left(\makecell{\sigma=0.1 \\ \eta=0.1}\right)$}}
      & Mode 1 & $1.53 \times 10^{-3}$ & $8.65 \times 10^{-3}$ & $2.74 \times 10^{-3}$ & $2.41 \times 10^{-3}$ & \underline{$2.03 \times 10^{-4}$} & $\mathbf{1.27 \times 10^{-4}}$ \\
      & & Mode 2 & $4.90 \times 10^{-3}$ & $1.31 \times 10^{-2}$ & $5.02 \times 10^{-3}$ & $4.77 \times 10^{-3}$ & \underline{$9.98 \times 10^{-4}$} & $\mathbf{3.23 \times 10^{-4}}$ \\
      & & Mode 3 & $8.78 \times 10^{-2}$ & $4.36 \times 10^{-2}$ & $7.53 \times 10^{-2}$ & $8.55 \times 10^{-2}$ & \underline{$2.21 \times 10^{-2}$} & $\mathbf{1.69 \times 10^{-3}}$ \\
      & & Mode 4 & $1.48 \times 10^{-1}$ & \underline{$5.01 \times 10^{-2}$} & $1.24 \times 10^{-1}$ & $1.43 \times 10^{-1}$ & $2.33 \times 10^{-1}$ & $\mathbf{8.02 \times 10^{-3}}$ \\
      & & Mode 5 & $2.53 \times 10^{-1}$ & $\mathbf{6.49 \times 10^{-2}}$ & $2.02 \times 10^{-1}$ & $2.27 \times 10^{-1}$ & $2.37 \times 10^{-1}$ & \underline{$7.93 \times 10^{-2}$} \\
      & & Mode 6 & $6.33 \times 10^{-1}$ & $1.18 \times 10^{-1}$ & $6.55 \times 10^{-1}$ & $6.58 \times 10^{-1}$ & \underline{$1.01 \times 10^{-1}$} & $\mathbf{9.50 \times 10^{-2}}$ \\
    \cmidrule(lr){2-9}
      & \multirow{6}{*}{\shortstack{High Noise\\$\left(\makecell{\sigma=0.15 \\ \eta=0.15}\right)$}}
      & Mode 1 & $2.65 \times 10^{-3}$ & $1.84 \times 10^{-2}$ & $3.57 \times 10^{-3}$ & $3.29 \times 10^{-3}$ & \underline{$4.39 \times 10^{-4}$} & $\mathbf{2.77 \times 10^{-4}}$ \\
      & & Mode 2 & $8.45 \times 10^{-3}$ & $1.82 \times 10^{-2}$ & $1.01 \times 10^{-2}$ & $9.06 \times 10^{-3}$ & \underline{$3.15 \times 10^{-3}$} & $\mathbf{4.58 \times 10^{-4}}$ \\
      & & Mode 3 & $1.25 \times 10^{-1}$ & \underline{$5.10 \times 10^{-2}$} & $1.24 \times 10^{-1}$ & $1.19 \times 10^{-1}$ & $1.20 \times 10^{-1}$ & $\mathbf{3.31 \times 10^{-3}}$ \\
      & & Mode 4 & $1.80 \times 10^{-1}$ & $\mathbf{5.18 \times 10^{-2}}$ & $1.89 \times 10^{-1}$ & $1.79 \times 10^{-1}$ & \underline{$1.44 \times 10^{-1}$} & $1.93 \times 10^{-1}$ \\
      & & Mode 5 & $2.45 \times 10^{-1}$ & $\mathbf{6.59 \times 10^{-2}}$ & $1.93 \times 10^{-1}$ & $2.34 \times 10^{-1}$ & \underline{$1.18 \times 10^{-1}$} & $2.07 \times 10^{-1}$ \\
      & & Mode 6 & $6.60 \times 10^{-1}$ & $\mathbf{8.99 \times 10^{-2}}$ & $6.85 \times 10^{-1}$ & $6.69 \times 10^{-1}$ & $1.14 \times 10^{-1}$ & \underline{$9.56 \times 10^{-2}$} \\
    \midrule
    \multirow{18}{*}{Channel Flow}
      & \multirow{6}{*}{\shortstack{Low Noise\\$\left(\makecell{\sigma=0.05 \\ \eta=0.05}\right)$}}
      & Mode 1 & $9.78 \times 10^{-5}$ & $9.16 \times 10^{-3}$ & $3.42 \times 10^{-4}$ & $1.05 \times 10^{-4}$ & \underline{$3.39 \times 10^{-5}$} & $\mathbf{9.02 \times 10^{-6}}$ \\
      & & Mode 2 & $6.53 \times 10^{-3}$ & $1.63 \times 10^{-2}$ & $5.71 \times 10^{-3}$ & \underline{$4.64 \times 10^{-3}$} & $1.10 \times 10^{-2}$ & $\mathbf{2.98 \times 10^{-3}}$ \\
      & & Mode 3 & $8.96 \times 10^{-3}$ & $1.92 \times 10^{-2}$ & $5.86 \times 10^{-3}$ & $5.07 \times 10^{-3}$ & \underline{$4.84 \times 10^{-3}$} & $\mathbf{2.12 \times 10^{-3}}$ \\
      & & Mode 4 & $8.90 \times 10^{-3}$ & $2.55 \times 10^{-2}$ & $7.40 \times 10^{-3}$ & $\mathbf{5.95 \times 10^{-3}}$ & \underline{$6.92 \times 10^{-3}$} & $1.48 \times 10^{-2}$ \\
      & & Mode 5 & $8.84 \times 10^{-3}$ & $3.03 \times 10^{-2}$ & $8.56 \times 10^{-3}$ & $\mathbf{6.25 \times 10^{-3}}$ & \underline{$7.12 \times 10^{-3}$} & $1.17 \times 10^{-2}$ \\
      & & Mode 6 & $1.10 \times 10^{-2}$ & $3.45 \times 10^{-2}$ & \underline{$9.18 \times 10^{-3}$} & $9.60 \times 10^{-3}$ & $\mathbf{7.32 \times 10^{-3}}$ & $1.47 \times 10^{-2}$ \\
    \cmidrule(lr){2-9}
      & \multirow{6}{*}{\shortstack{Medium Noise\\$\left(\makecell{\sigma=0.1 \\ \eta=0.1}\right)$}}
      & Mode 1 & $8.90 \times 10^{-5}$ & $9.80 \times 10^{-3}$ & $2.81 \times 10^{-4}$ & $1.02 \times 10^{-4}$ & \underline{$4.01 \times 10^{-5}$} & $\mathbf{1.15 \times 10^{-5}}$ \\
      & & Mode 2 & $5.80 \times 10^{-3}$ & $1.23 \times 10^{-2}$ & $5.93 \times 10^{-3}$ & \underline{$4.35 \times 10^{-3}$} & $7.14 \times 10^{-3}$ & $\mathbf{3.37 \times 10^{-3}}$ \\
      & & Mode 3 & $7.51 \times 10^{-3}$ & $1.79 \times 10^{-2}$ & $7.21 \times 10^{-3}$ & \underline{$5.15 \times 10^{-3}$} & $5.16 \times 10^{-3}$ & $\mathbf{2.58 \times 10^{-3}}$ \\
      & & Mode 4 & $3.42 \times 10^{-2}$ & $2.99 \times 10^{-2}$ & $2.34 \times 10^{-2}$ & $\mathbf{2.16 \times 10^{-2}}$ & \underline{$2.16 \times 10^{-2}$} & $2.40 \times 10^{-2}$ \\
      & & Mode 5 & $2.19 \times 10^{-2}$ & $3.43 \times 10^{-2}$ & $\mathbf{2.00 \times 10^{-2}}$ & \underline{$2.04 \times 10^{-2}$} & $4.36 \times 10^{-2}$ & $3.16 \times 10^{-2}$ \\
      & & Mode 6 & $2.78 \times 10^{-2}$ & $4.13 \times 10^{-2}$ & $\mathbf{2.11 \times 10^{-2}}$ & \underline{$2.28 \times 10^{-2}$} & $1.40 \times 10^{-1}$ & $3.59 \times 10^{-2}$ \\
    \cmidrule(lr){2-9}
      & \multirow{6}{*}{\shortstack{High Noise\\$\left(\makecell{\sigma=0.15 \\ \eta=0.15}\right)$}}
      & Mode 1 & $1.43 \times 10^{-4}$ & $1.45 \times 10^{-2}$ & $3.22 \times 10^{-4}$ & $9.00 \times 10^{-5}$ & \underline{$4.22 \times 10^{-5}$} & $\mathbf{9.80 \times 10^{-6}}$ \\
      & & Mode 2 & $5.78 \times 10^{-3}$ & $1.94 \times 10^{-2}$ & $5.93 \times 10^{-3}$ & \underline{$4.64 \times 10^{-3}$} & $6.55 \times 10^{-3}$ & $\mathbf{4.21 \times 10^{-3}}$ \\
      & & Mode 3 & $1.93 \times 10^{-2}$ & $2.26 \times 10^{-2}$ & $1.65 \times 10^{-2}$ & $1.39 \times 10^{-2}$ & \underline{$1.37 \times 10^{-2}$} & $\mathbf{3.70 \times 10^{-3}}$ \\
      & & Mode 4 & $8.68 \times 10^{-2}$ & \underline{$3.14 \times 10^{-2}$} & $5.89 \times 10^{-2}$ & $7.73 \times 10^{-2}$ & $5.97 \times 10^{-2}$ & $\mathbf{1.57 \times 10^{-2}}$ \\
      & & Mode 5 & $2.36 \times 10^{-2}$ & $3.90 \times 10^{-2}$ & $2.15 \times 10^{-2}$ & \underline{$1.79 \times 10^{-2}$} & $1.08 \times 10^{-1}$ & $\mathbf{1.05 \times 10^{-2}}$ \\
      & & Mode 6 & $3.36 \times 10^{-2}$ & $3.75 \times 10^{-2}$ & $\mathbf{2.96 \times 10^{-2}}$ & \underline{$3.02 \times 10^{-2}$} & $1.64 \times 10^{-1}$ & $1.07 \times 10^{-1}$ \\
    \bottomrule
  \end{tabularx}
\end{table*}
\section{Experiments}
\label{sec:experiments}
To evaluate the performance of our proposed CR-DMD framework, we conducted experiments in two sequential stages:
mode extraction and dimensional reduction under various noise conditions.
We compared our method against state-of-the-art method: RPCA \cite{scherl2020robust}, 
TLS-DMD \cite{hemati2017biasing}, OptDMD \cite{askham2018variable}, CDMD \cite{azencot2019consistent}, 
and PiDMD \cite{baddoo2023physics} for mode extraction.
Subsequently, in the dimensional reduction stage, in addition to these methods, we further included standard DMD \cite{schmid2010dynamic}, DMD with criteria (DMDc) \cite{kou2017improved}, and SpDMD \cite{jovanovic2014sparsity} to specifically evaluate the effectiveness of the proposed amplitude estimation.
For each method, we used the publicly available implementations provided by the respective authors\footnote{Source codes for RPCA, TLS-DMD, OptDMD, CDMD, PiDMD, and SpDMD are available at \url{https://github.com/ischerl/RPCA-PIV}, \url{https://github.com/cwrowley/dmdtools}, \url{https://github.com/duqbo/optdmd}, \url{https://github.com/azencot/nld.cdmd}, \url{https://github.com/baddoo/piDMD}, and \url{http://www.umn.edu/~mihailo/software/dmdsp/}, respectively.}.

As ground-truth data, we adopted two fluid flow datasets with distinct dynamic characteristics, as summarized below:
\begin{enumerate}
    \item \textbf{Cylinder Wake}: A simulation of fluid flow past a cylinder at a Reynolds number\footnote{The Reynolds number \(Re\) is a dimensionless quantity describing the ratio of inertial forces to viscous forces in fluid dynamics.} of \(Re=100\) \cite{zhang2015mechanism}, characterized by periodic vortex shedding.
    The vorticity field was cropped to \(N=449 \times 199\) with \(M=151\) snapshots and normalized to \([-0.5, 0.5]\).
    \item \textbf{Channel Flow}: A high-dimensional turbulent channel flow simulation from the Johns Hopkins Turbulence Databases \cite{li2008public} at a friction Reynolds number\footnote{The friction Reynolds number \(Re_{\tau}\) is defined based on the friction velocity at the wall, providing a measure of turbulence intensity near the boundary.} of \(Re_{\tau} \approx 1000\), representing a highly turbulent regime with chaotic fluctuations.
    A subset of the velocity field was cropped to \(N=512 \times 512\) with \(M=300\) snapshots and normalized to \([0, 1]\).
\end{enumerate}

To simulate realistic data corruption, we introduced a mixture of white Gaussian noise with a standard deviation
\(\sigma\) and sparse outliers with a corruption ratio \(p_s\).
We tested three noise levels: low noise (\(\sigma=0.05, p_s=0.05\)), medium noise (\(\sigma=0.1, p_s=0.1\)), and high noise (\(\sigma=0.15, p_s=0.15\)).
The type of sparse outliers varied by dataset.
For the cylinder wake, we applied salt-and-pepper noise, randomly setting a fraction \(p_s\) of data points to either the minimum or maximum intensity value.
For the channel flow, we introduced missing values by randomly setting a fraction \(p_s\) of data points to zero.

\subsection{Performance on Mode Extraction}
\label{subsec:mode_extraction}
In this subsection, we evaluate the accuracy and stability of the extracted dynamics and spatial modes under various noise conditions.
We conducted \(K=100\) trials for each noise level to assess statistical reliability.

For the proposed method, the constraint radii \(\varepsilon\) and \(\eta\) in Prob.~\eqref{eq:erpgn} were tuned based on the statistical properties of the noise.
For the cylinder wake, corrupted by additive white Gaussian noise and salt-and-pepper noise, the radii were set as follows:
\begin{align}
  \varepsilon = \alpha \sigma \sqrt{(1-p_s)NM}, \:
  \eta = \alpha \frac{p_s NM}{2},
  \label{eq:params_cylinder}
\end{align}
where the parameter \(\alpha\) was set to \(0.95, 0.91\), and \(0.89\) for low, medium, and high noise levels, respectively.
The balancing parameter \(w\) was set to \(0.9\) for all noise levels.
For the channel flow dataset, considering that the missing values correspond to data points randomly set to zero, the sparse noise radius \(\eta\) was set as follows:
\begin{align}
  \eta = \alpha \frac{p_s\|\dataovs\|_1}{1-p_s}.
  \label{eq:params_channel}
\end{align}
The data-fidelity radius \(\varepsilon\) was set in the same manner as in Eq.~\eqref{eq:params_cylinder}.
The parameter \(\alpha\) was set to \(0.93, 0.91\), and \(0.89\) for low, medium, and high noise levels, respectively.
The balancing parameter \(w\) was set to \(0.3\) for all noise levels.
The stopping criterion of Alg.~\ref{algo_ERPGN} was set as follows:
\begin{align}
  \frac{\|\dataclean^{(\IndexAlg+1)} - \dataclean^{(\IndexAlg)}\|_2}{\|\dataclean^{(\IndexAlg)}\|_2} < 10^{-4}, \frac{\|\outliers^{(\IndexAlg+1)} - \outliers^{(\IndexAlg)}\|_2}{\|\outliers^{(\IndexAlg)}\|_2} < 10^{-4}.
\end{align}

\begin{figure*}[t!]
  \centering
  \includegraphics[width=1\linewidth]{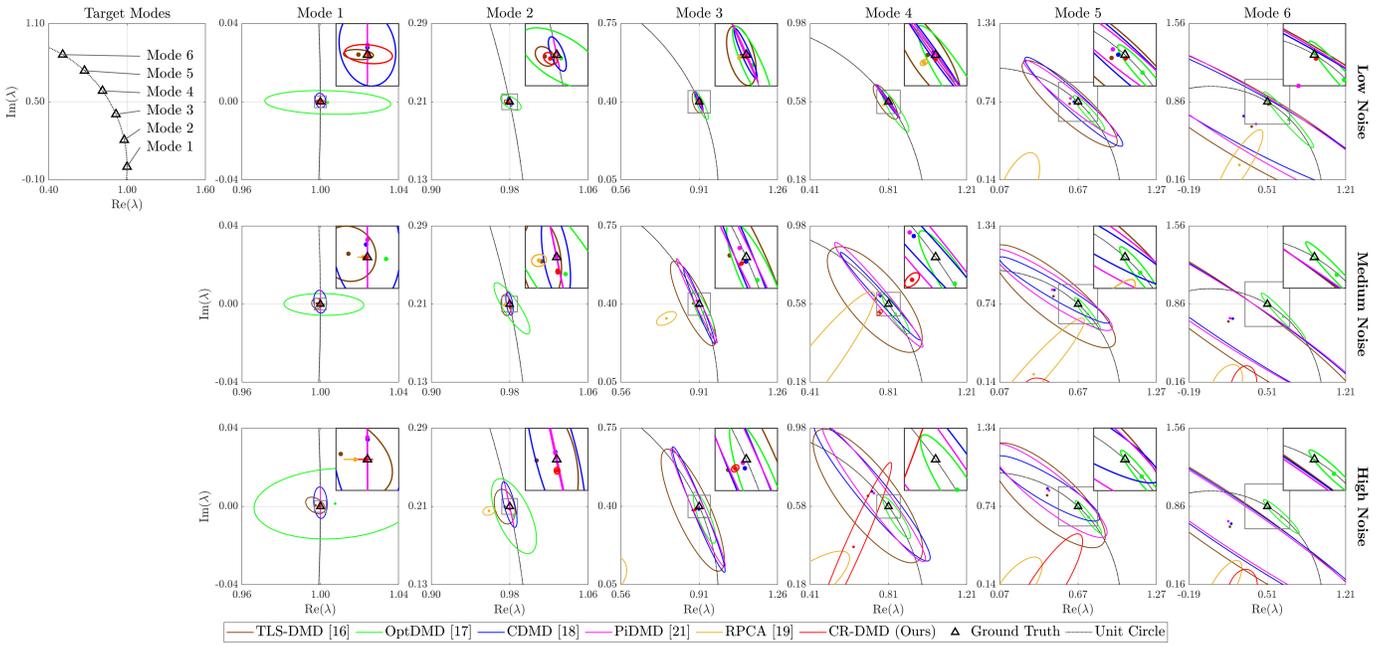}
  \caption{Estimated dynamics results for the noisy cylinder flow. The target modes indicate the ground truth dynamics associated with each mode for comparison. The results are arranged by noise condition in rows and by the eigenvalue corresponding to each mode in columns. In each plot, the 95\% confidence ellipses and the mean predictions for each method are displayed.}  \label{fig:dynamics_cylinder}
\end{figure*}
\begin{figure*}[t!]
  \centering
  \includegraphics[width=1\linewidth]{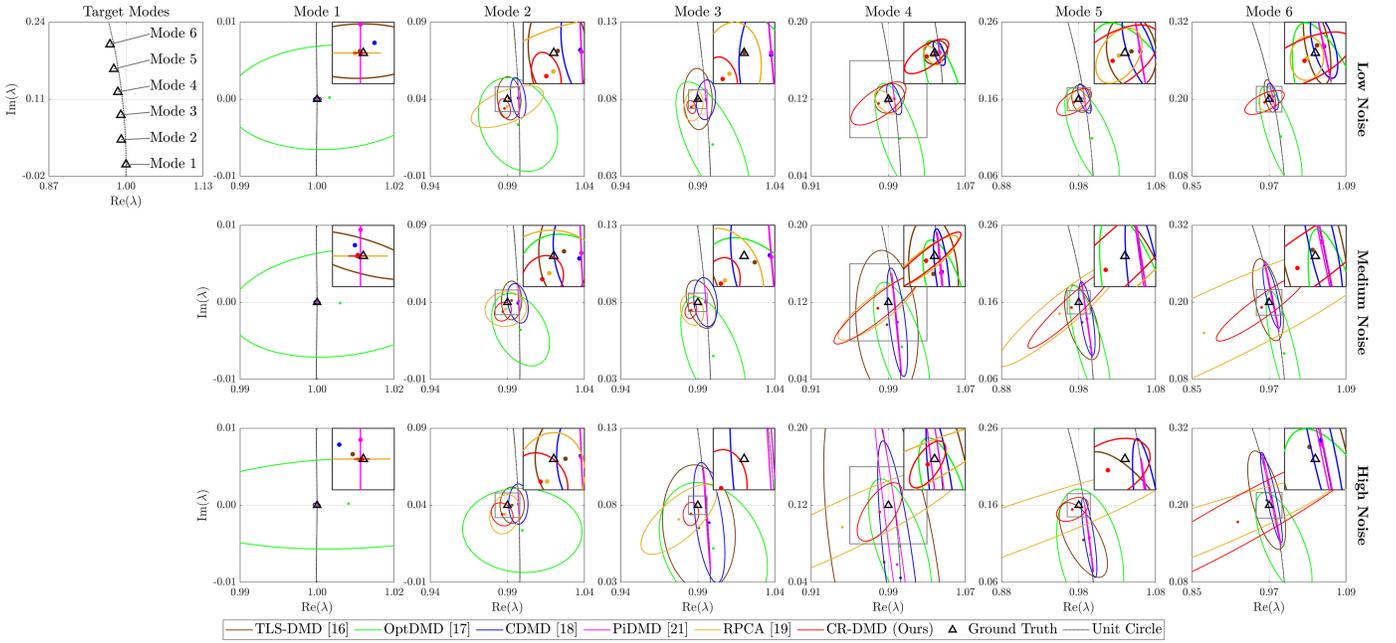}
  \caption{Estimated dynamics results for the noisy turbulent channel flow. The target modes indicate the ground truth dynamics associated with each mode for comparison. The results are arranged by noise condition in rows and by the eigenvalue corresponding to each mode in columns. In each plot, the 95\% confidence ellipses and the mean predictions for each method are displayed.}  \label{fig:dynamics_channel}
\end{figure*}
To quantitatively evaluate the accuracy and stability of the mode extraction,
we focus on the discrete-time eigenvalues \(\dynamics\) corresponding to the dominant modes.
As shown in \eqref{eq:dmd_full}, these eigenvalues govern the temporal evolution of each mode.
Thus, accurate eigenvalue estimation is equivalent to capturing the true system dynamics.
We evaluate the performance for each target mode individually using the Mean Squared Error (MSE);
\begin{align}
  \mathrm{MSE} = \frac{1}{K}\sum_{k=1}^K \|\dynamics_k - \dynamics_{\text{gt}}\|_2^2,
\end{align}
and the Standard Deviation (STD);
\begin{align}
  \mathrm{STD} = \sqrt{\frac{1}{K}\sum_{k=1}^K \|\dynamics_k - \dynamics_{\text{avg}}\|_2^2},
\end{align}
where \(\dynamics_k\), \(\dynamics_{\text{gt}}\), and \(\dynamics_{\text{avg}}\) are the estimated eigenvalue in the \(k\)-th trial,
the ground-truth eigenvalue, and the average of \(\dynamics_k\) over \(K\) trials, respectively.
We evaluated these metrics for each target mode individually.
Generally, lower MSE and STD values correspond to more accurate and stable estimations, respectively.

\begin{table*}[t]
  \centering
  \scriptsize
  \caption{Dimensional Reduction Results for Noisy Cylinder Wake Data (MPSNR[dB] / MSSIM)}
  \label{table:dr_cylinder}
  \setlength{\tabcolsep}{2pt}
  \renewcommand{\arraystretch}{1.08}
  \begin{tabularx}{\linewidth}{c c | *{5}{>{\centering\arraybackslash}X} | *{4}{>{\centering\arraybackslash}X}}
    \toprule
    & & \multicolumn{5}{c|}{Existing Mode Extraction Methods} & \multicolumn{4}{c}{Methods using Mode by Proposed Preprocessing} \\
    \cmidrule(lr){3-7} \cmidrule(lr){8-11}
    Noise & Modes & TLS-DMD\cite{hemati2017biasing} & OptDMD\cite{askham2018variable} & CDMD\cite{azencot2019consistent} & PiDMD\cite{baddoo2023physics} & RPCA\cite{scherl2020robust} & DMD\cite{schmid2010dynamic} & DMDc\cite{kou2017improved} & SpDMD\cite{jovanovic2014sparsity} & CR-DMD \\
    \midrule
    \multirow{3}{*}{\shortstack{Low Noise\\$\left(\makecell{\sigma=0.05 \\ \eta=0.05}\right)$}}
      & 5 & 28.42 / 0.7339 & 24.90 / 0.6768 & 27.38 / 0.6165 & 27.31 / 0.6339 & 31.23 / 0.8698 & \underline{35.27} / \textbf{0.9262} & 35.27 / \underline{0.9262} & 32.79 / 0.8951 & \textbf{36.84} / 0.9203 \\
      & 11 & 27.18 / 0.6145 & 24.81 / 0.6631 & 27.15 / 0.6041 & 26.68 / 0.5834 & 31.24 / 0.8708 & 35.86 / 0.9438 & \underline{35.86} / 0.9438 & 35.83 / \underline{0.9438} & \textbf{39.01} / \textbf{0.9651} \\
      & 21 & 27.05 / 0.6093 & 24.99 / 0.6844 & 27.04 / 0.5990 & 26.61 / 0.5791 & 31.24 / 0.8708 & \underline{35.85} / 0.9437 & 35.85 / 0.9437 & 35.85 / \underline{0.9437} & \textbf{38.98} / \textbf{0.9649} \\
    \cmidrule(lr){1-11}
    \multirow{3}{*}{\shortstack{Medium Noise\\$\left(\makecell{\sigma=0.1 \\ \eta=0.1}\right)$}}
      & 5 & 25.99 / 0.6086 & 25.07 / 0.7136 & 25.18 / 0.4892 & 25.36 / 0.4955 & 28.10 / 0.7764 & 34.38 / 0.8905 & \underline{34.97} / \underline{0.9095} & 34.78 / 0.9084 & \textbf{36.39} / \textbf{0.9119} \\
      & 11 & 25.19 / 0.4970 & 25.00 / 0.7083 & 24.99 / 0.4776 & 25.19 / 0.4857 & 28.10 / 0.7765 & 35.02 / 0.9108 & 35.02 / 0.9108 & \underline{35.02} / \underline{0.9109} & \textbf{36.54} / \textbf{0.9160} \\
      & 21 & 25.13 / 0.4941 & 23.82 / 0.5452 & 24.90 / 0.4728 & 25.16 / 0.4836 & 28.10 / 0.7765 & 35.03 / 0.9108 & 35.03 / 0.9108 & \underline{35.03} / \underline{0.9108} & \textbf{36.54} / \textbf{0.9160} \\
    \cmidrule(lr){1-11}
    \multirow{3}{*}{\shortstack{High Noise\\$\left(\makecell{\sigma=0.15 \\ \eta=0.15}\right)$}}
      & 5 & 23.27 / 0.3808 & 23.59 / 0.5483 & 23.27 / 0.3786 & 23.24 / 0.3746 & 26.38 / 0.7112 & 31.69 / 0.8627 & \underline{31.87} / \textbf{0.8744} & 31.83 / \underline{0.8740} & \textbf{33.69} / 0.8577 \\
      & 11 & 23.13 / 0.3746 & 23.62 / 0.5560 & 22.98 / 0.3631 & 23.10 / 0.3667 & 26.38 / 0.7112 & 31.86 / 0.8742 & \underline{31.87} / \textbf{0.8746} & 31.87 / \underline{0.8746} & \textbf{33.70} / 0.8589 \\
      & 21 & 22.98 / 0.3666 & 20.19 / 0.3617 & 22.88 / 0.3577 & 23.03 / 0.3628 & 26.38 / 0.7112 & 31.87 / \underline{0.8746} & 31.87 / 0.8746 & \underline{31.87} / \textbf{0.8746} & \textbf{33.72} / 0.8573 \\
    \bottomrule
  \end{tabularx}
\end{table*}
\begin{table*}[t]
  \centering
  \scriptsize
  \caption{Dimensional Reduction Results for Noisy Channel Flow Data (MPSNR[\(\text{dB}\)] / MSSIM)}
  \label{table:dr_channel}
  \setlength{\tabcolsep}{2pt}
  \begin{tabularx}{\linewidth}{c c | *{5}{>{\centering\arraybackslash}X} | *{4}{>{\centering\arraybackslash}X}}
    \toprule
    & & \multicolumn{5}{c|}{Existing Mode Extraction Methods} & \multicolumn{4}{c}{Methods using Mode by Proposed Preprocessing} \\
    \cmidrule(lr){3-7} \cmidrule(lr){8-11}
    Noise & Modes & TLS-DMD\cite{hemati2017biasing} & OptDMD\cite{askham2018variable} & CDMD\cite{azencot2019consistent} & PiDMD\cite{baddoo2023physics} & RPCA\cite{scherl2020robust} & DMD\cite{schmid2010dynamic} & DMDc\cite{kou2017improved} & SpDMD\cite{jovanovic2014sparsity} & CR-DMD \\
    \midrule
    \multirow{4}{*}{\shortstack{Low Noise\\$\left(\makecell{\sigma=0.05 \\ \eta=0.05}\right)$}}
      & 5 & 23.55 / 0.5676 & - / - & 15.79 / 0.0832 & 25.53 / 0.6930 & \textbf{29.82} / 0.8722 & 23.85 / 0.8187 & 17.89 / 0.7146 & 26.89 / \underline{0.8969} & \underline{27.61} / \textbf{0.9018} \\
      & 11 & 24.21 / 0.5077 & - / - & 15.22 / 0.0725 & 25.48 / 0.5447 & \textbf{32.32} / 0.8850 & 21.37 / 0.7937 & 21.31 / 0.7929 & 28.35 / \underline{0.9106} & \underline{29.28} / \textbf{0.9169} \\
      & 21 & 24.94 / 0.4510 & - / - & 15.11 / 0.0701 & 25.00 / 0.4512 & \textbf{33.15} / 0.8861 & 29.65 / 0.8911 & 29.10 / 0.8889 & 30.11 / \underline{0.9265} & \underline{31.62} / \textbf{0.9367} \\
      & 100 & 24.70 / 0.4245 & 21.74 / 0.2249 & 15.05 / 0.0699 & 24.72 / 0.4179 & 33.47 / 0.8951 & \underline{39.08} / \underline{0.9745} & 39.08 / 0.9745 & 39.08 / 0.9745 & \textbf{39.57} / \textbf{0.9749} \\
    \cmidrule(lr){1-11}
    \multirow{4}{*}{\shortstack{Medium Noise\\$\left(\makecell{\sigma=0.1 \\ \eta=0.1}\right)$}}
      & 5 & 20.24 / 0.5384 & - / - & 19.15 / 0.2664 & 21.10 / 0.5570 & 24.95 / 0.7149 & 23.34 / 0.8578 & 20.35 / 0.8240 & \underline{25.60} / \underline{0.8862} & \textbf{26.25} / \textbf{0.8921} \\
      & 11 & 20.63 / 0.4075 & - / - & 19.10 / 0.2657 & 20.85 / 0.4042 & \textbf{27.97} / 0.7653 & 18.52 / 0.7431 & 20.00 / 0.8150 & 26.88 / \underline{0.9002} & \underline{27.02} / \textbf{0.9003} \\
      & 21 & 20.66 / 0.3595 & - / - & 19.08 / 0.2705 & 20.64 / 0.3470 & \underline{28.73} / 0.8127 & 28.11 / 0.8843 & 25.63 / 0.8992 & 28.10 / \underline{0.9141} & \textbf{29.47} / \textbf{0.9267} \\
      & 100 & 20.56 / 0.3416 & 17.99 / 0.1399  & 19.17 / 0.2558 & 20.55 / 0.3307 & 28.75 / 0.8145 & \underline{32.82} / \textbf{0.9605} & 32.82 / \underline{0.9605} & 32.82 / 0.9605 & \textbf{32.84} / 0.9553 \\
    \cmidrule(lr){1-11}
    \multirow{4}{*}{\shortstack{High Noise\\$\left(\makecell{\sigma=0.15 \\ \eta=0.15}\right)$}}
      & 5 & 17.82 / 0.4215 & - / - & 14.26 / 0.0676 & 18.04 / 0.4639 & 18.04 / 0.7389 & 16.22 / 0.7600 & 16.22 / 0.7600 & \underline{24.00} / \underline{0.8946} & \textbf{24.43} / \textbf{0.8984} \\
      & 11 & 17.77 / 0.3216 & - / - & 15.63 / 0.1016 & 17.80 / 0.3179 & 17.80 / 0.7385 & 19.80 / 0.7913 & 18.04 / 0.7967 & \underline{24.93} / \underline{0.9066} & \textbf{25.32} / \textbf{0.9107} \\
      & 21 & 17.65 / 0.2830 & - / - & 15.55 / 0.0989 & 17.64 / 0.2741 & 17.64 / 0.7368 & 24.87 / 0.8901 & \underline{26.22} / 0.9194 & 25.92 / \underline{0.9224} & \textbf{26.65} / \textbf{0.9358} \\
      & 100 & 17.59 / 0.2725 & 15.48 / 0.1019 & 15.52 / 0.0976 & 17.59 / 0.2647 & 17.59 / 0.7426 & \textbf{26.98} / \textbf{0.9481} & \underline{26.98} / \underline{0.9481} & 26.98 / 0.9481 & 26.89 / 0.9411 \\
    \bottomrule
  \end{tabularx}
\end{table*}
\subsubsection{Quantitative Comparison}
\label{subsubsec:tab_dynamics}
Tables~\ref{table:mse_eigen} and \ref{table:std_eigen}
show MSEs and STDs in the experiments on mode extraction for fluid flow data contaminated by mixed noise, respectively.
The best and second results are highlighted in bold and underlined, respectively.
TLS-DMD exhibits suboptimal performance in most cases, except for specific modes in the channel flow dataset.
The results of OptDMD and CDMD show high accuracy, particularly in Modes 4, 5, and 6 of the cylinder wake dataset for MSEs.
However, their stability is significantly compromised, as indicated by the high STDs.
This instability is likely due to the non-convex nature of the optimization problem,
which makes these methods sensitive to initialization and susceptible to local minima.
PiDMD shows improved accuracy and stability compared to OptDMD and CDMD due to its convex formulation. 
However, its performance degrades significantly as noise levels increase, 
indicating limited robustness against severe corruption.
In contrast, RPCA exhibits high stability in most cases, owing to its convex preprocessing approach.
Nevertheless, its accuracy for less dominant modes (e.g., Modes 4-6) deteriorates rapidly as noise levels increase.
On the other hand, CR-DMD achieves the best MSEs and STDs in most cases, demonstrating its effectiveness 
in accurately and stably extracting modes from fluid flow data corrupted by mixed noise.

\subsubsection{Qualitative Comparison}
\label{subsubsec:fig_dynamics}
Fig.~\ref{fig:dynamics_cylinder} and \ref{fig:dynamics_channel} show the estimated eigenvalues 
for the cylinder wake and turbulent channel flow datasets, respectively.
In these figures, the results are arranged by noise condition in rows and by the eigenvalue representing the temporal dynamics of each mode in columns. 
The black triangles denote the ground-truth eigenvalues corresponding to the dynamics of the target modes selected for comparison.
The colored markers indicate the mean of the estimated eigenvalues, 
and the ellipses represent the 95\% confidence intervals.

Fig. \ref{fig:dynamics_cylinder} shows the estimated eigenvalues for the noisy cylinder wake dataset.
The 95\% confidence ellipses of the operator-centric methods (i.e., TLS-DMD, OptDMD, CDMD, and PiDMD) elongate along the direction of the unit circle. 
While this indicates that these methods extract eigenvalues near the unit circle, it simultaneously highlights instability in the phase of the eigenvalues, as reflected in the large variance perpendicular to the radial direction.
In particular, since PiDMD constrains the operator to be a unitary matrix, this characteristic is clearly manifested, especially in the results for Mode 1 and 2. 
Among these methods, TLS-DMD shows accurate mean estimation but suffers from large phase variance.
CDMD exhibits a similar trend but outperforms it with improved mean accuracy and reduced variance.
OptDMD demonstrates consistent performance across all modes, maintaining accuracy and ellipse sizes, even for lower-ranking modes.
In contrast, RPCA shows high stability with compact confidence ellipses for dominant modes (Modes 1--3).
However, its accuracy is lower, and performance drops for posterior modes (Modes 4--6).
The proposed CR-DMD achieves both high accuracy and stability in most cases except for posterior modes under high noise conditions. 
Even in the enlarged views, the confidence ellipses are significantly smaller than those of competing methods. 
This advantage is clearly observed in the results for Mode 4 under medium noise, where the proposed method exhibits distinct localization.

Fig. \ref{fig:dynamics_channel} shows the estimated eigenvalues for the turbulent channel flow. 
While the overall trends resemble those in the cylinder wake case, distinct behaviors are observed for specific methods due to the increased complexity of the flow.
Notably, TLS-DMD exhibits relatively high mean estimation accuracy under low and medium noise conditions.
In contrast, the performance of OptDMD deteriorates significantly compared to other methods, as it demonstrates both larger variances and lower mean accuracy.
This suggests that OptDMD is more susceptible to the high-dimensional and chaotic channel flow data.
Despite these challenges, the proposed CR-DMD maintains the highest accuracy and smallest confidence ellipses across most conditions, achieving robust mode extraction even in complex turbulent flows.

\begin{figure*}[t!]
    \centering
    \newcommand{\imgwidth}{0.18\linewidth} 

    \begin{minipage}[t]{\imgwidth}
        \centering
        \includegraphics[width=\linewidth]{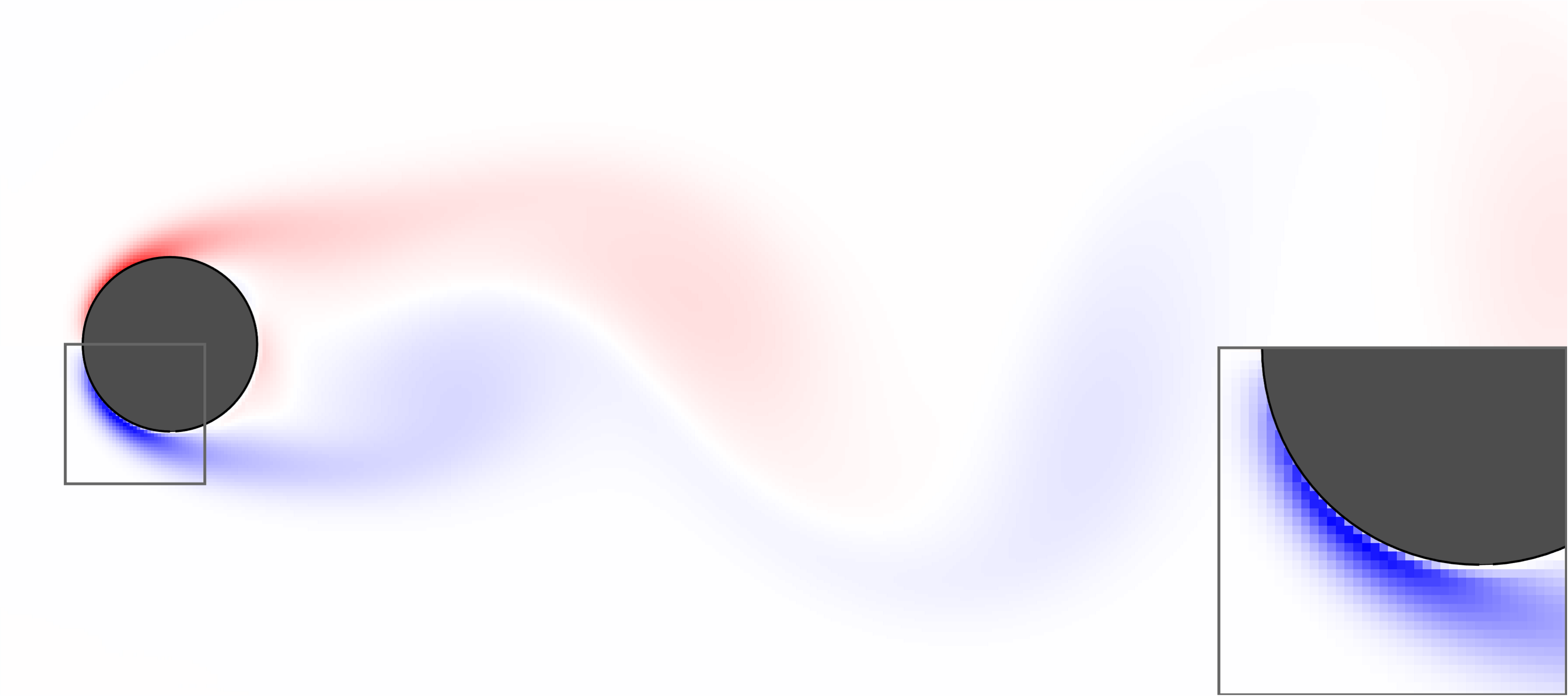}\\[1mm]
        \phantom{\includegraphics[width=\linewidth]{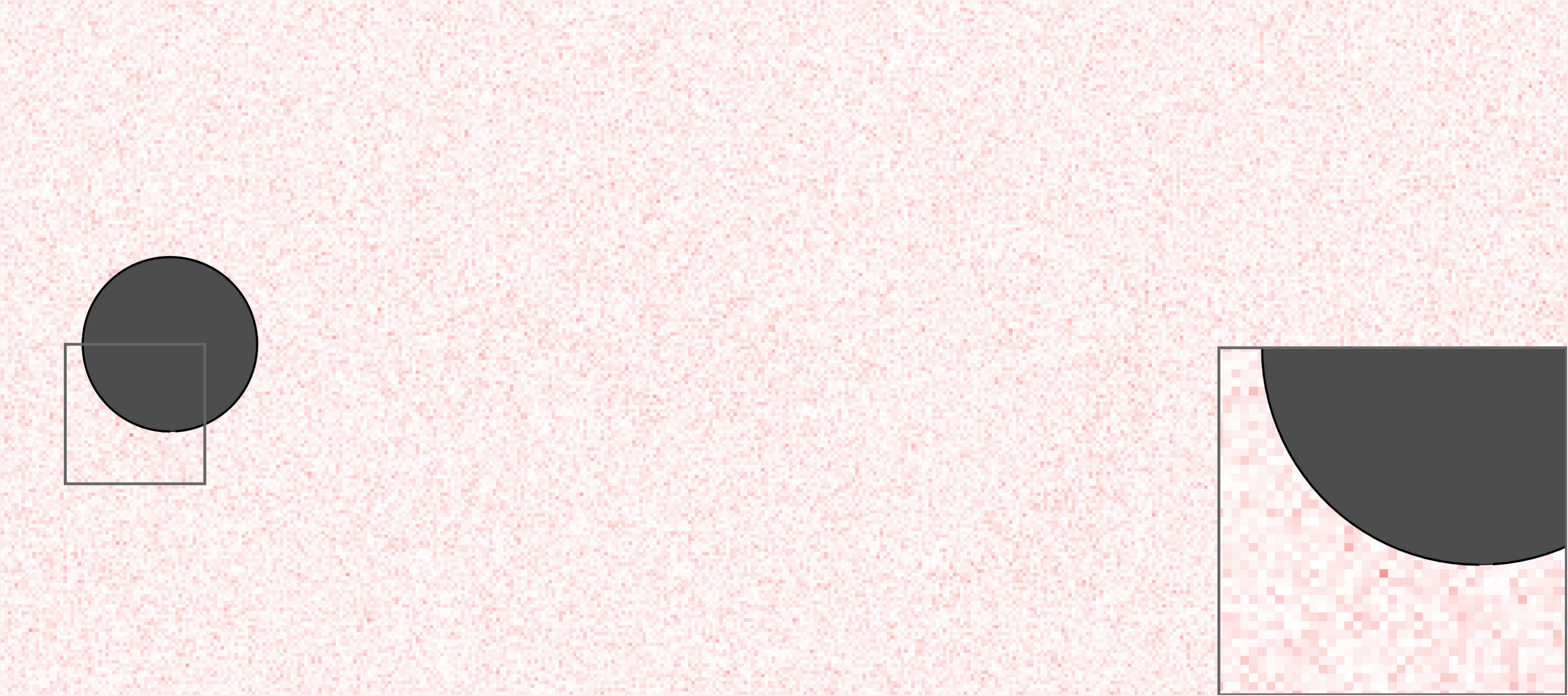}}
        \par\vspace{1mm}
        {\footnotesize (a) Ground-truth}
        \label{fig:dr_cylinder_gt}
    \end{minipage}\hfill
    \begin{minipage}[t]{\imgwidth}
        \centering
        \includegraphics[width=\linewidth]{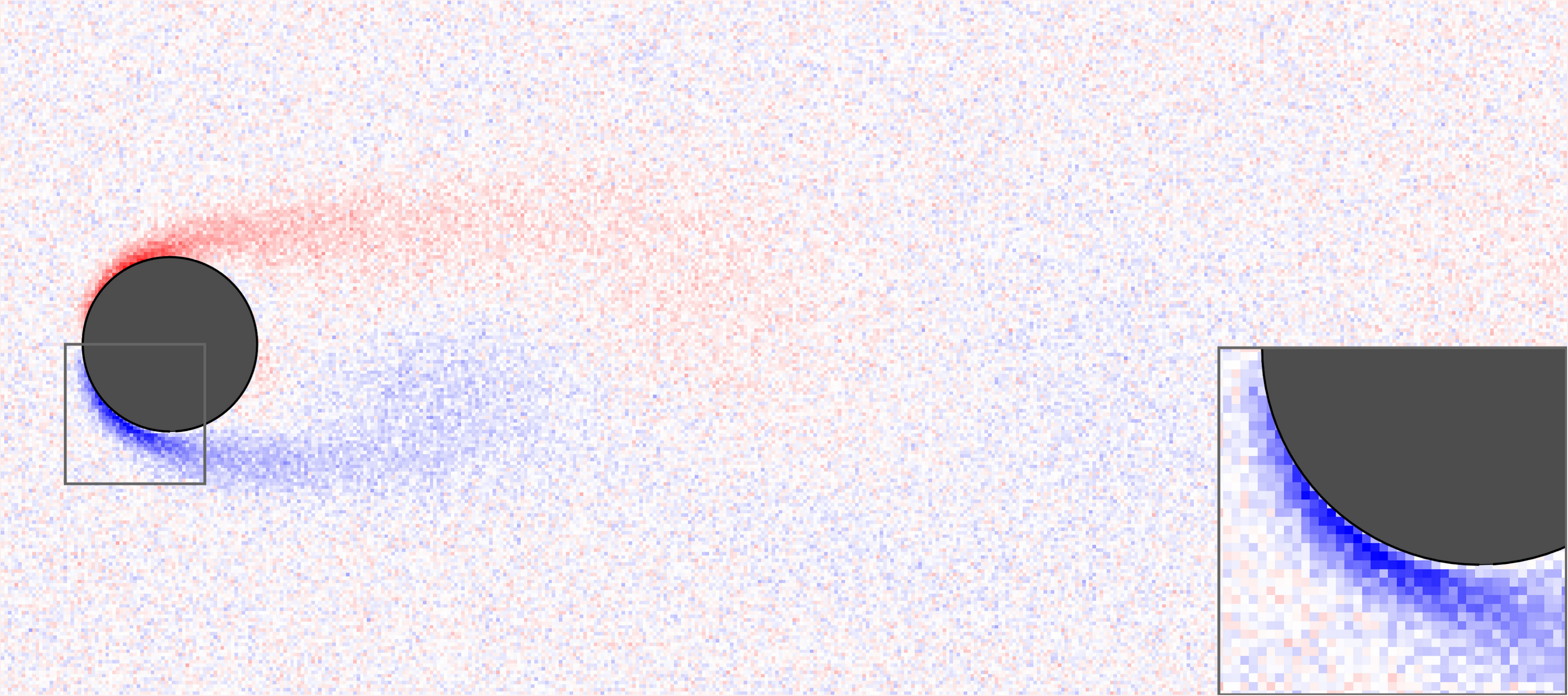}\\[1mm]
        \includegraphics[width=\linewidth]{fig/Dr_cylinder/TLSDMD_Error-eps-converted-to.pdf}
        \par\vspace{1mm}
        {\footnotesize (b) TLS-DMD}
        \label{fig:dr_cylinder_tlsdmd}
    \end{minipage}\hfill
    \begin{minipage}[t]{\imgwidth}
        \centering
        \includegraphics[width=\linewidth]{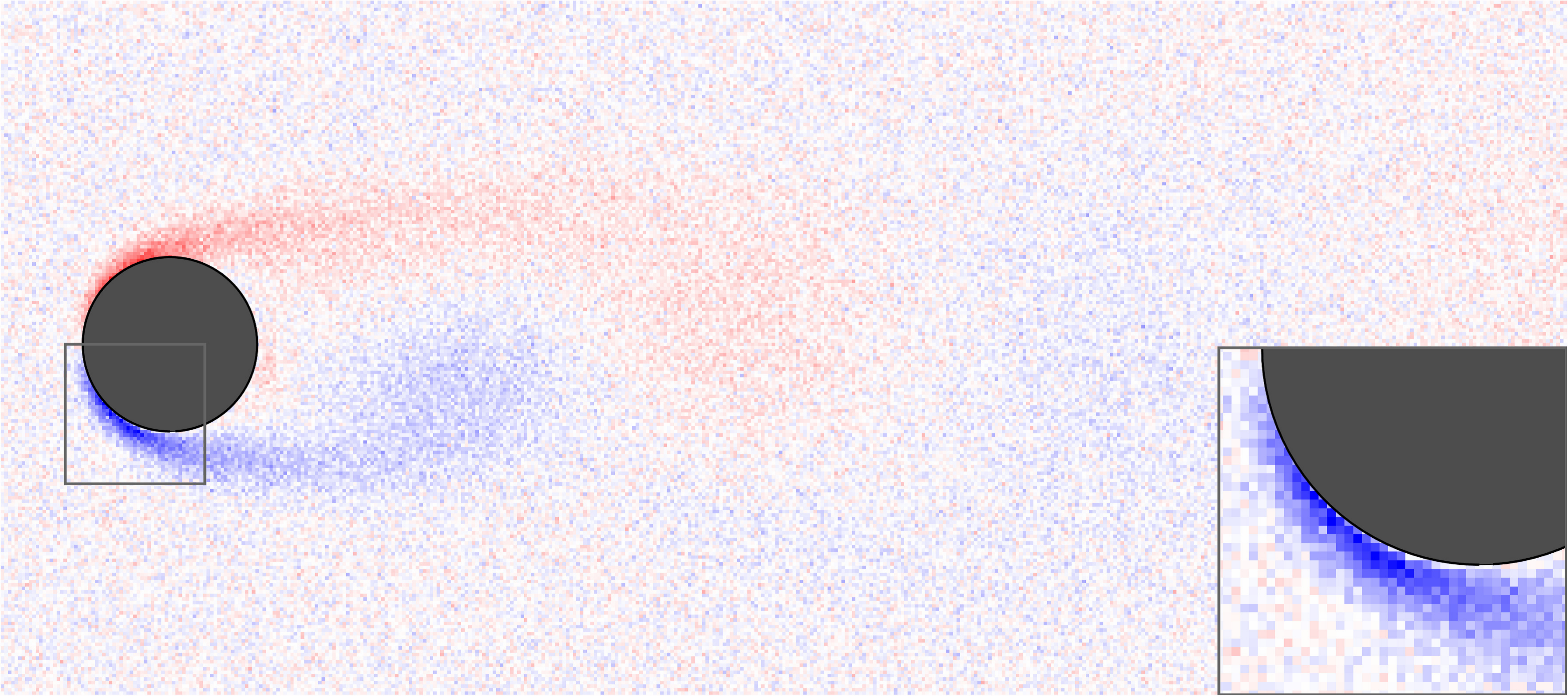}\\[1mm]
        \includegraphics[width=\linewidth]{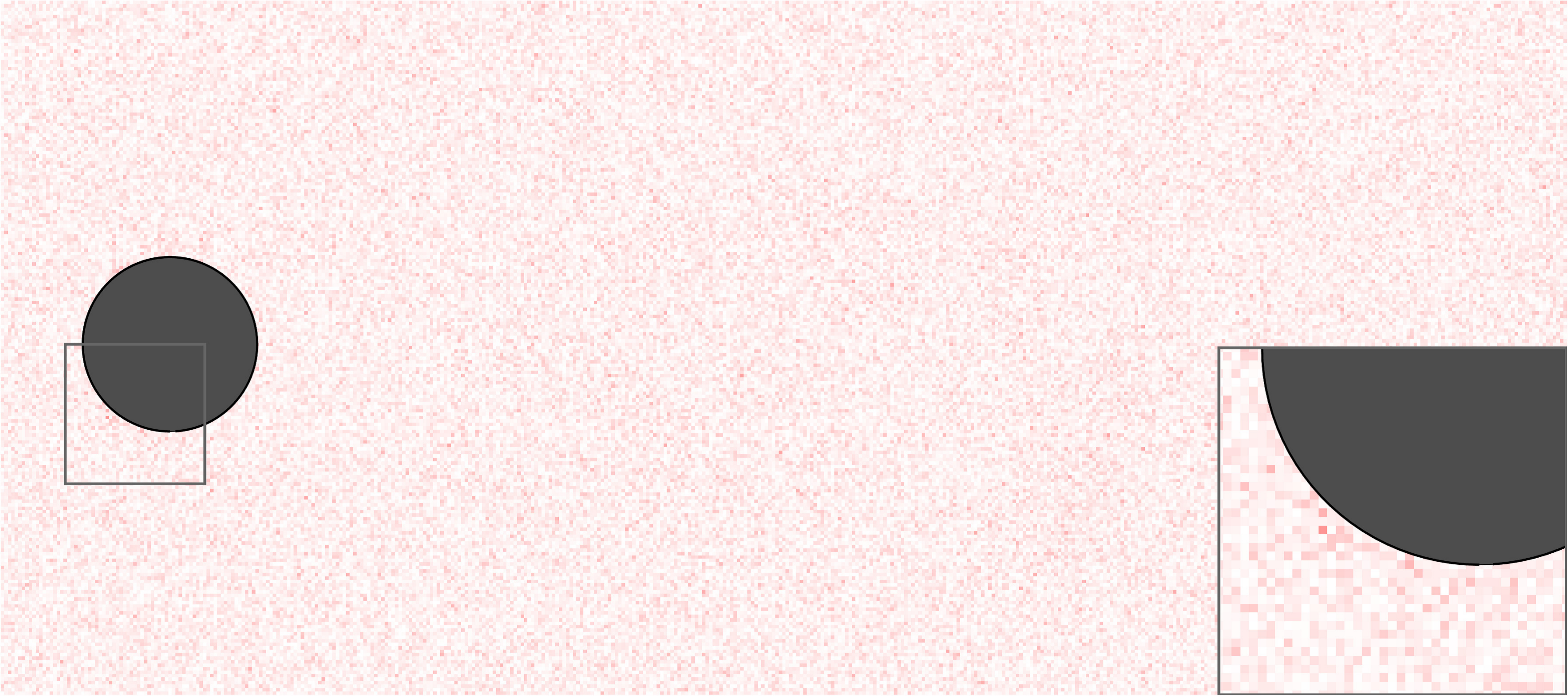}
        \par\vspace{1mm}
        {\footnotesize (c) OptDMD}
        \label{fig:dr_cylinder_optdmd}
    \end{minipage}\hfill
    \begin{minipage}[t]{\imgwidth}
        \centering
        \includegraphics[width=\linewidth]{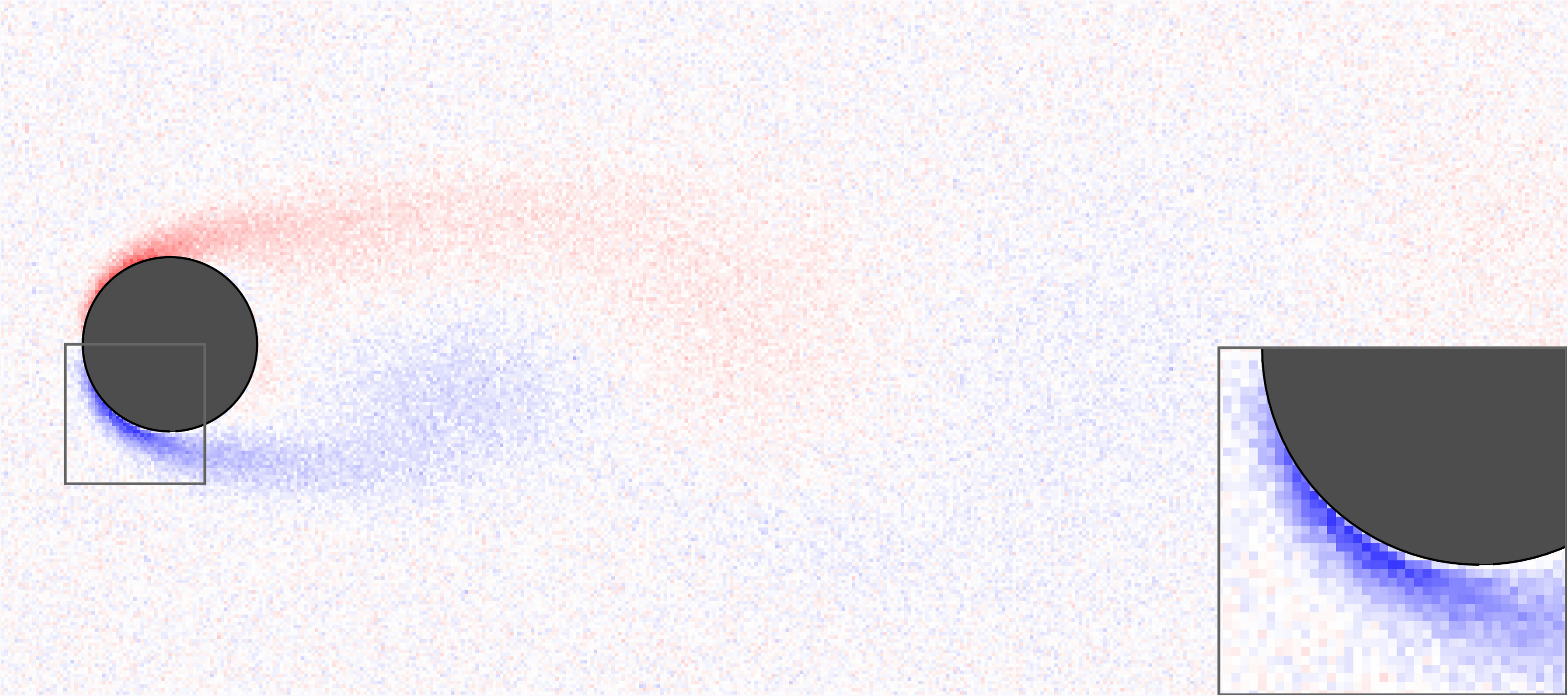}\\[1mm]
        \includegraphics[width=\linewidth]{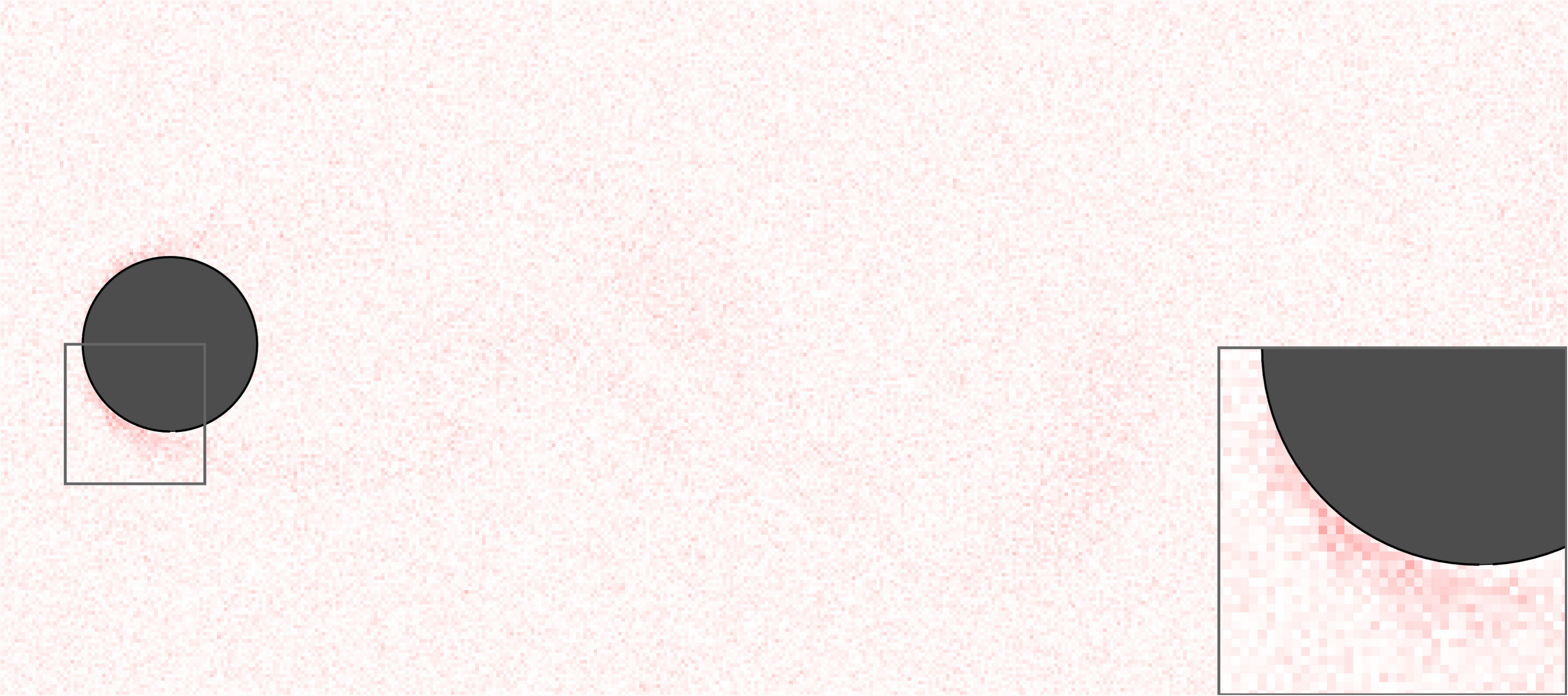}
        \par\vspace{1mm}
        {\footnotesize (d) CDMD}
        \label{fig:dr_cylinder_cdmd}
    \end{minipage}\hfill
    \begin{minipage}[t]{\imgwidth}
        \centering
        \includegraphics[width=\linewidth]{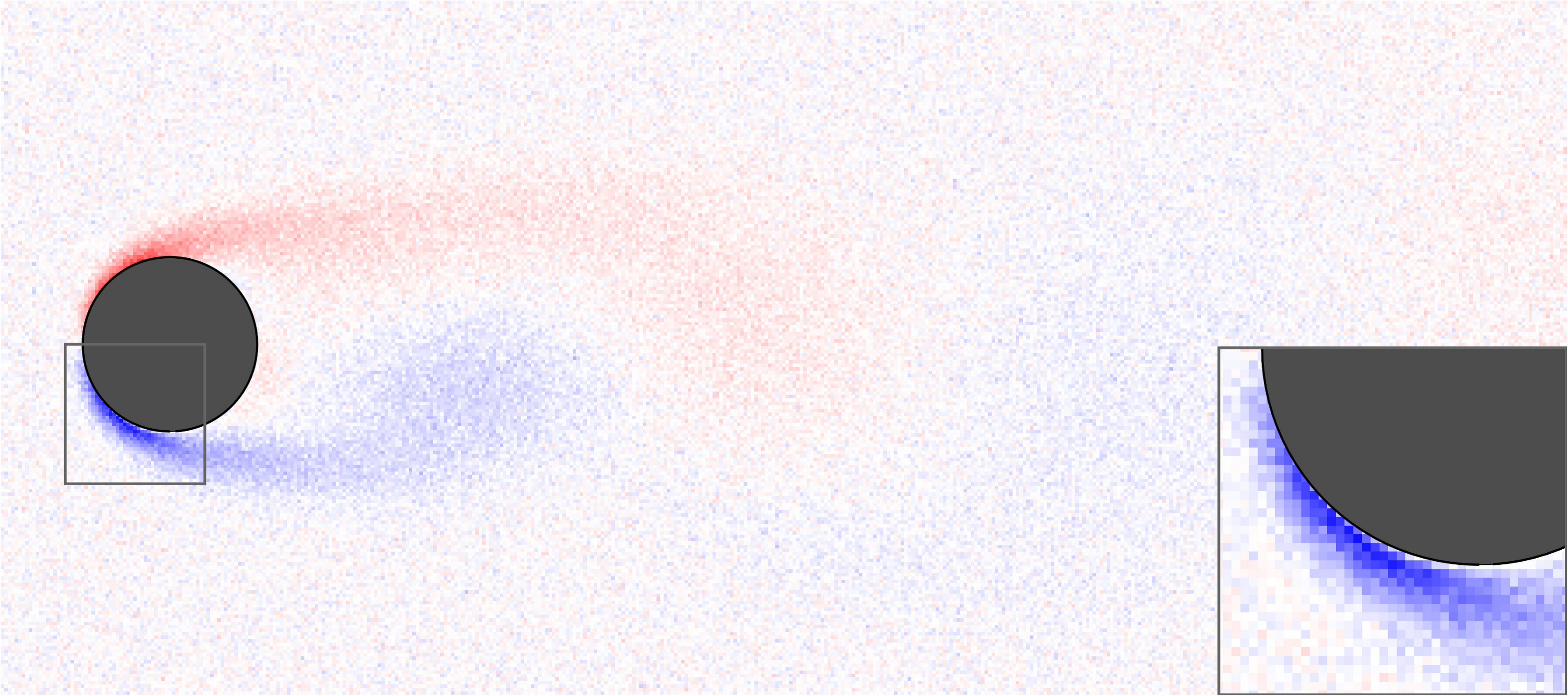}\\[1mm]
        \includegraphics[width=\linewidth]{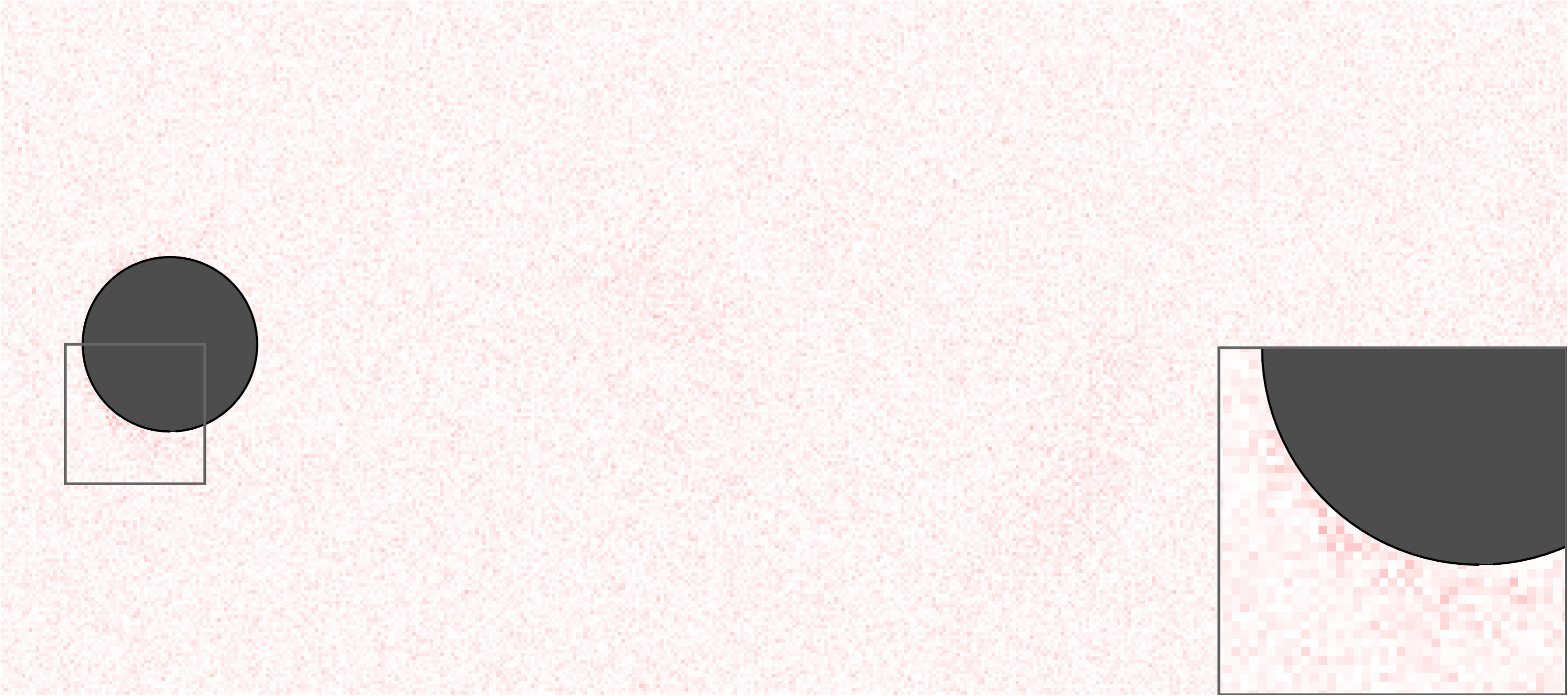}
        \par\vspace{1mm}
        {\footnotesize (e) PiDMD}
        \label{fig:dr_cylinder_pidmd}
    \end{minipage}\hfill
    \begin{minipage}[t]{0.05\linewidth}
        \centering
        \includegraphics[width=\linewidth]{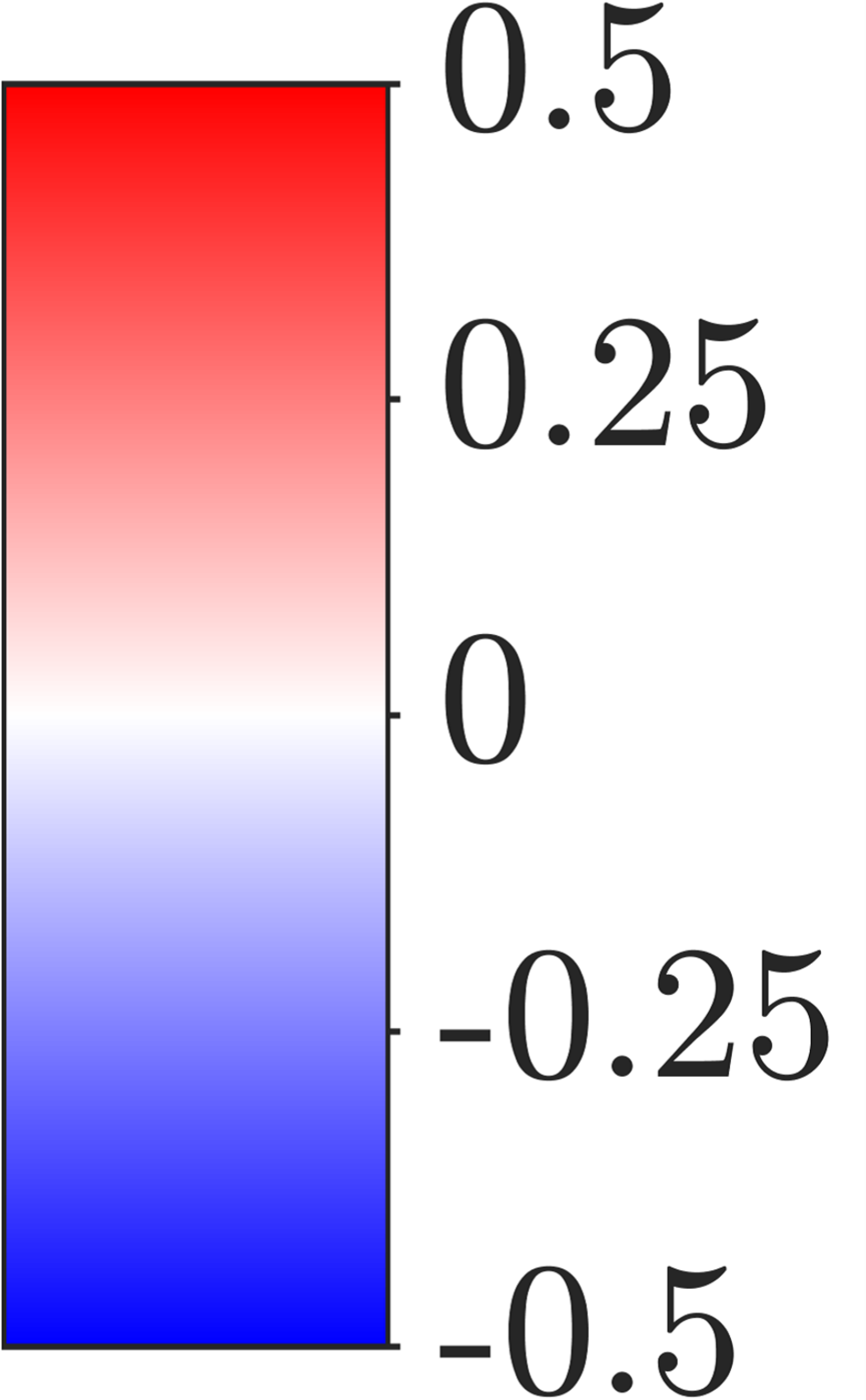}\\[1mm]
        \includegraphics[width=\linewidth]{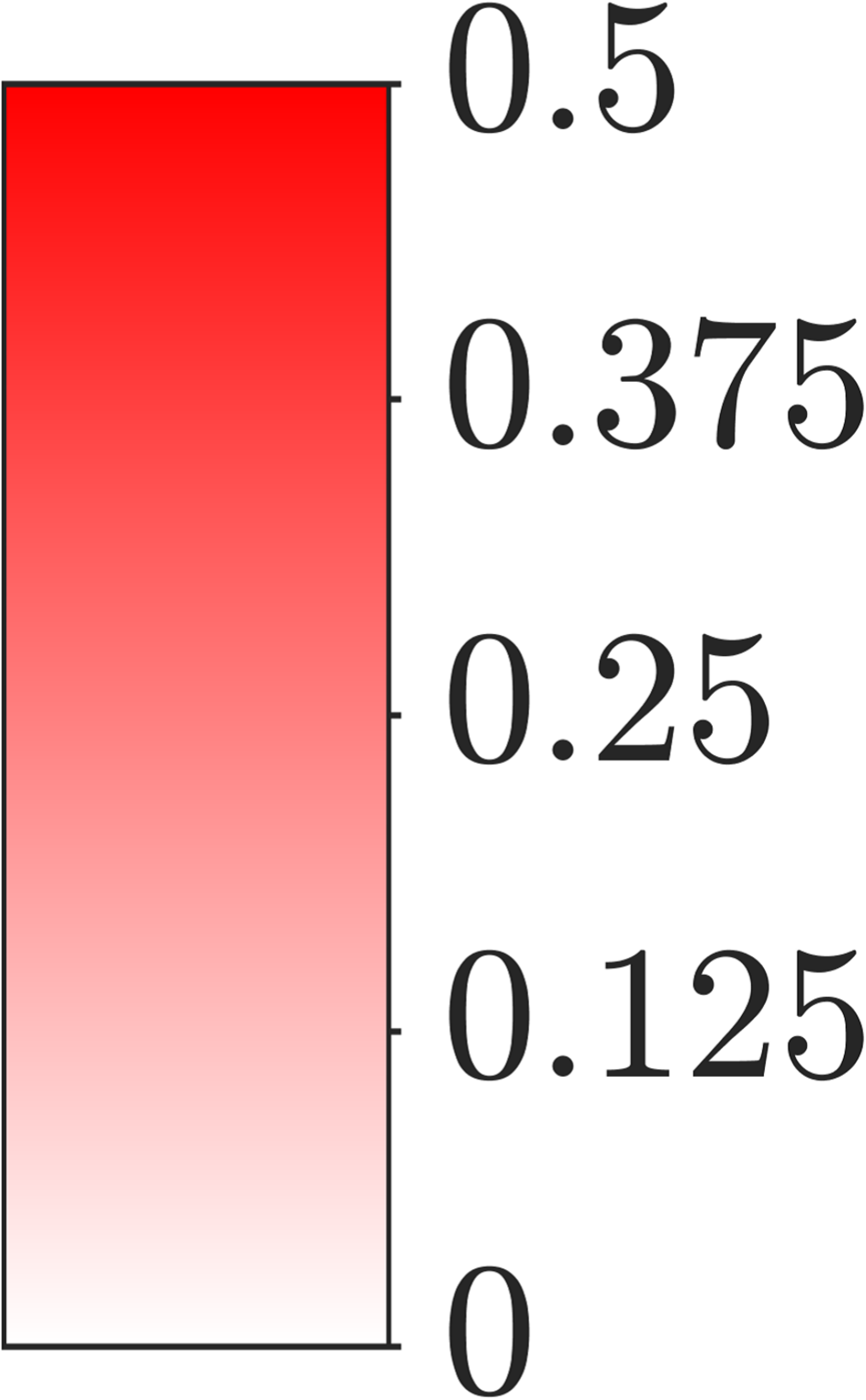}
        \par\vspace{1mm}
    \end{minipage}

    \vspace{1em}

    \begin{minipage}[t]{\imgwidth}
        \centering
        \includegraphics[width=\linewidth]{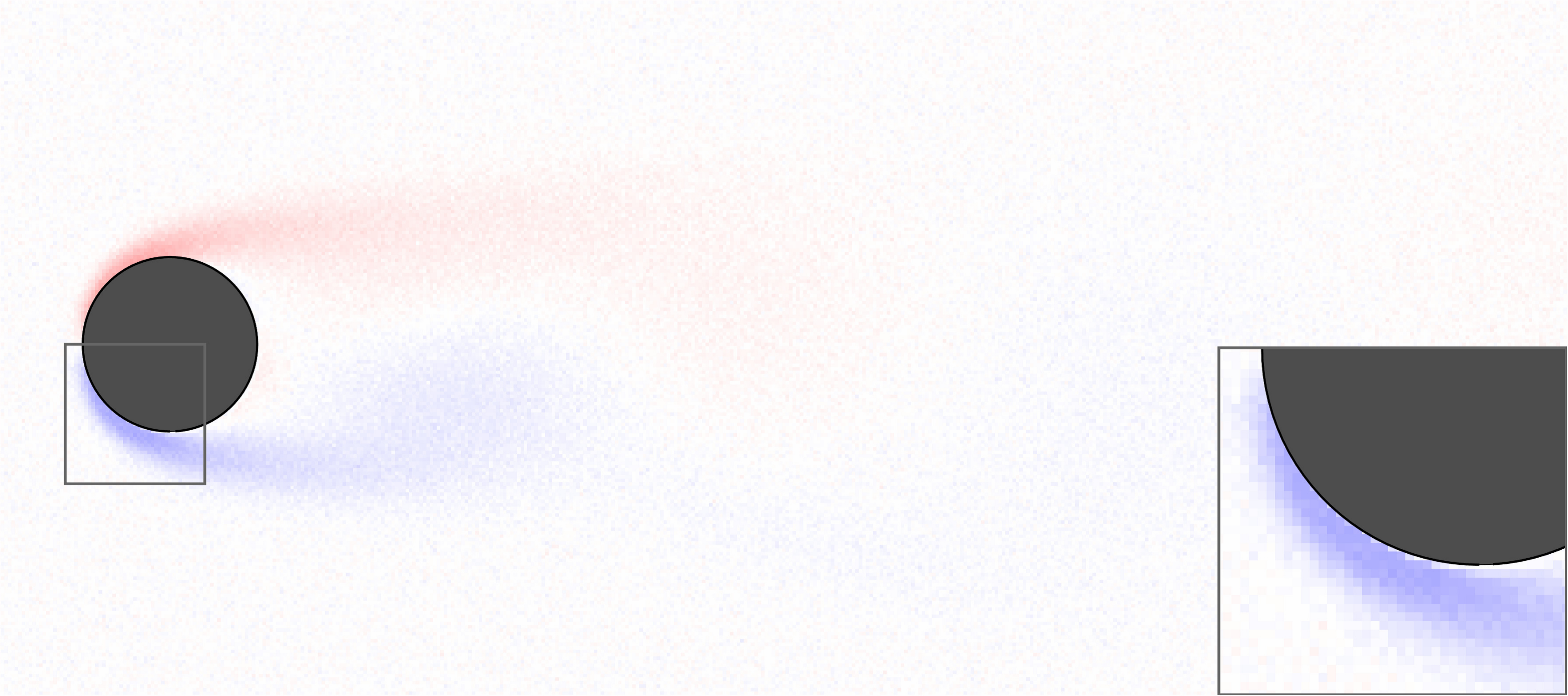}\\[1mm]
        \includegraphics[width=\linewidth]{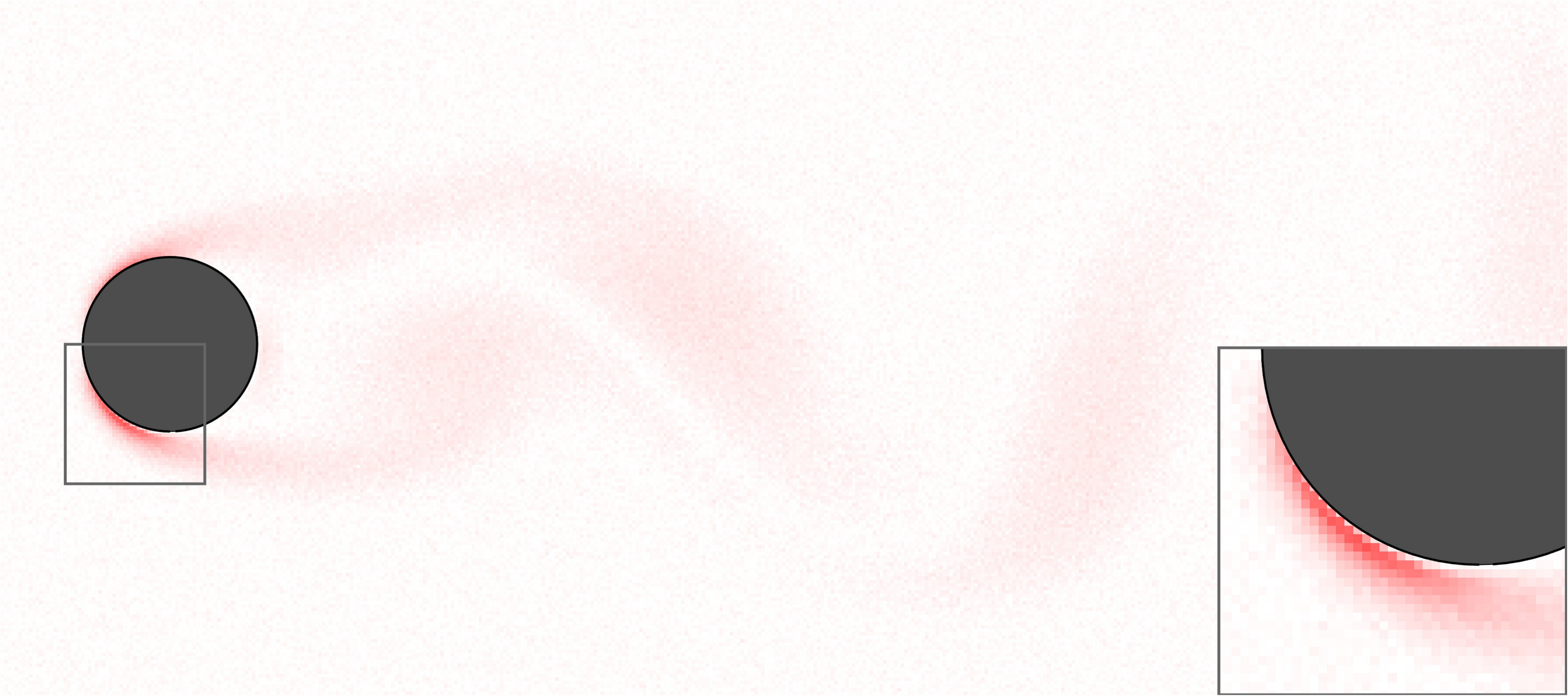}
        \par\vspace{1mm}
        {\footnotesize (f) RPCA}
        \label{fig:dr_cylinder_rpca}
    \end{minipage}\hfill
    \begin{minipage}[t]{\imgwidth}
        \centering
        \includegraphics[width=\linewidth]{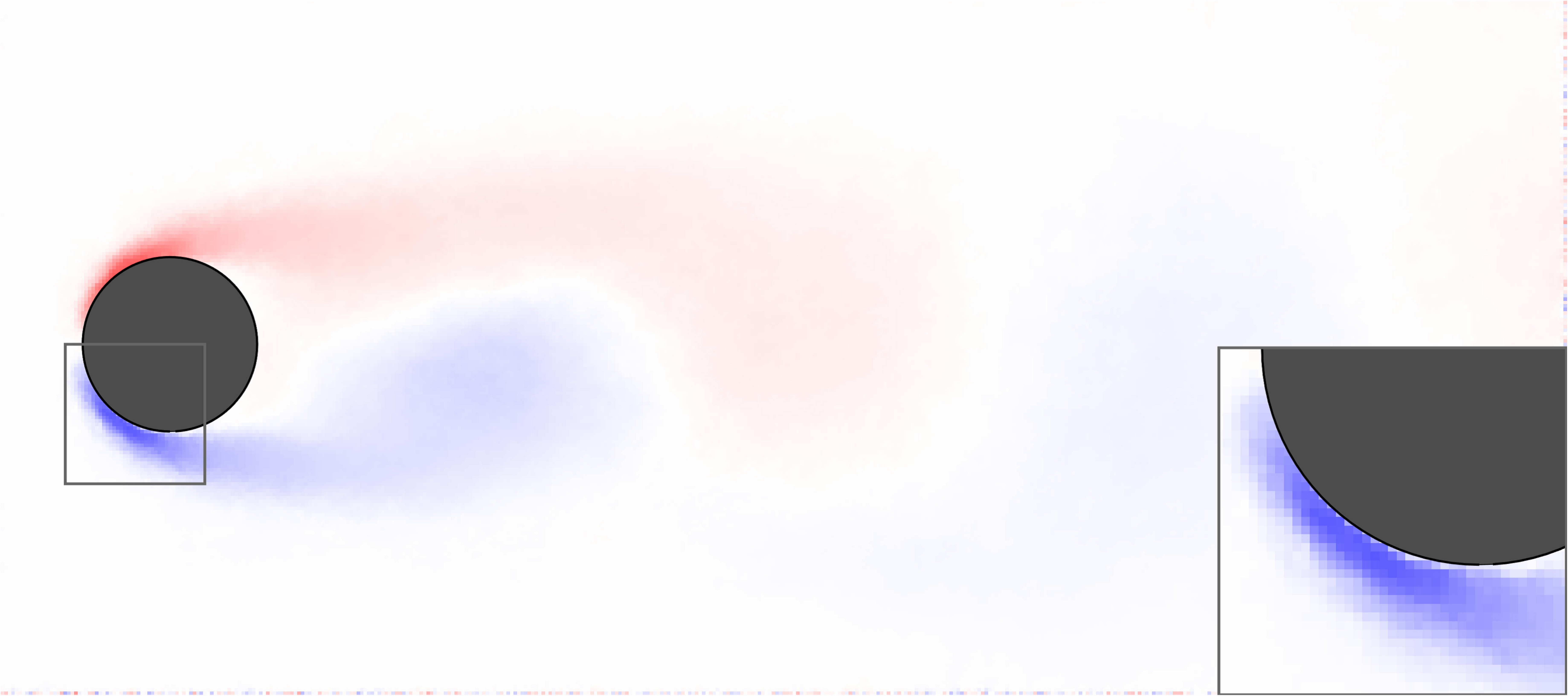}\\[1mm]
        \includegraphics[width=\linewidth]{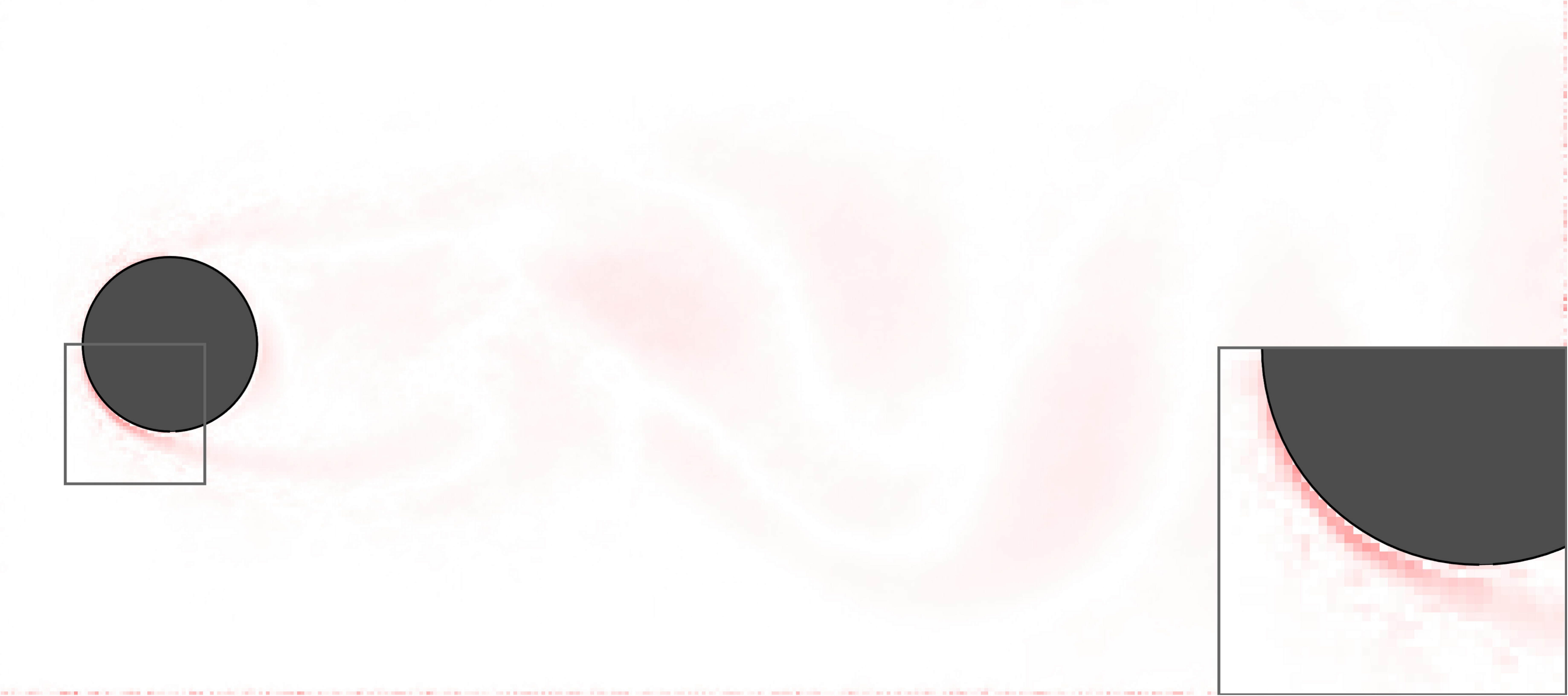}
        \par\vspace{1mm}
        {\footnotesize (g) DMD}
        \label{fig:dr_cylinder_dmd}
    \end{minipage}\hfill
    \begin{minipage}[t]{\imgwidth}
        \centering
        \includegraphics[width=\linewidth]{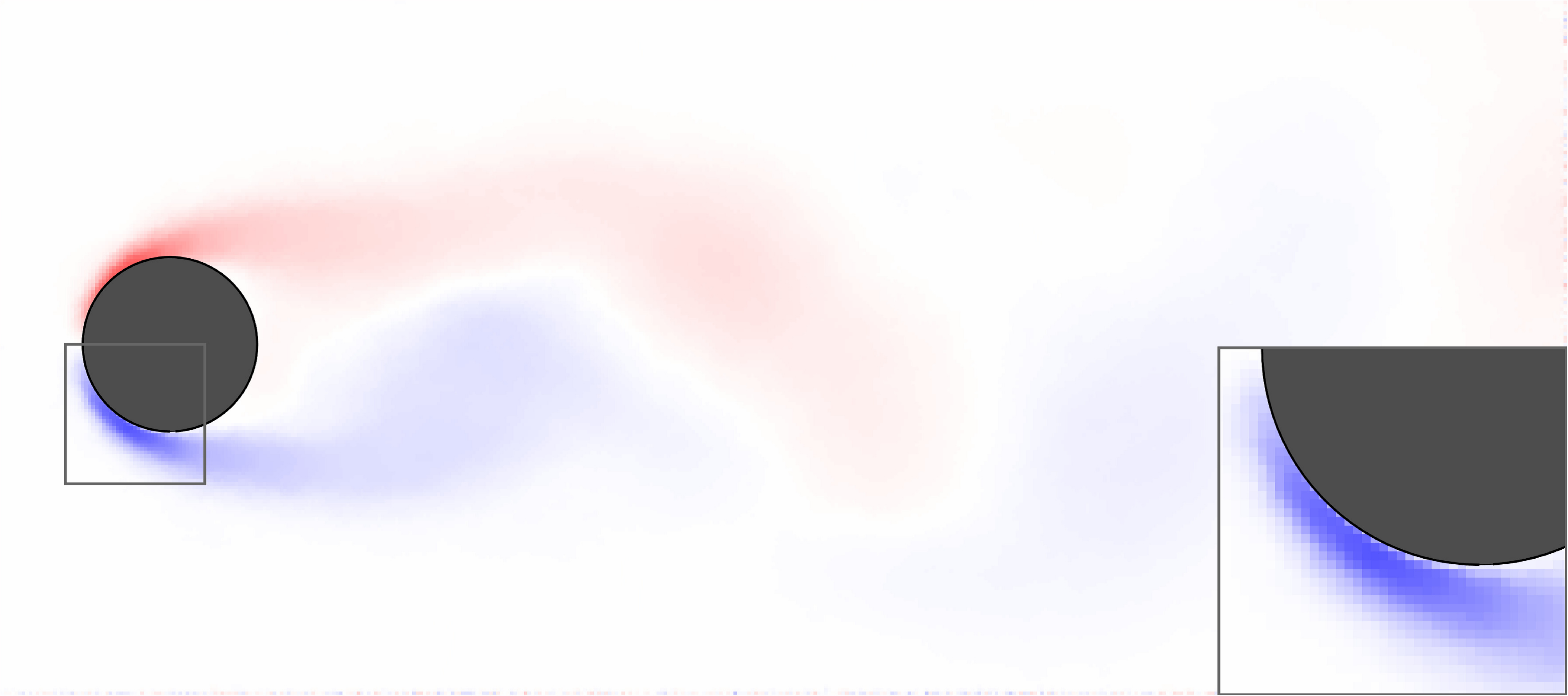}\\[1mm]
        \includegraphics[width=\linewidth]{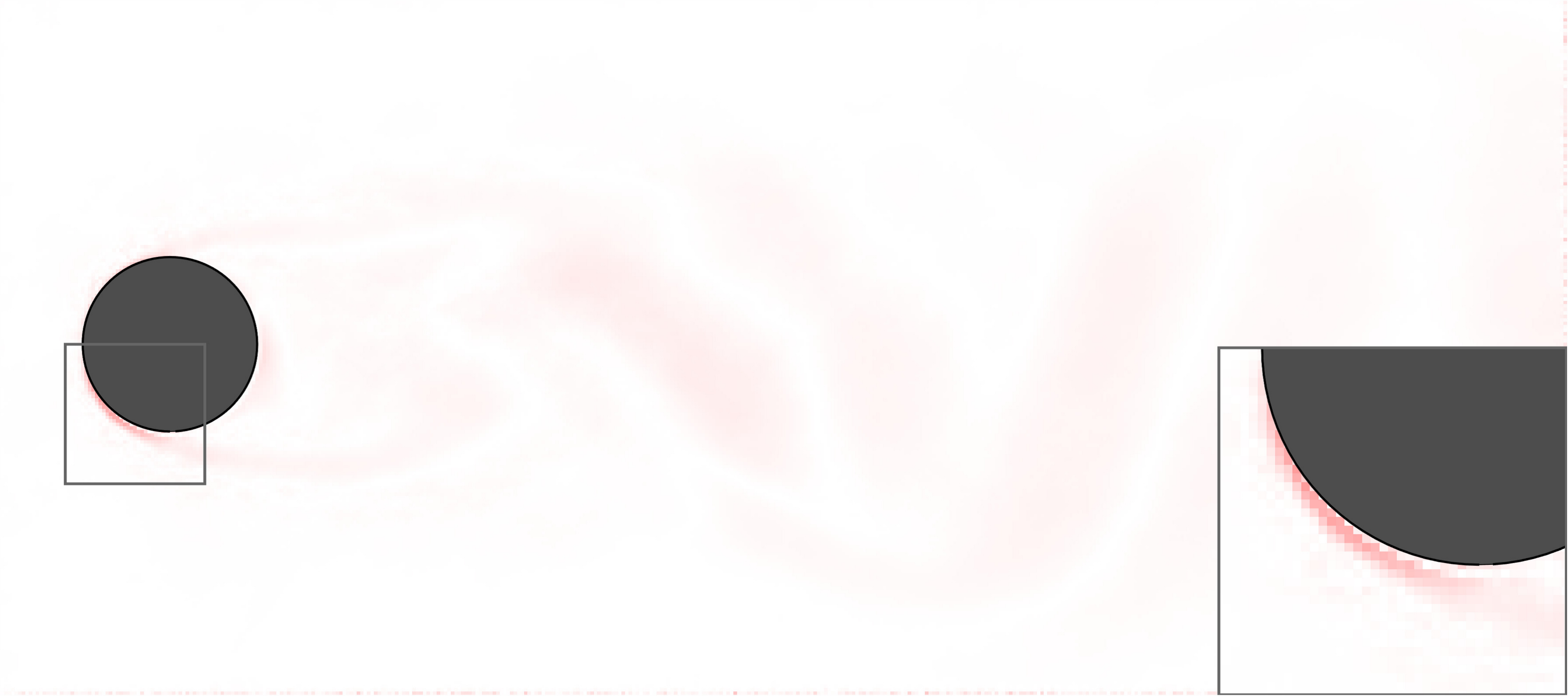}
        \par\vspace{1mm}
        {\footnotesize (h) DMDc}
        \label{fig:dr_cylinder_dmdc}
    \end{minipage}\hfill
    \begin{minipage}[t]{\imgwidth}
        \centering
        \includegraphics[width=\linewidth]{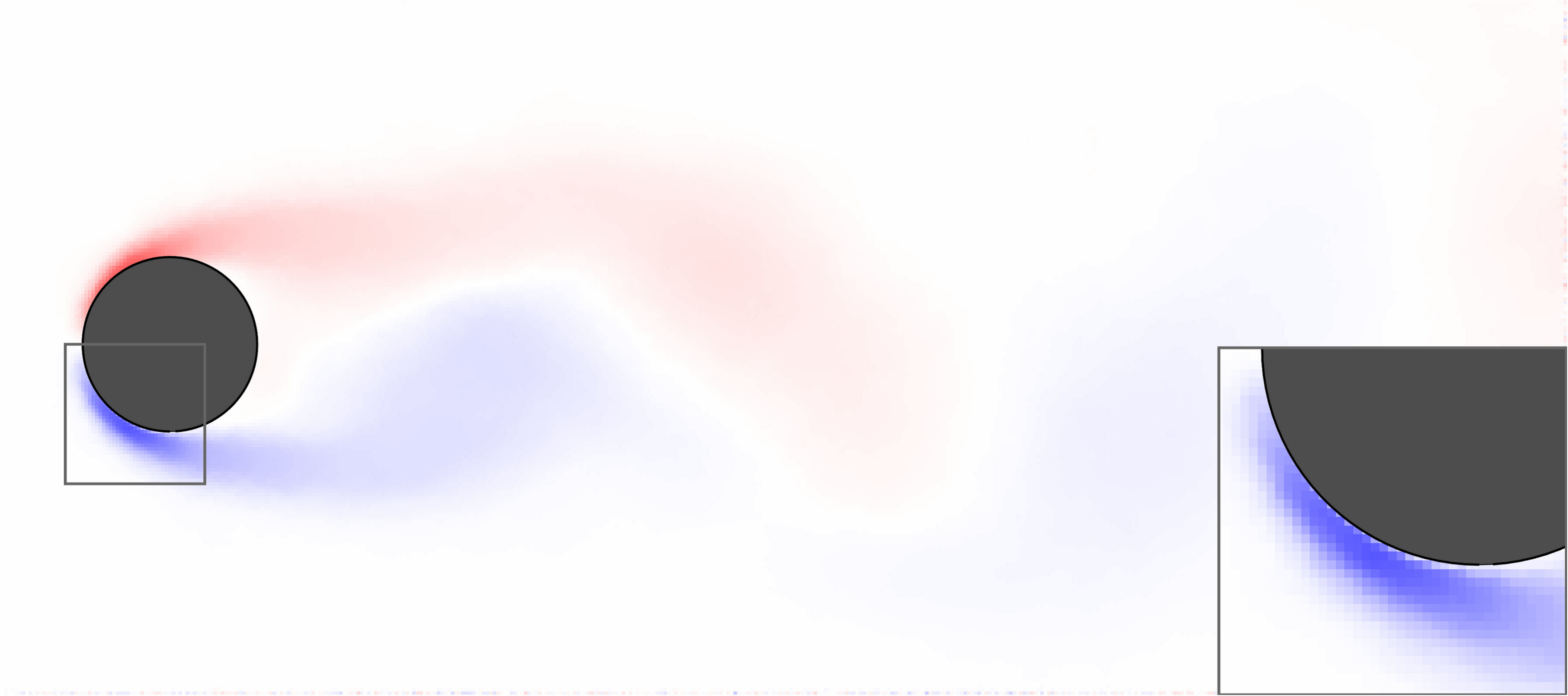}\\[1mm]
        \includegraphics[width=\linewidth]{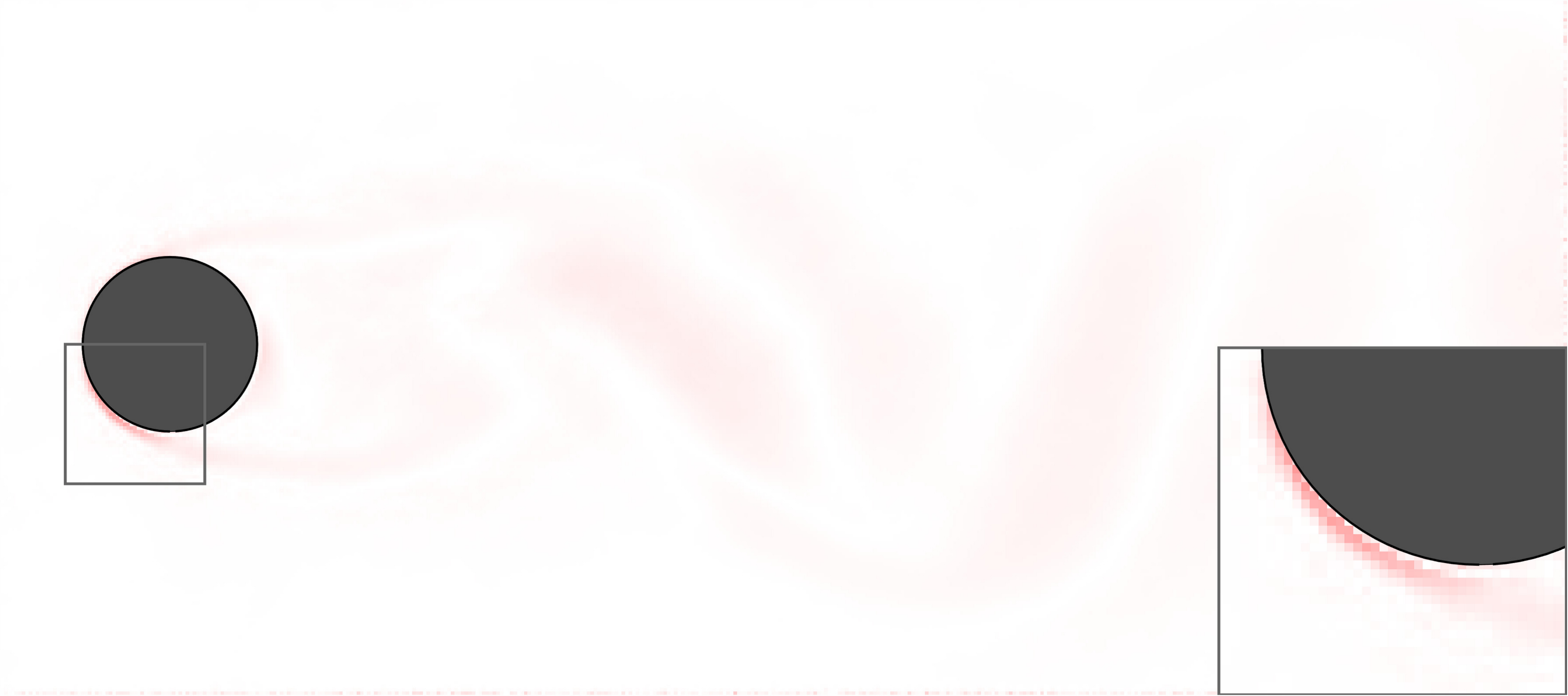}
        \par\vspace{1mm}
        {\footnotesize (i) SpDMD}
        \label{fig:dr_cylinder_spdmd}
    \end{minipage}\hfill
    \begin{minipage}[t]{\imgwidth}
        \centering
        \includegraphics[width=\linewidth]{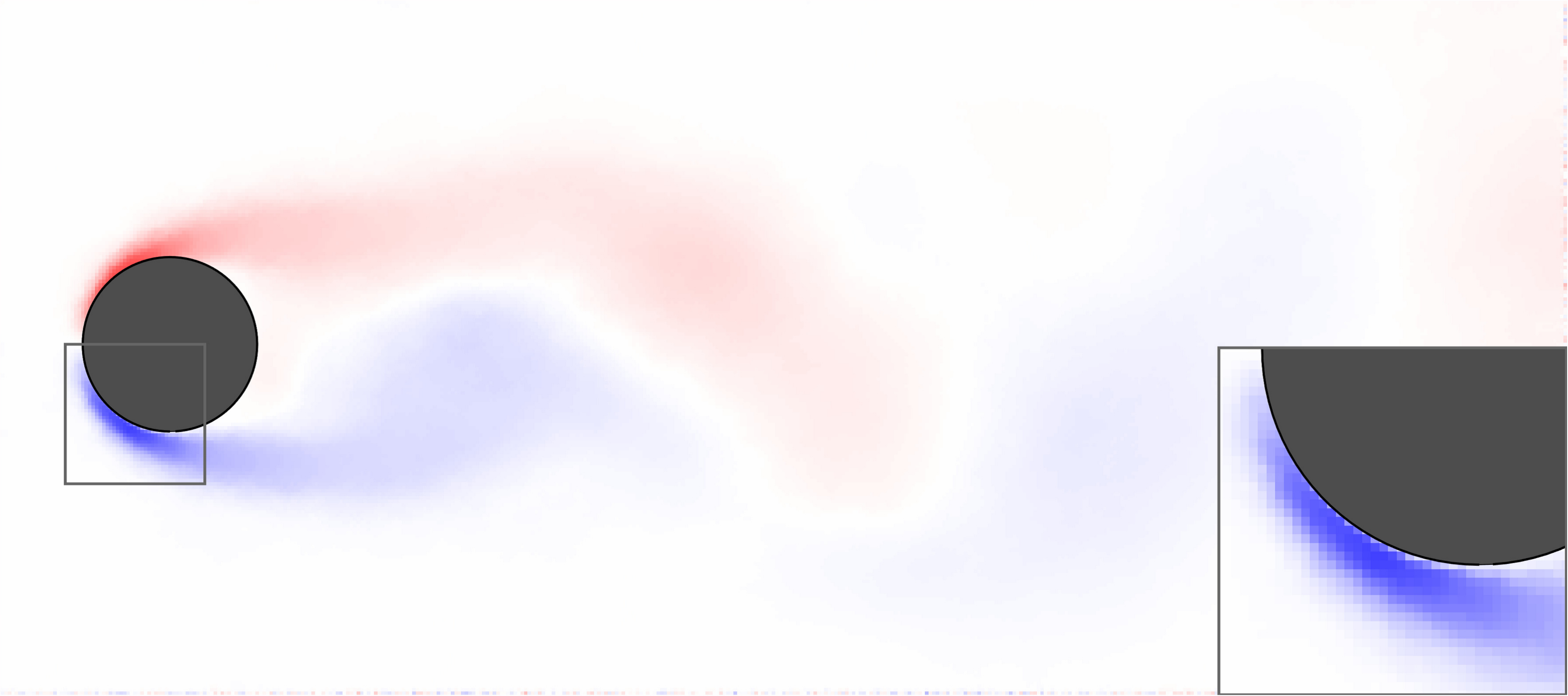}\\[1mm]
        \includegraphics[width=\linewidth]{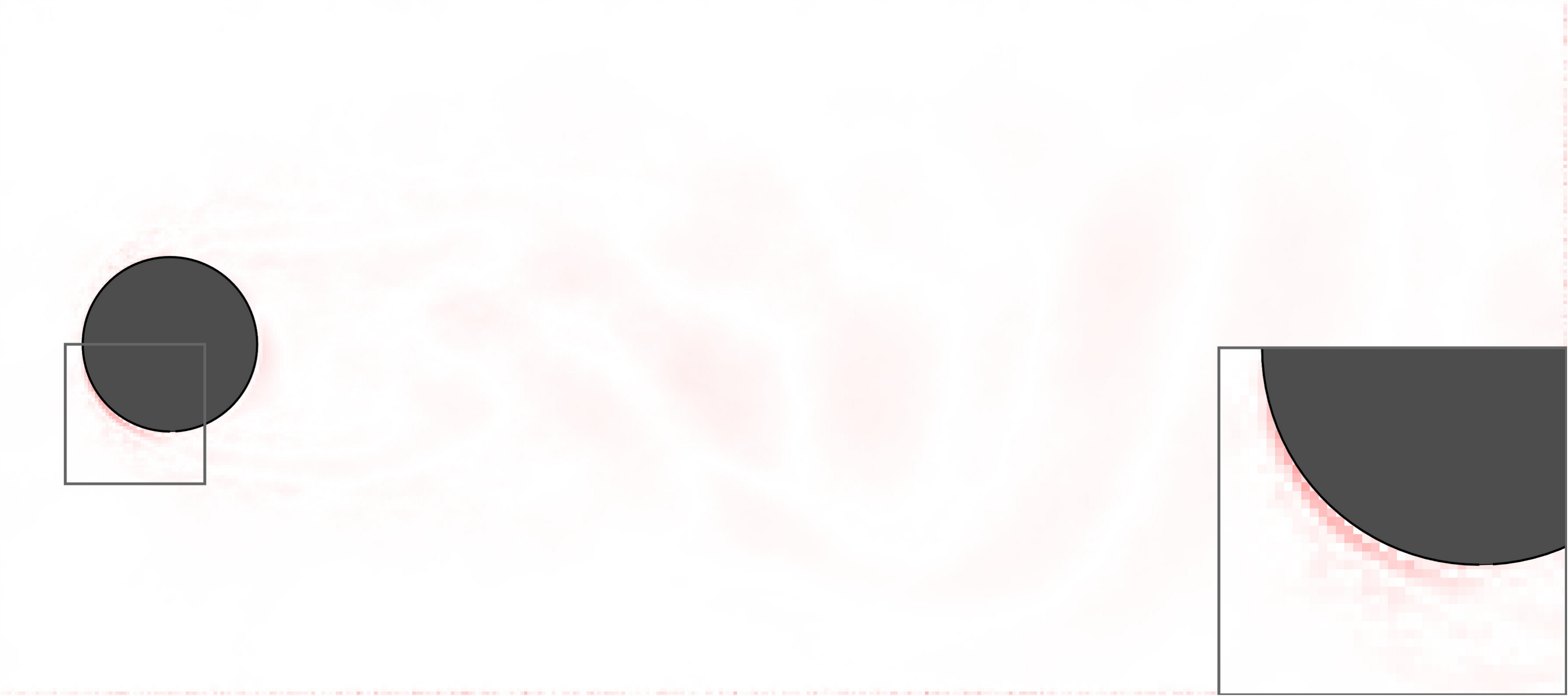}
        \par\vspace{1mm}
        {\footnotesize (j) CR-DMD (ours)}
    \label{fig:dr_cylinder_crdmd}
    \end{minipage}\hfill
    \begin{minipage}[t]{0.05\linewidth}
        \centering
        \includegraphics[width=\linewidth]{fig/Dr_cylinder/Colorbar_Recon-eps-converted-to.pdf}\\[1mm]
        \includegraphics[width=\linewidth]{fig/Dr_cylinder/Colorbar_Error-eps-converted-to.pdf}
        \par\vspace{1mm}
    \end{minipage}
    \caption{Dimensional reduction results for Cylinder Wake with 5 modes (the 1st snapshot). In each subfigure, the upper image is the reconstructed vorticity field, and the lower image is the absolute difference between the ground truth and the reconstructed image. The magnified region in the bottom-right corner highlights detailed vortex structures.}
    \label{fig:dr_cylinder}
\end{figure*}
\subsection{Performance on Dimensional Reduction}
\label{subsec:data_reconstruction}
In this subsection, we evaluate the quality of the low-dimensional representations obtained through dimensional reduction under various noise conditions.
For the proposed method, the constraint radii \(\varepsilon\) and \(\eta\) were tuned based on the noise statistics, similar to the mode extraction stage.
For the cylinder wake, the radii were set according to Eq.~\eqref{eq:params_cylinder}.
The parameter \(\alpha\) was set to \(0.86\) for all noise levels, and the balancing parameter \(w\) was set to \(0.9\).
For the channel flow dataset, the sparse noise radius \(\eta\) was set according to Eq.~\eqref{eq:params_channel}, while the data-fidelity radius \(\varepsilon\) followed Eq.~\eqref{eq:params_cylinder}.
The parameter \(\alpha\) was set to \(0.95, 0.92\), and \(0.89\) for low, medium, and high noise levels, respectively.
The balancing parameter \(w\) was set to \(0.3\) for all noise levels.
The stopping criterion of Alg.~\ref{algo_dimensional_reduction} was set as follows:
\begin{align}
  \frac{\|\Amplitude^{(\IndexAlg+1)} - \Amplitude^{(\IndexAlg)}\|_2}{\|\Amplitude^{(\IndexAlg)}\|_2} < 10^{-4}, \frac{\|\outliers^{(\IndexAlg+1)} - \outliers^{(\IndexAlg)}\|_2}{\|\outliers^{(\IndexAlg)}\|_2} < 10^{-4}.
\end{align}

To quantitatively evaluate the quality of the low-dimensional representations, we employed the mean peak signal-to-noise ratio (MPSNR):
\begin{align}
  \mathrm{MPSNR} = \frac{1}{M}\sum_{m=1}^M 10 \log_{10} \frac{N}{\|\dataclean_m - \hat{\dataclean}_m\|_2^2},
\end{align}
and the mean structural similarity index (MSSIM) \cite{wang2004image}:
\begin{align}
  \mathrm{MSSIM} = \frac{1}{M}\sum_{m=1}^M \mathrm{SSIM}(\dataclean_m, \hat{\dataclean}_m),
\end{align}
where \(\dataclean_m\) and \(\hat{\dataclean}_m\) are the snapshots at the \(m\)-th time step of the ground-truth and the estimated low-dimensional representation, respectively.
Generally, higher MPSNR and MSSIM values correspond to better reconstruction performance.
\begin{figure*}[t!]
    \centering
    \newcommand{\imgwidth}{0.16\linewidth} 
    \newcommand{\cbarwidth}{0.04\linewidth} 
    \newcommand{\gap}{1mm}
    \newcommand{\cbarheight}{3.8cm} 

    \begin{minipage}[t]{\imgwidth}
        \centering
        \includegraphics[width=0.99\linewidth]{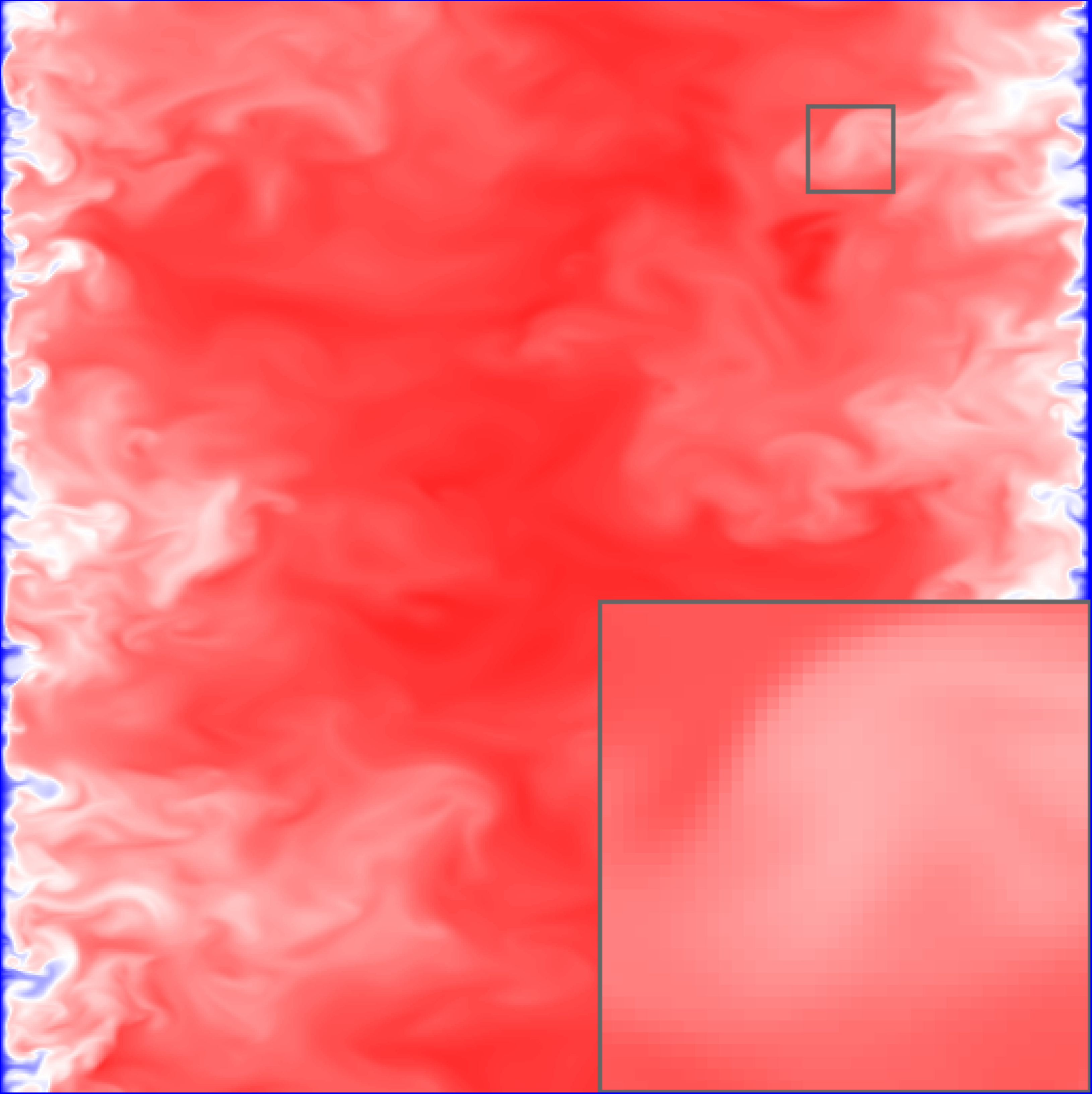}\\[1mm]
        \phantom{\includegraphics[width=0.99\linewidth]{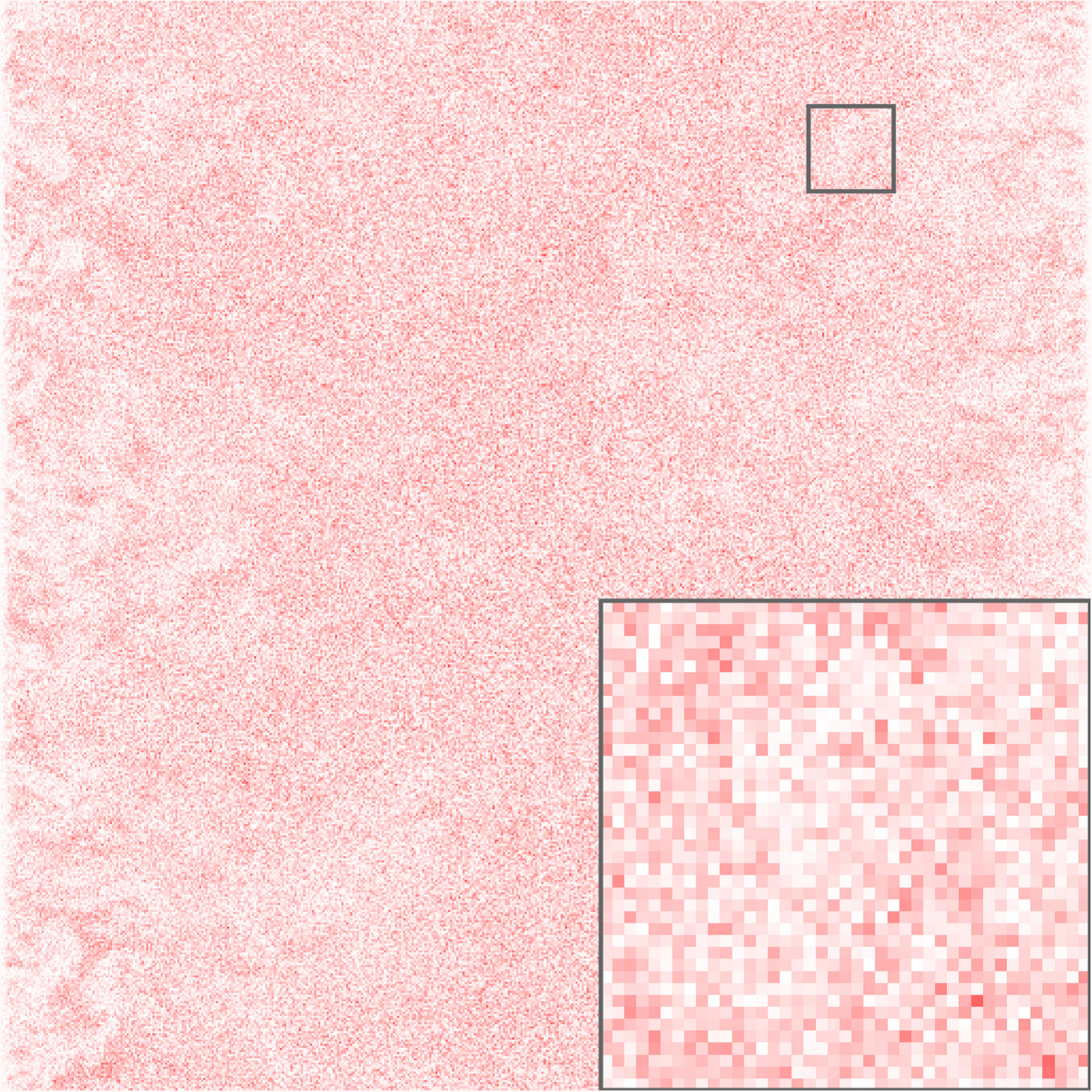}}
        \par\vspace{0mm}
        {\footnotesize (a) Ground-truth}
        \label{fig:dr_channel_gt}
    \end{minipage}\hspace{\gap}
    \begin{minipage}[t]{\imgwidth}
        \centering
        \includegraphics[width=0.99\linewidth]{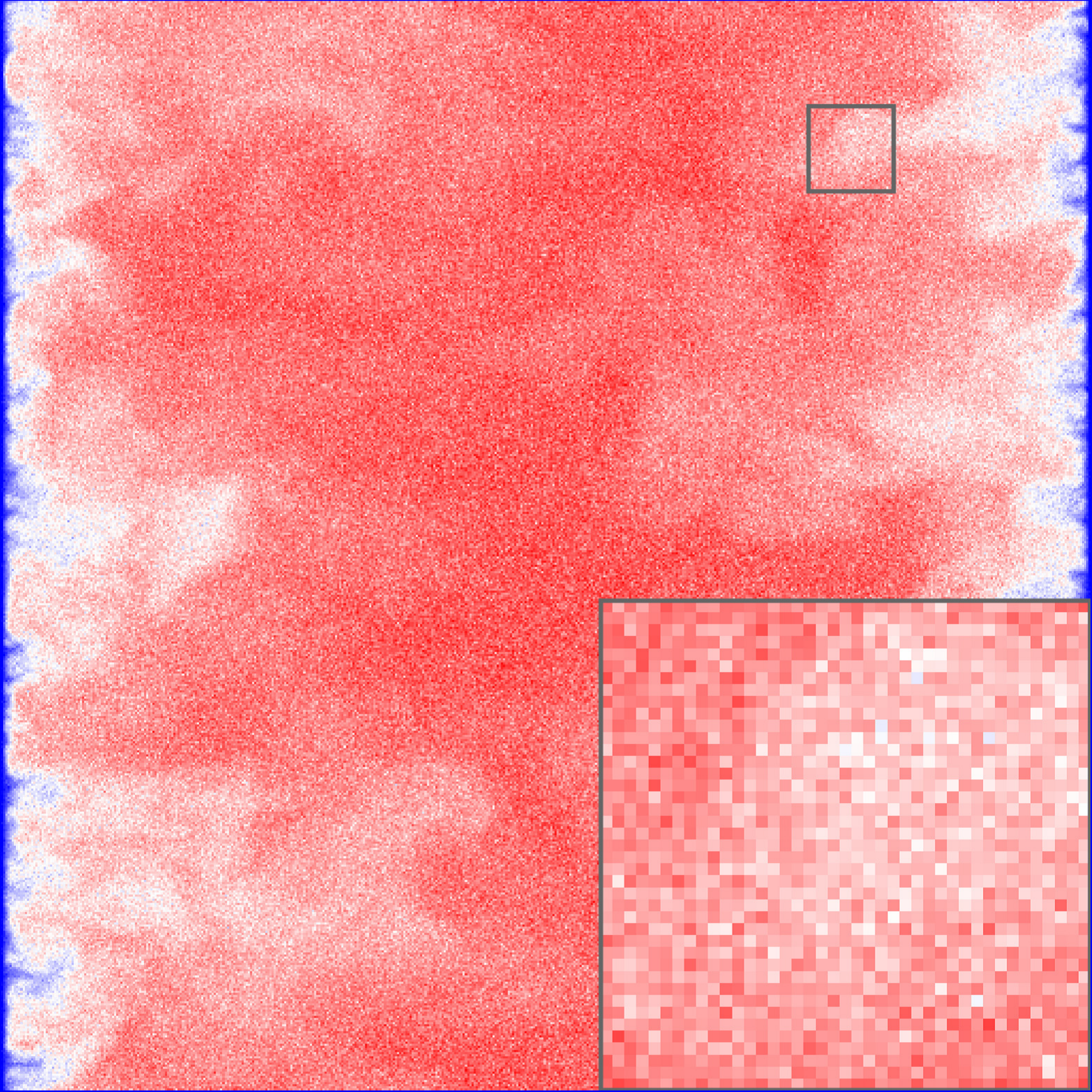}\\[1mm]
        \includegraphics[width=0.99\linewidth]{fig/Dr_channel/TLSDMD_Error-eps-converted-to.pdf}
        \par\vspace{0mm}
        {\footnotesize (b) TLS-DMD}
        \label{fig:dr_channel_tlsdmd}
    \end{minipage}\hspace{\gap}
    \begin{minipage}[t]{\imgwidth}
        \centering
        \includegraphics[width=0.99\linewidth]{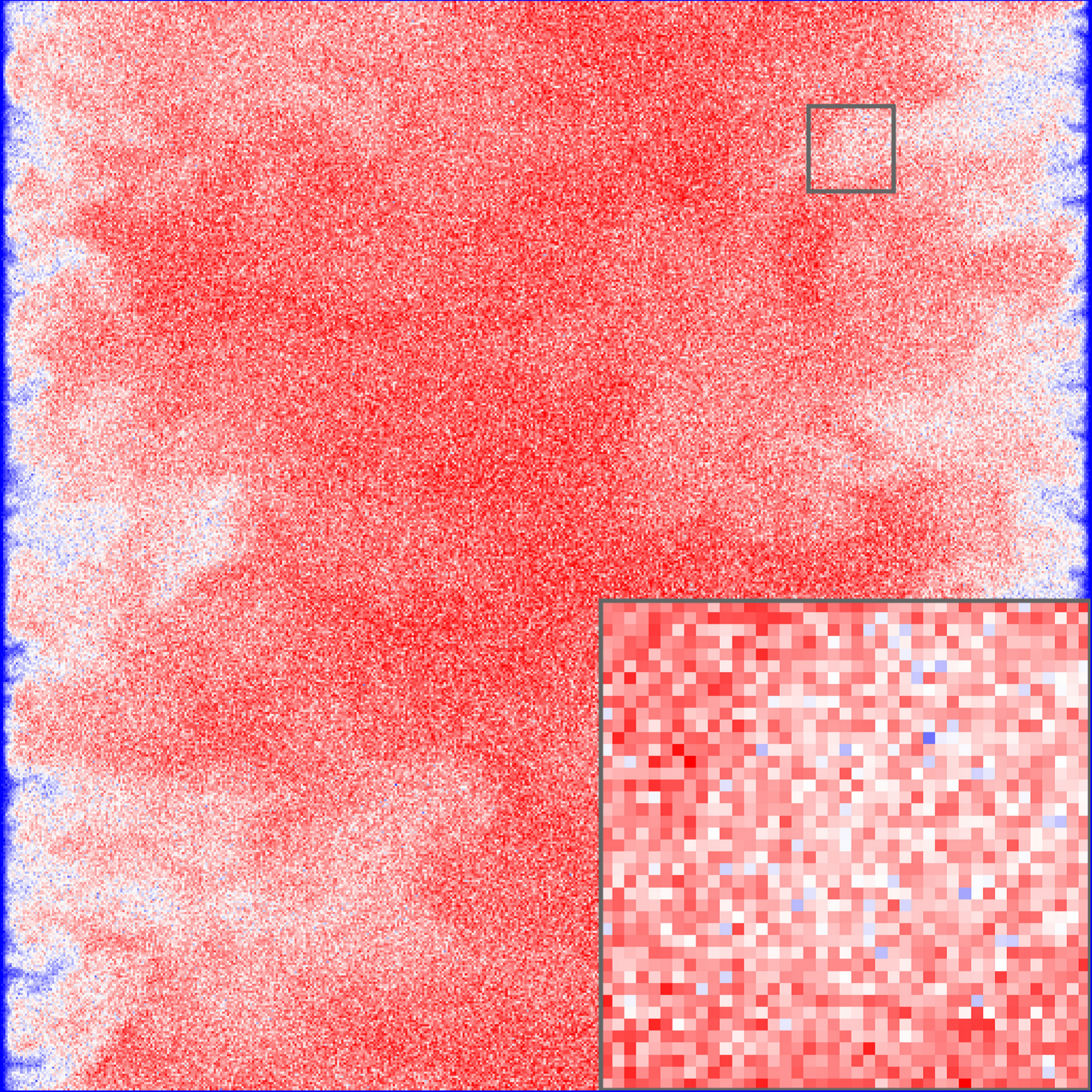}\\[1mm]
        \includegraphics[width=0.99\linewidth]{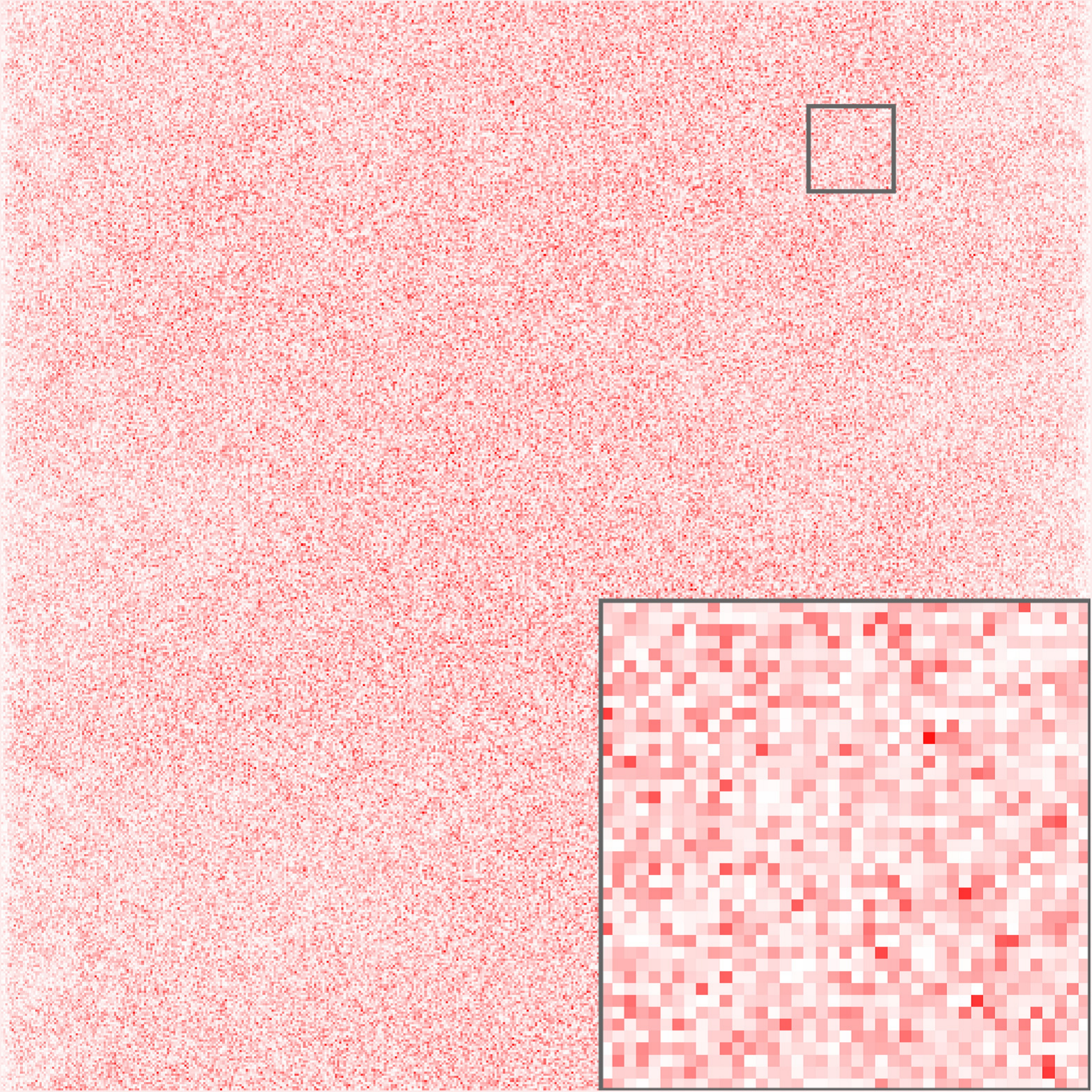}
        \par\vspace{0mm}
        {\footnotesize (c) OptDMD}
        \label{fig:dr_channel_optdmd}
    \end{minipage}\hspace{\gap}
    \begin{minipage}[t]{\imgwidth}
        \centering
        \includegraphics[width=0.99\linewidth]{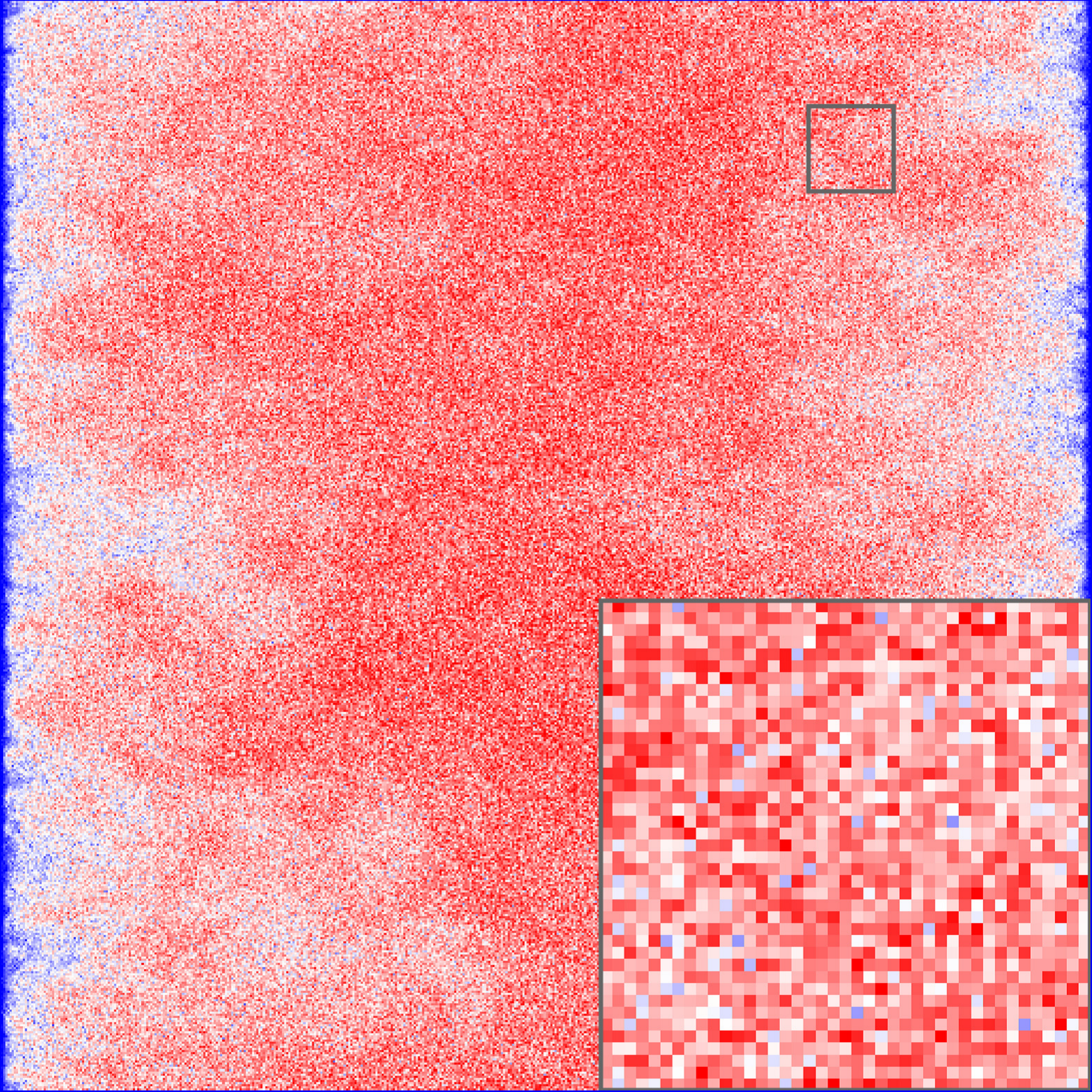}\\[1mm]
        \includegraphics[width=0.99\linewidth]{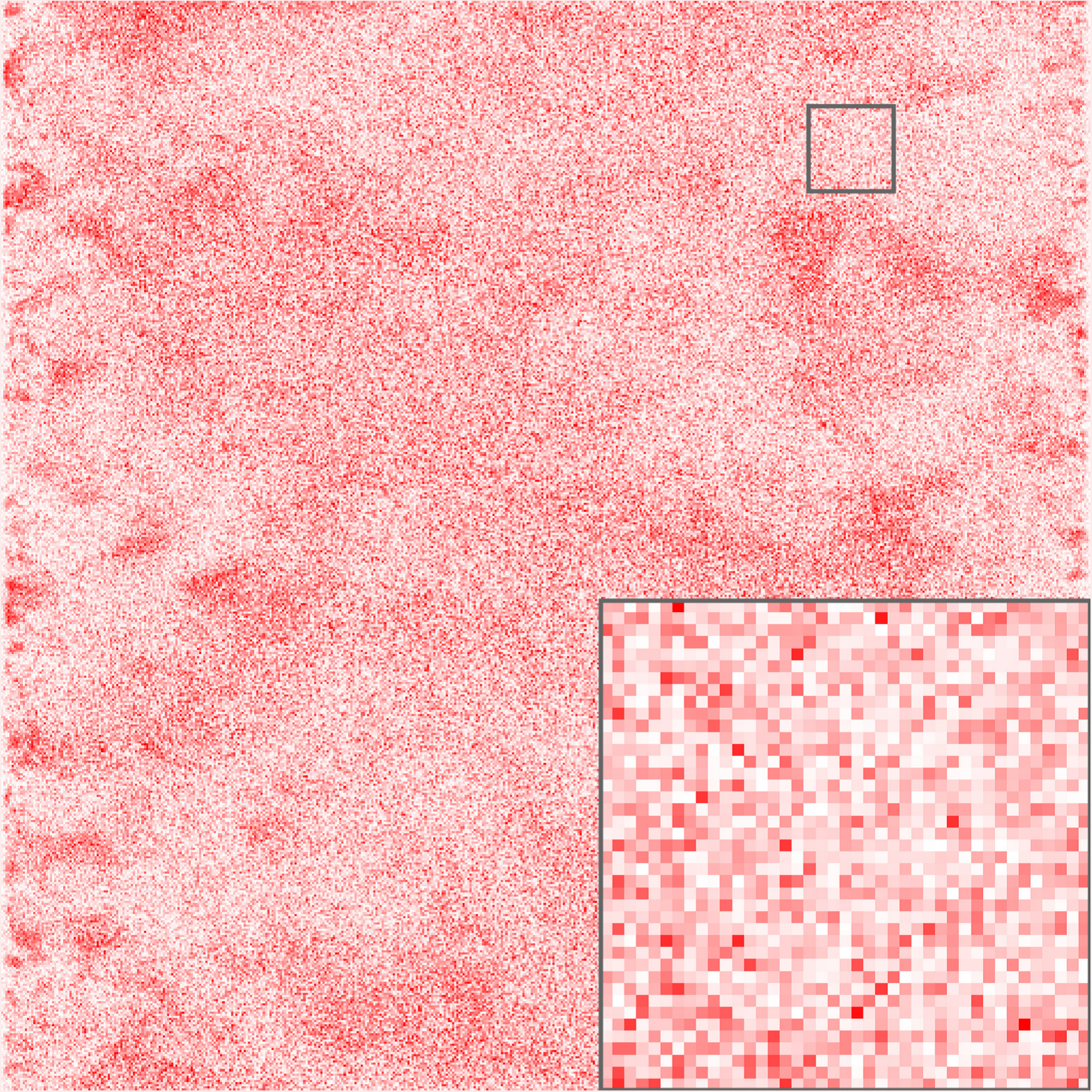}
        \par\vspace{0mm}
        {\footnotesize (d) CDMD}
        \label{fig:dr_channel_cdmd}
    \end{minipage}\hspace{\gap}
    \begin{minipage}[t]{\imgwidth}
        \centering
        \includegraphics[width=0.99\linewidth]{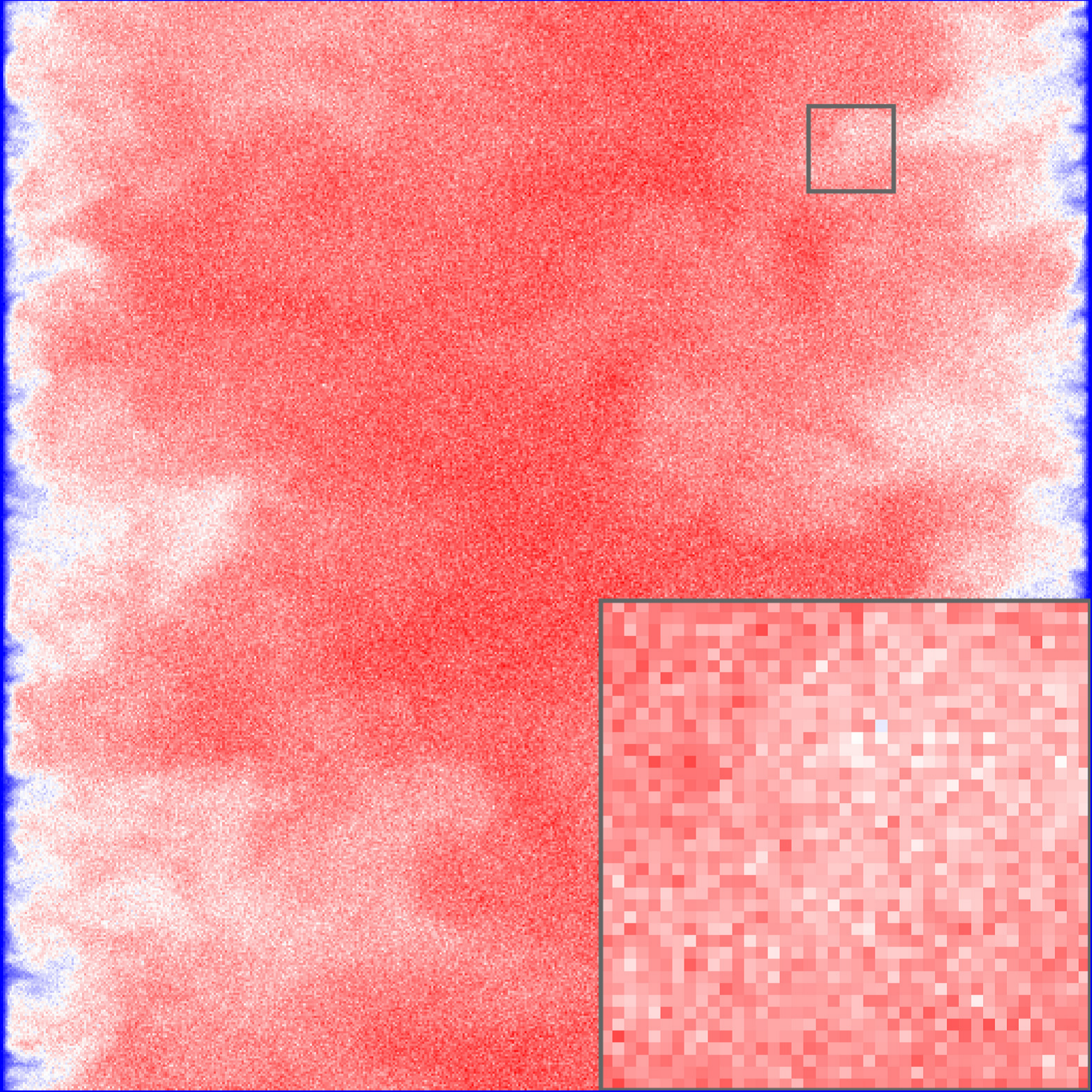}\\[1mm]
        \includegraphics[width=0.99\linewidth]{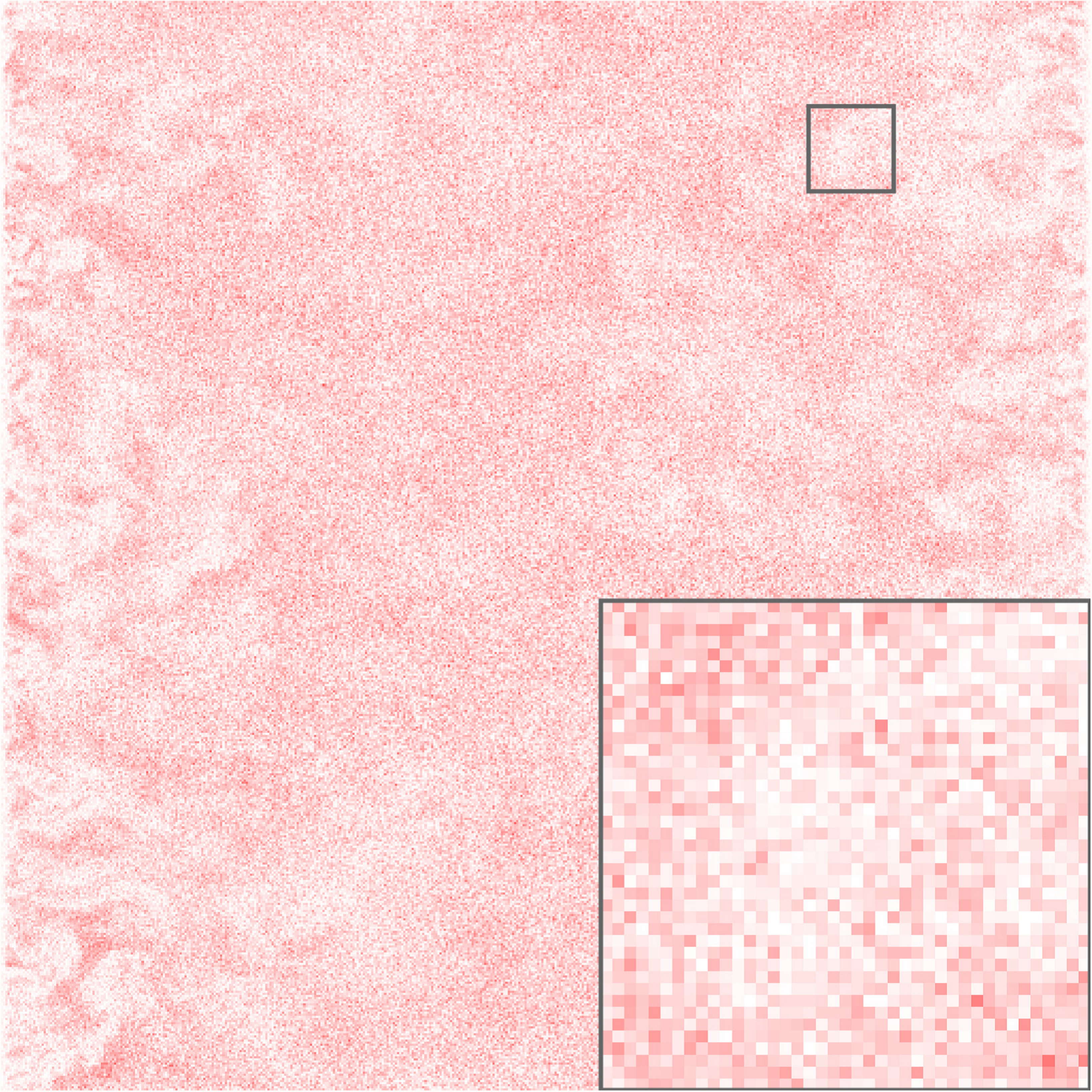}
        \par\vspace{0mm}
        {\footnotesize (e) PiDMD}
        \label{fig:dr_channel_pidmd}
    \end{minipage}\hspace{\gap}
    \begin{minipage}[t]{0.05\linewidth}
        \centering
        \includegraphics[width=\linewidth]{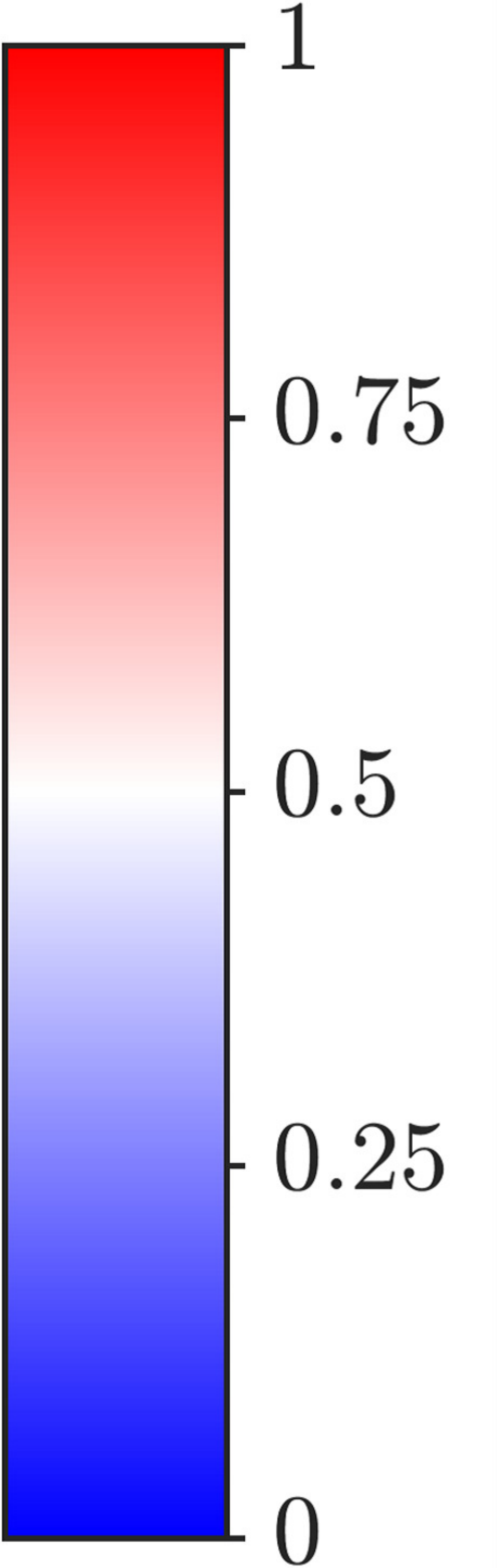}\\[1mm]
        \includegraphics[width=\linewidth]{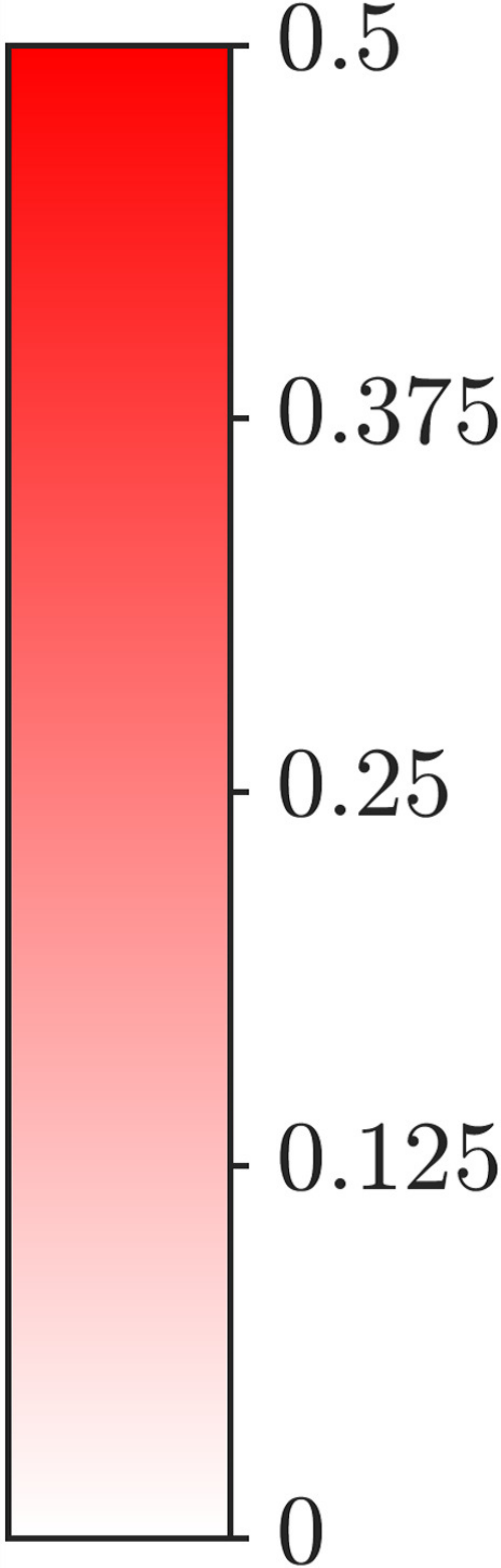}
        \par\vspace{1mm}
    \end{minipage}

    \vspace{0.5em}

    \begin{minipage}[t]{\imgwidth}
        \centering
        \includegraphics[width=0.99\linewidth]{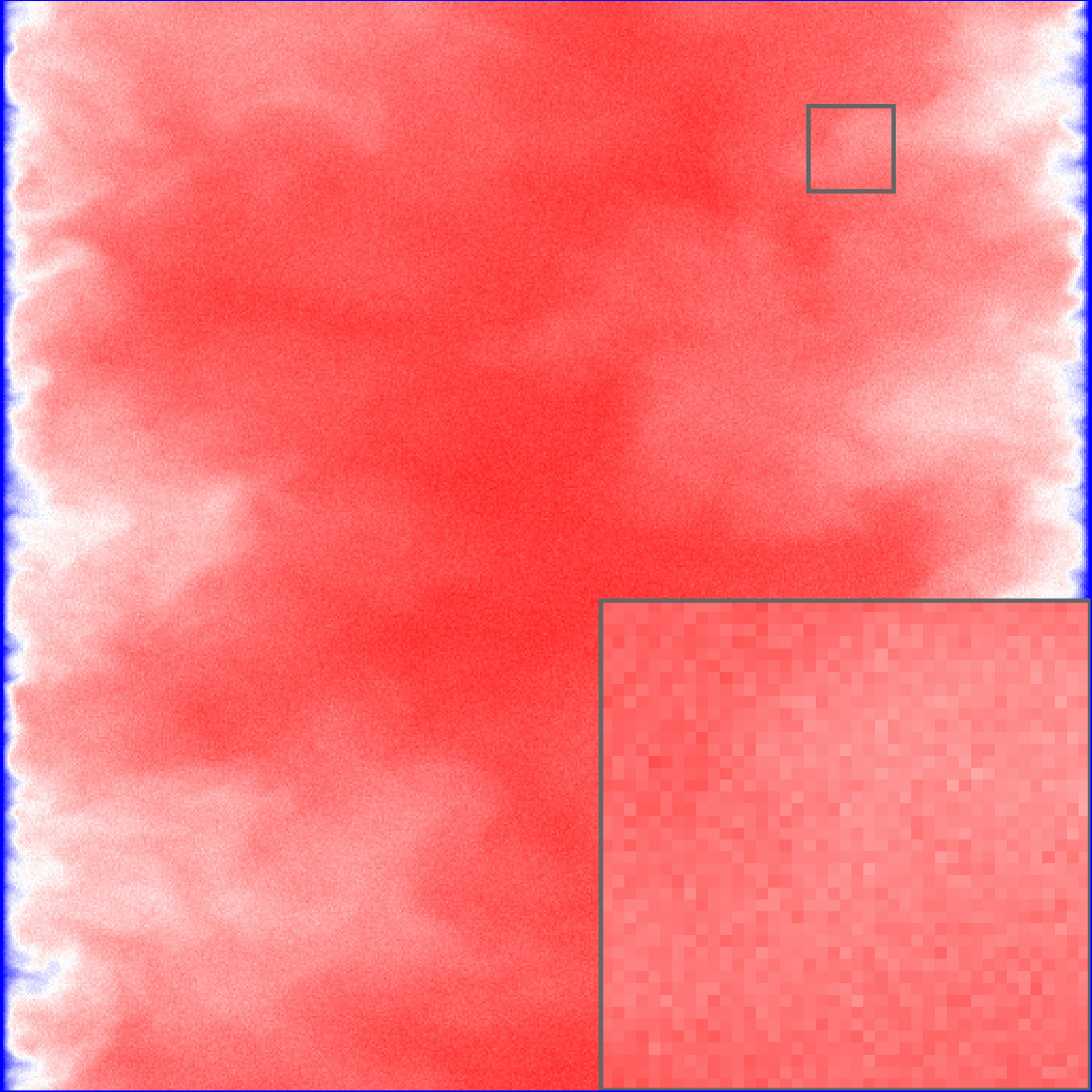}\\[1mm]
        \includegraphics[width=0.99\linewidth]{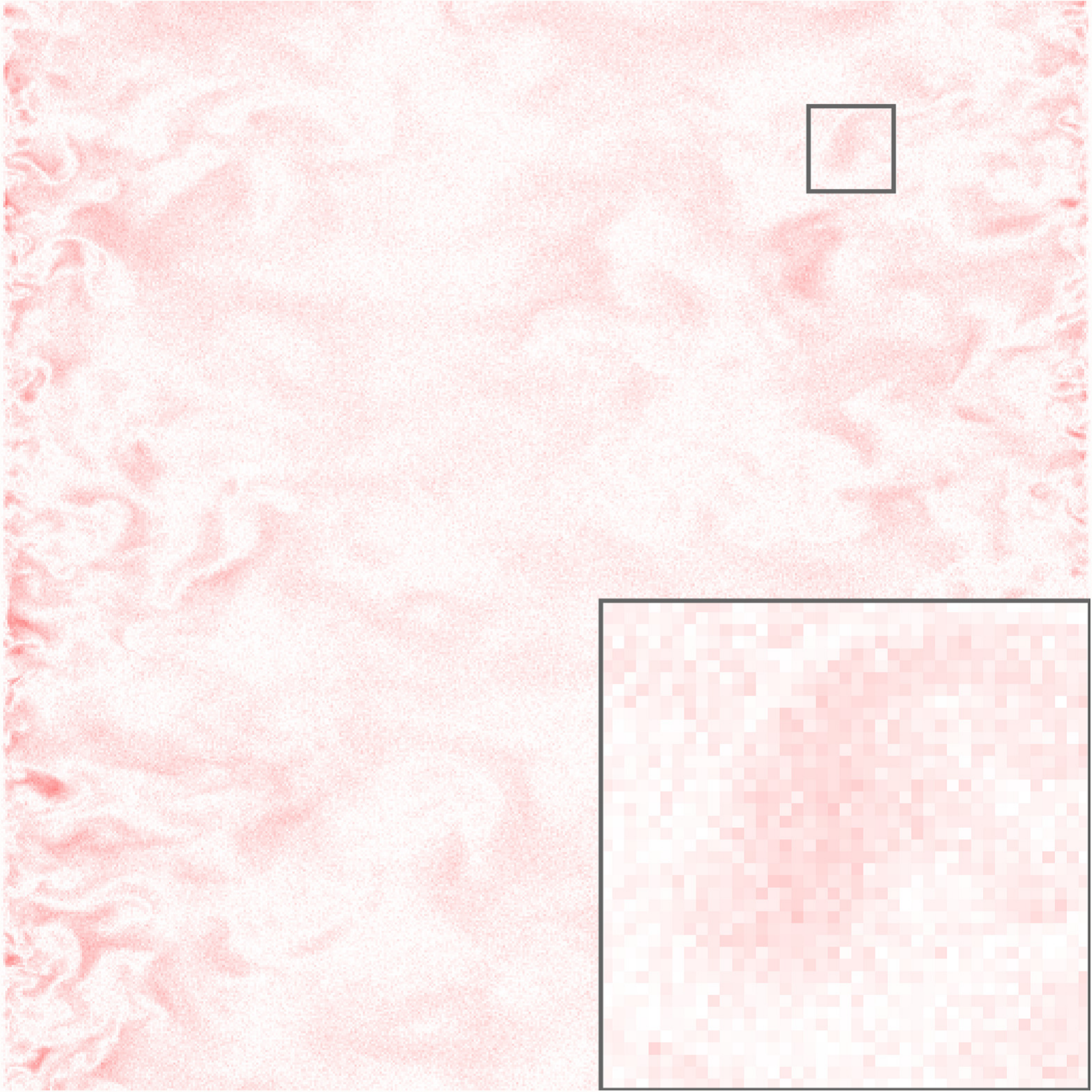}
        \par\vspace{0mm}
        {\footnotesize (f) RPCA}
        \label{fig:dr_channel_rpca}
    \end{minipage}\hspace{\gap}
    \begin{minipage}[t]{\imgwidth}
        \centering
        \includegraphics[width=0.99\linewidth]{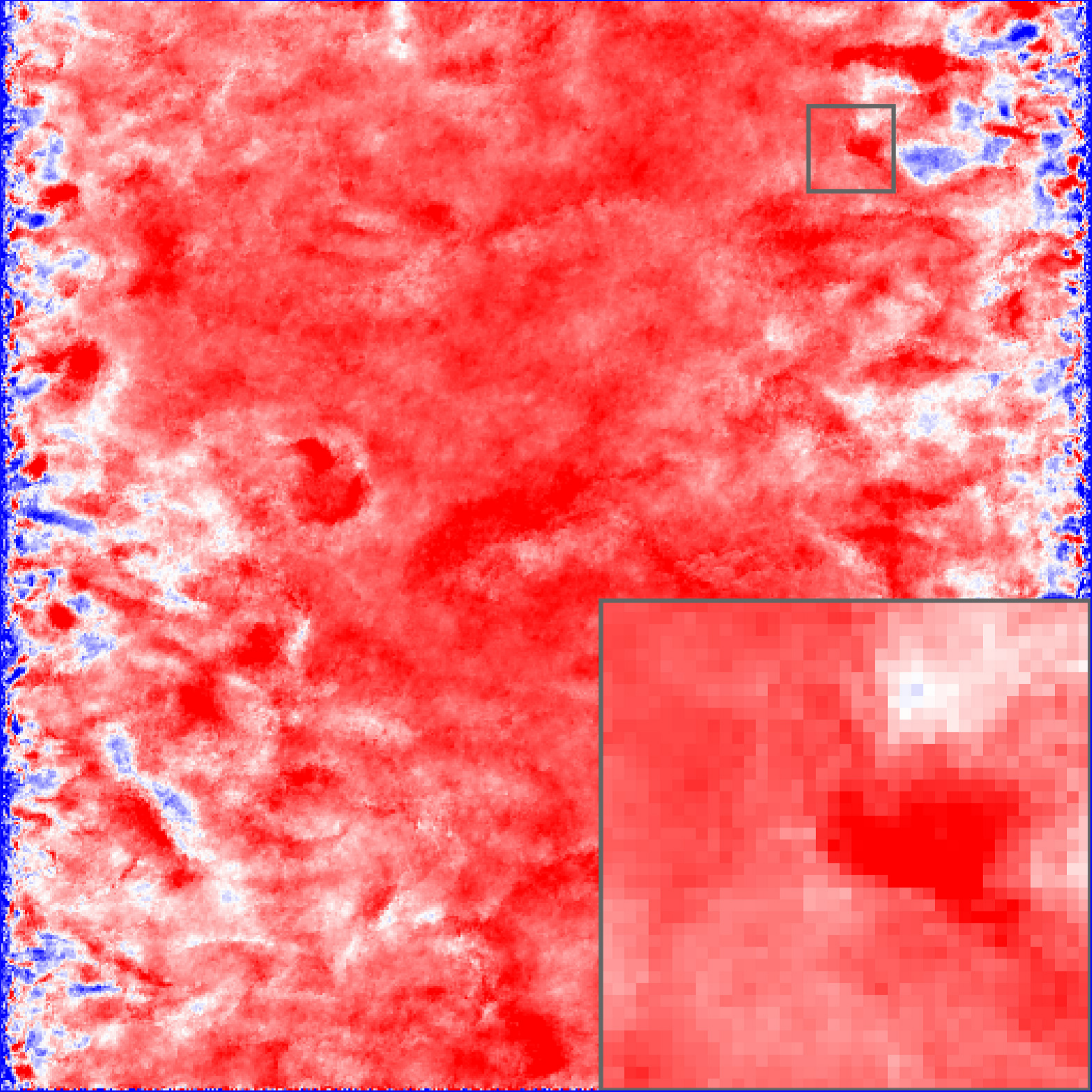}\\[1mm]
        \includegraphics[width=0.99\linewidth]{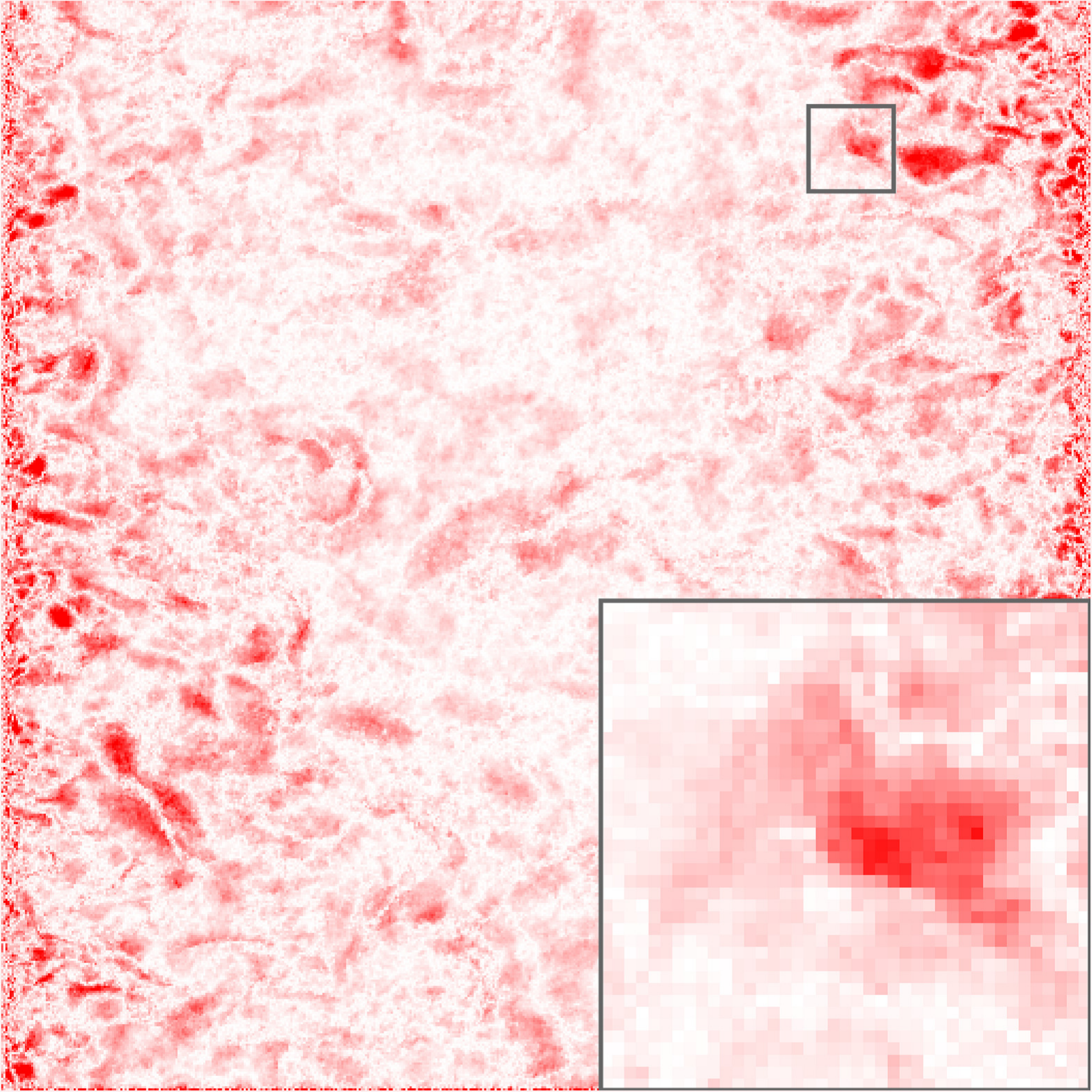}
        \par\vspace{0mm}
        {\footnotesize (g) DMD}
        \label{fig:dr_channel_dmd}
    \end{minipage}\hspace{\gap}
    \begin{minipage}[t]{\imgwidth}
        \centering
        \includegraphics[width=0.99\linewidth]{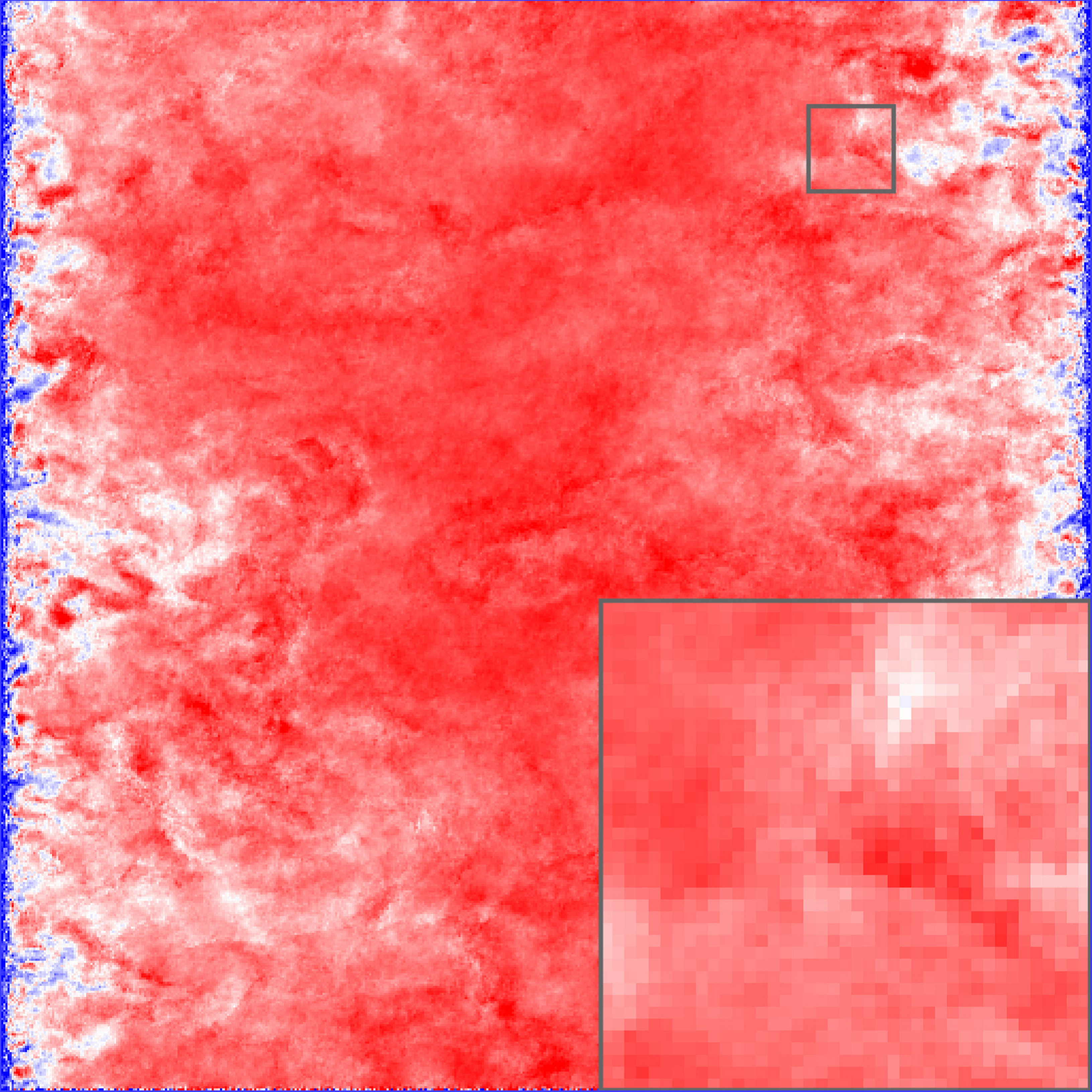}\\[1mm]
        \includegraphics[width=0.99\linewidth]{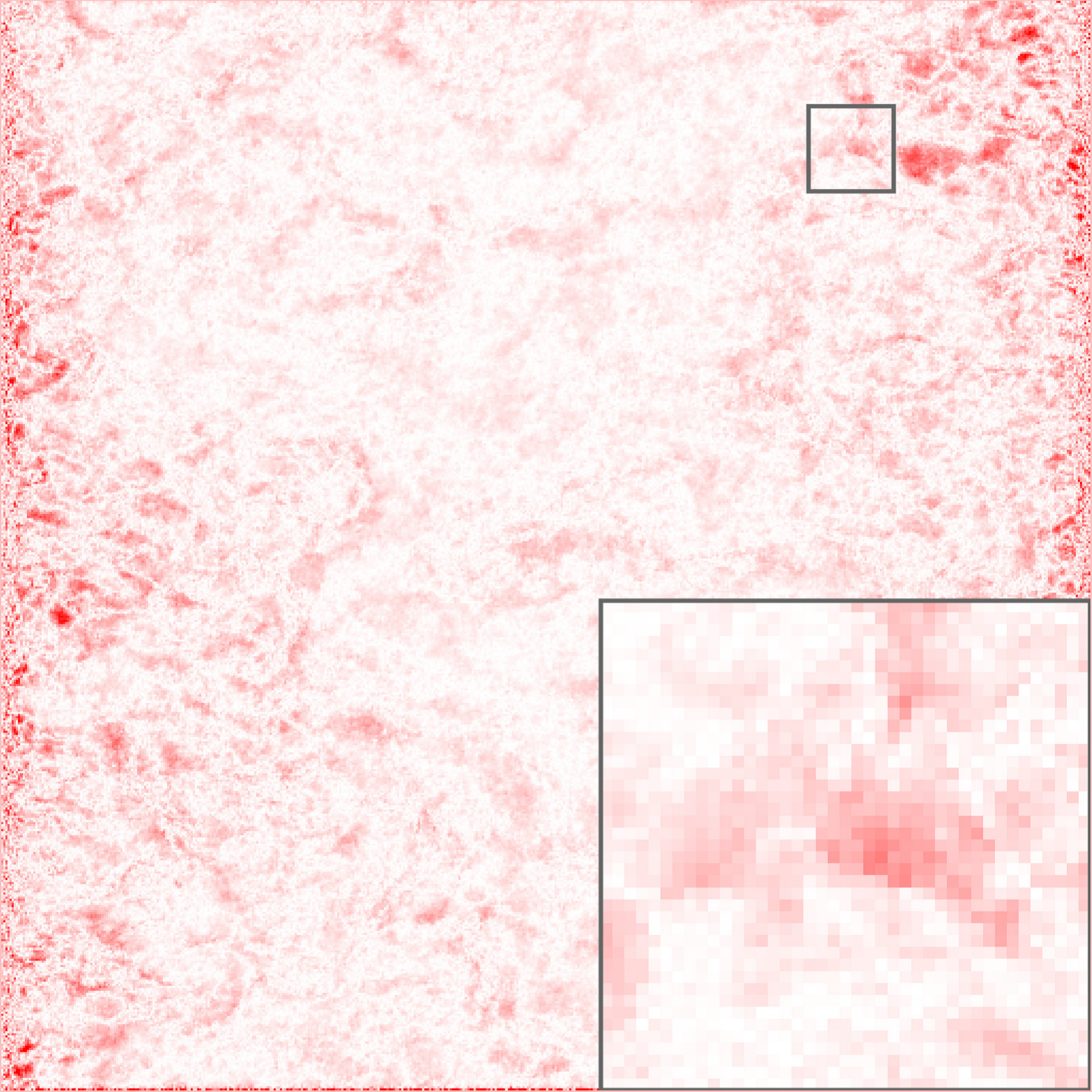}
        \par\vspace{0mm}
        {\footnotesize (h) DMDc}
        \label{fig:dr_channel_dmdc}
    \end{minipage}\hspace{\gap}
    \begin{minipage}[t]{\imgwidth}
        \centering
        \includegraphics[width=0.99\linewidth]{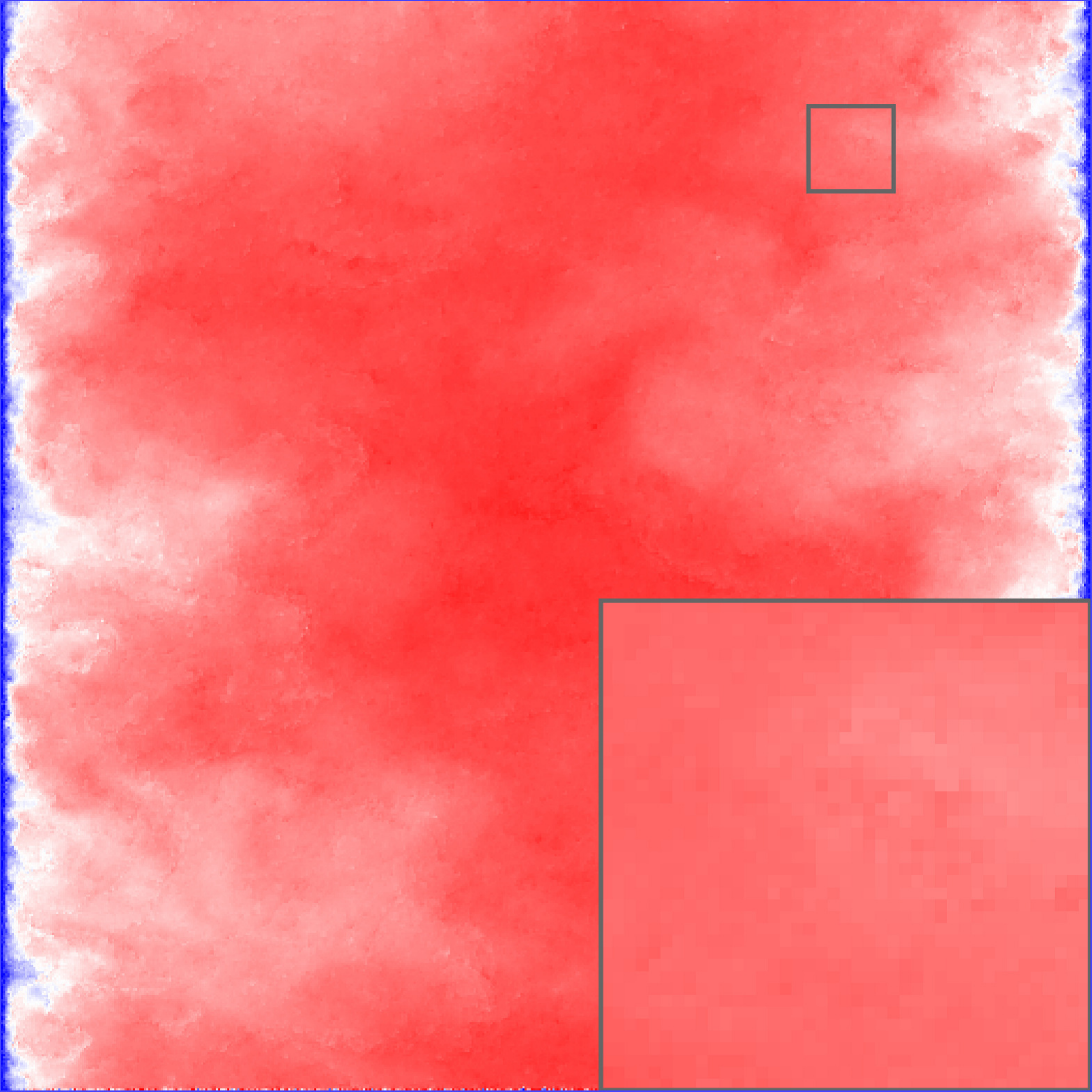}\\[1mm]
        \includegraphics[width=0.99\linewidth]{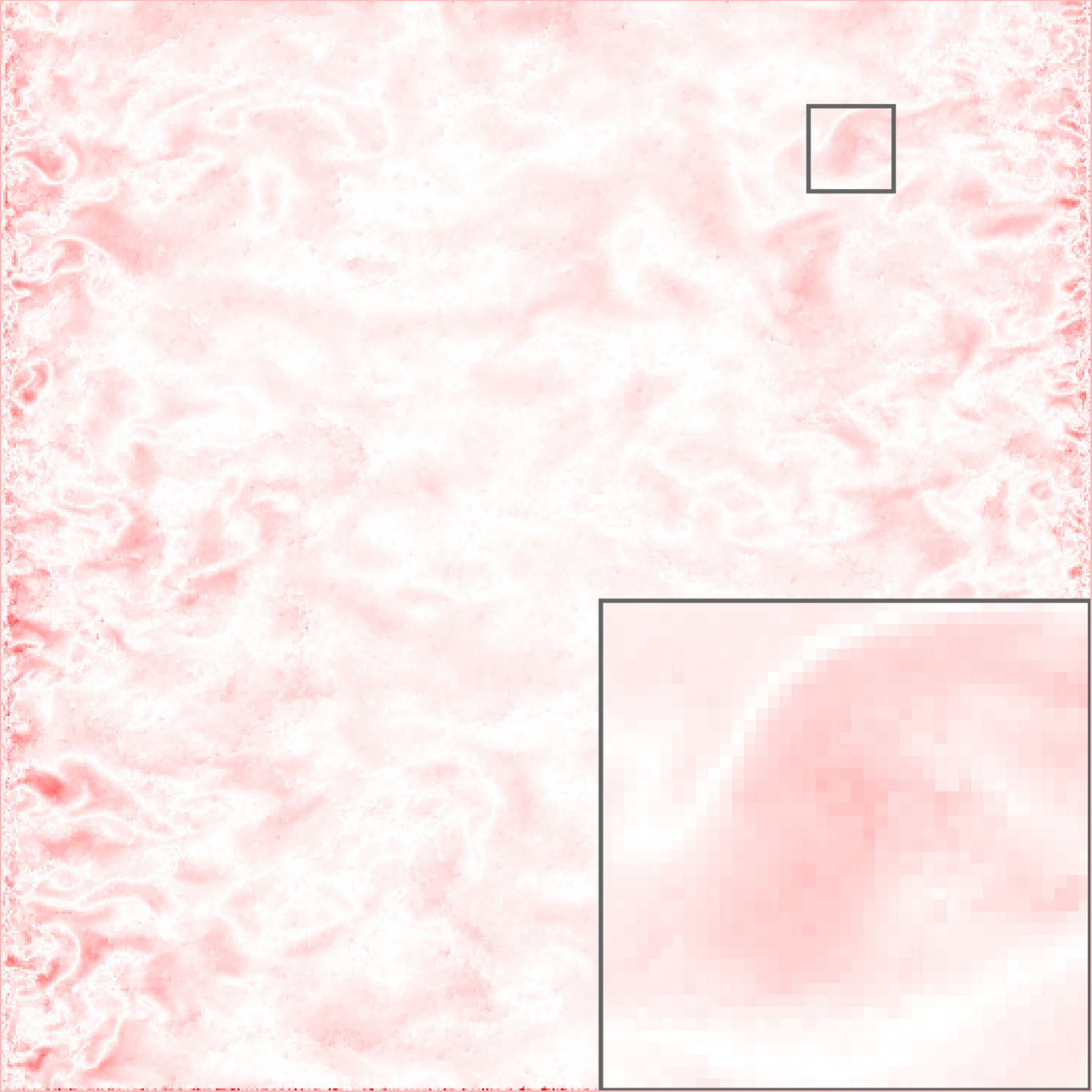}
        \par\vspace{0mm}
        {\footnotesize (i) SpDMD}
        \label{fig:dr_channel_spdmd}
    \end{minipage}\hspace{\gap}
    \begin{minipage}[t]{\imgwidth}
        \centering
        \includegraphics[width=0.99\linewidth]{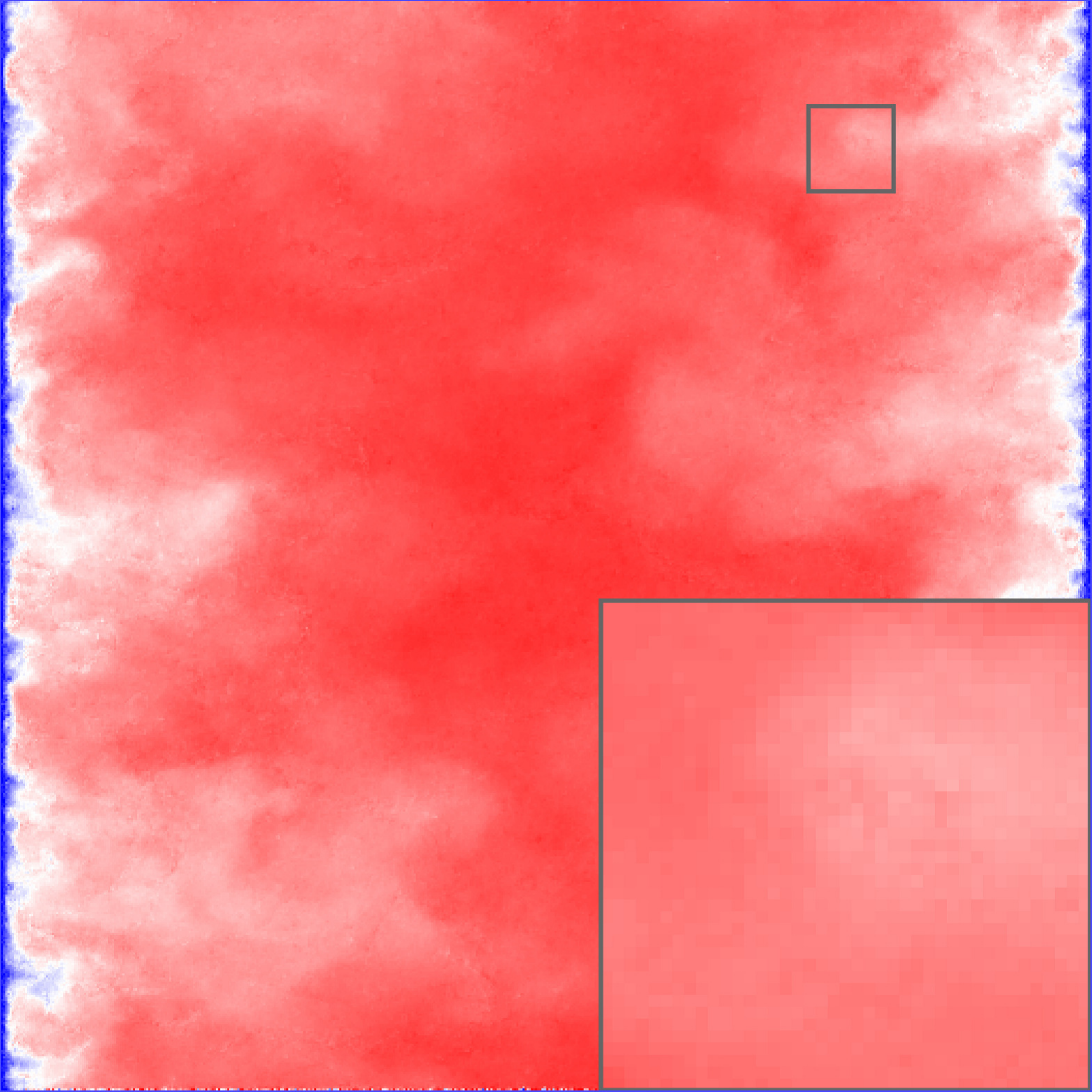}\\[1mm]
        \includegraphics[width=0.99\linewidth]{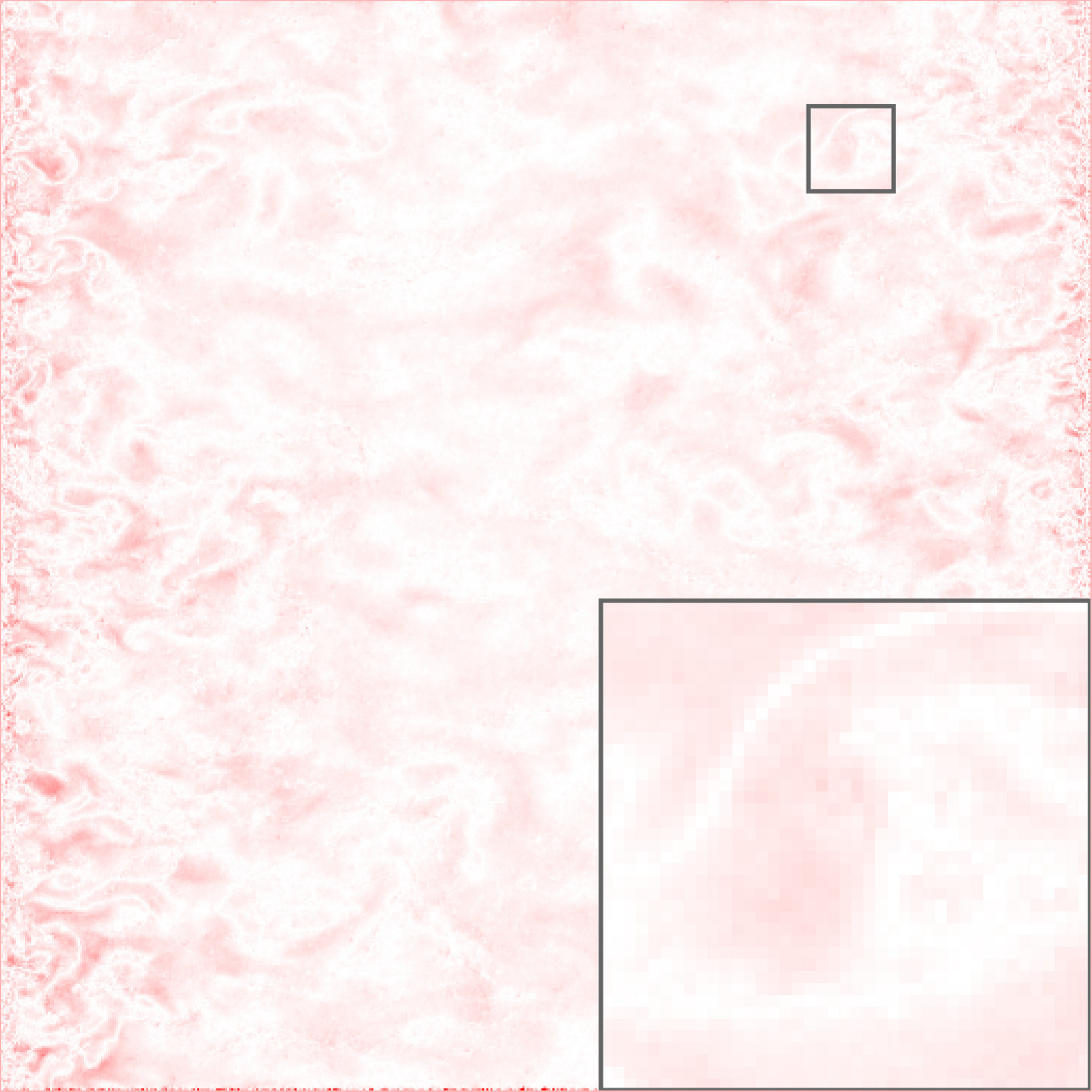}
        \par\vspace{0mm}
        {\footnotesize (j) CR-DMD (ours)}
    \label{fig:dr_channel_crdmd}
    \end{minipage}\hspace{\gap}
    \begin{minipage}[t]{0.05\linewidth}
        \centering
            \includegraphics[width=\linewidth]{fig/Dr_channel/Colorbar_Recon-eps-converted-to.pdf}\\[1mm]
            \includegraphics[width=\linewidth]{fig/Dr_channel/Colorbar_Error-eps-converted-to.pdf}
        \par\vspace{1mm}
    \end{minipage}

    \caption{Dimensional reduction results for Channel Flow with 21 modes (except OptDMD using 39 modes for stability) (the \(10\)-th snapshot). In each subfigure, the upper image is the reconstructed vorticity field, and the lower image is the absolute difference between the ground truth and the reconstructed image. The magnified region in the bottom-right corner highlights detailed vortex structures.}
    \label{fig:dr_channel}
\end{figure*}
\subsubsection{Quantitative Comparison}
\label{subsubsec:tab_dr}
Tables~\ref{table:dr_cylinder} and \ref{table:dr_channel} show MPSNRs and MSSIMs with respect to the number of retained modes in the dimensional reduction experiments for the cylinder wake and turbulent channel flow datasets contaminated by mixed noise, respectively.
The best and second results are highlighted in bold and underlined, respectively.
In both datasets, the operator-centric methods generally exhibit limited performance across all noise levels and numbers of retained modes.
The RPCA results show high MPSNRs in low noise for the channel flow. However, its performance degrades
when the noise level increases especially in the channel flow dataset.
Although standard DMD and DMDc share the same amplitude calculation mechanism, they yield different rankings.
Both methods, however, result in relatively low reconstruction accuracy when only a small number of modes are utilized.
SpDMD improves upon standard DMD and DMDc by promoting sparsity in mode selection.
As for our method, CR-DMD consistently achieves the highest MPSNRs and MSSIMs across almost all conditions.
This can be attributed to the fact that other dimensional reduction techniques rely solely on preprocessed data.
In contrast, the proposed method leverages the information-rich noisy observation data directly, incorporating
the structures that may have been lost during preprocessing.

\subsubsection{Visual Quality Comparison for Dimensional Reduction}
\label{subsubsec:fig_dr}
Figs.~\ref{fig:dr_cylinder} and \ref{fig:dr_channel} show the low-dimensional representations with the first snapshot by each method for the cylinder wake and turbulent channel flow datasets, respectively, in the medium noise intensity case.
The results of the operator-centric methods in (b), (c), (d), and (e) are heavily contaminated by noise, failing to recover the underlying structures of the fluid flows.
In contrast, preprocessing-based methods in (f), (g), (h), and (i) yield representations that are visually closer to the noise-free ground-truth.
However, for RPCA, the magnified region in Fig.~\ref{fig:dr_cylinder} indicates that fine-scale structures with high vorticity magnitude are not fully captured, and Gaussian noise persists in the reconstructed field. Regarding DMD and DMDc, Fig.~\ref{fig:dr_channel} highlights that they misrepresent detailed vortex structures. Similarly, SpDMD exhibits deviations in the detailed regions, showing relatively large errors in the magnified absolute difference plot in Fig.~\ref{fig:dr_channel}. Unlike these methods, CR-DMD in (j) achieves the most faithful representation of the data. It captures fine-scale structures that are over-smoothed by other preprocessing-based methods. This demonstrates the effectiveness of explicitly referencing the original observation data to recover detailed information lost during preprocessing.

\section{Conclusion}
\label{sec:conclusion}
We have proposed a new robust DMD framework, named CR-DMD, for mode extraction and dimensional reduction of fluid flow data corrupted by mixed noise.
CR-DMD is composed of two stages: preprocessing for noise removal to ensure accurate mode extraction, and dimensional reduction to explicitly link the modes to the original noisy observations for a faithful low-dimensional representation.
In both stages, we have formulated the denoising and dimensional reduction problems as convex optimization problems, and developed the optimization algorithms based on P-PDS.
Experiments on two fluid flow datasets contaminated by mixed noise have demonstrated the superiority of CR-DMD over existing methods. For future work, we will develop a unified framework that incorporates the proposed spatial regularization strategies into operator-centric mode extraction methods to further enhance their robustness and applicability.

\bibliographystyle{IEEEtran}
% Generated by IEEEtran.bst, version: 1.14 (2015/08/26)

\end{document}